\documentclass[apjl]{emulateapj}
\usepackage{apjfonts}

\newcommand{\Mo}{\ensuremath{\mathrm{M}_\sun}}
\newcommand{\Ro}{\ensuremath{\mathrm{R}_\sun}}
\newcommand{\Rs}{\ensuremath{r_\mathrm{S}}}
\newcommand{\rg}{\ensuremath{r_\mathrm{g}}}
\newcommand{\ths}{\ensuremath{\theta_\mathrm{S}}}
\newcommand{\muas}{\ensuremath{\mu\mathrm{as}}}

\begin{document}

\submitted{Submitted on October 27, 2008}
\journalinfo{Submitted to The Astrophysical Journal}

\title{Prospects for testing the nature of Sgr~A*'s NIR flares on the basis of current VLT- and future VLTI-observations}

\author{N.~Hamaus\altaffilmark{1}, T.~Paumard\altaffilmark{1,2}, T.~M\"uller\altaffilmark{3}, S.~Gillessen\altaffilmark{1}, F.~Eisenhauer\altaffilmark{1}, S.~Trippe\altaffilmark{1}, and R.~Genzel\altaffilmark{1,4}}

\altaffiltext{1}{Max-Planck-Institut f\"ur Extraterrestrische Physik (MPE), Postfach 1312, D-85741 Garching, Germany}
\altaffiltext{2}{Laboratoire d'Etudes Spatiales et d'Instrumentation en Astrophysique (LESIA), Observatoire de Paris, CNRS, UPMC, Universit\'e Paris Diderot, 5 Place Jules Janssen, 92190 Meudon, France}
\altaffiltext{3}{Universit\"at Stuttgart, Visualisierungsinstitut der Universit\"at Stuttgart (VISUS), D-70569 Stuttgart, Germany}
\altaffiltext{4}{Department of Physics, University of California, CA 94720, Berkeley, USA}

\begin{abstract}
Sagittarius~A*, the supermassive compact object at the center of the Galaxy, exhibits outbursts in the near infrared and X-ray domains. These \emph{flares} are likely due to energetic events very close to the central object, on a scale of a few Schwarzschild radii. Optical interferometry will soon be able to provide astrometry with an accuracy of this order ($\simeq10$~\muas). In this article we use recent photometric near infrared data observed with the adaptive optics system NACO at the Very Large Telescope combined with simulations in order to deploy a method to test the nature of the flares and to predict the possible outcome of observations with the Very Large Telescope Interferometer. To accomplish this we implement a \emph{hot spot model} and investigate its appearance for a remote observer in terms of light curves and centroid tracks, based on general relativistic ray tracing methods. First, we use a simplified model of a small steady source in order to investigate the relativistic effects qualitatively. A more realistic scenario is then being developed by fitting our model to existing flare data. While indications for the spin of the black hole and multiple images due to lensing effects are marginal in the light curves, astrometric measurements offer the possibility to reveal these high-order general relativistic effects. This study makes predictions on these astrometric measurements and leads us to the conclusion that future infrared interferometers will be able to detect proper motion of hot spots in the vicinity of Sagittarius~A*.
\end{abstract}


\section{INTRODUCTION}

The center of our Galaxy is by  far the closest galactic nucleus to Earth, about $100$  times closer than the center of Andromeda. The radio source Sagittarius~A* (Sgr~A*) lies motionless \citep[with a motion constraint of less than $45$~km/s,][]{reid07} at the center of gravity of the nuclear cluster. It has  variable counterparts  in the infrared and  X-ray domains \citep{baganoff01,Porquet03,genzel03nat,eckart04}. Dynamical evidence shows that this source is powered by a supermassive object, most likely a black hole \citep{schoedel02,schoedel03,ghez03,eisenhauer05sinfoni}. Its mass $M_0=3.61\pm0.32\times10^6\:\Mo$ and distance $R_0=7.62\pm0.32$~kpc can be deduced from the motion of the innermost stars, which are dominated by a point mass potential \citep{eisenhauer05sinfoni}. The corresponding Schwarzschild radius is  given by $\Rs=2\mathcal  G  M /c^2\simeq15\:\Ro$, yielding an apparent size of $\ths=\Rs/R_0\simeq9.3$~\muas\ on sky. In contrast,  for a  stellar black hole ($M\simeq1\Mo$) at  $1$~kpc, $\ths\simeq2\times10^{-5}$~\muas. Indeed,  Sgr~A* is  the  black hole candidate with the largest apparent  size,  and therefore  the candidate  of choice for resolving effects at the Schwarzschild scale.

In the near infrared (NIR), Sgr~A* is studied using 10m-class telescopes such as the ESO  Very Large Telescope  (VLT) and  the Keck  telescopes with a diffraction limit of a few 10~mas. Both the VLT and Keck feature several telescopes on the same site (4 and 2, respectively).  They offer an interferometric mode in which the  light collected  by two  or more  of these  telescopes can be combined  in a coherent fashion, allowing phase-referenced measurements to be conducted. Used with a proper data acquisition  and  reduction schemes,  these  interferometers can  deliver reconstructed images  with resolutions as good as a few mas.  Even in the case of Sgr~A*, this  is  still  two orders  of  magnitude larger than \Rs. However,  these interferometers can  also provide  differential astrometry between  the target and a reference source with a very high accuracy. The \emph{Phase-Referenced Imaging and Micro-arcsecond Astrometry} \citep[PRIMA,][]{glindemann00} facility on the VLT Interferometer  (VLTI)  will  reach  $10$~\muas\  accuracy  for  a  typical observational   setup:   $t=30$~min   integration   time,  a   separation   of $\Theta=10\arcsec$  between  the reference  star  and  the  science object,  and  a (projected) baseline $b=200$~m.  This accuracy scales as \citep{shao92}:
\begin{equation}
\sigma\propto b^{-2/3}\Theta\, t^{-1/2} \label{eq:accuracy scaling}
\end{equation}

We have started  studying the possible outcome of  high-accuracy astrometry of matter very close to  Sgr~A* as part of the PRIMA  Reference Missions ESO document \citep{primaRM}. Further study \citep{paumard05vltiws} has led us to propose a second-generation  instrument  for  the   VLTI: \mbox{GRAVITY},  a  general purpose instrument optimized for the Galactic Center (GC) science cases \citep{eisenhauer05gravity}. The GRAVITY-consortium has completed a phase A study for this instrument, which was recommended to ESO by the scientific advisory committee in late 2007.

\subsection{Observational facts about Sgr~A*}

Sgr~A* is very faint compared to its Eddington luminosity \citep[$L\la10^{-8}L_{\mathrm{Edd}}$,][]{baganoff03}. However, it shows variability  at  various  timescales. In particular, it  exhibits outbursts  of energy called \emph{flares} in both X-rays \citep[e.g.][]{baganoff01,baganoff03} and NIR \citep[e.g.][]{genzel03nat,ghez04,clenet05}. These flares typically last for a few hours and occur several times a day. On these occasions,  the luminosity of Sgr~A* increases by a factor of up to $100$ in X-rays and up to $20$ in K-band, making Sgr~A*  a $m_K\simeq15$ source  \citep{genzel03nat}. Superposed upon these bright, hour-scale flares, quasi-periodic variations with a typical timescale around $20$~min are observed in the NIR \citep{genzel03nat,eckart06,trippe07}.

\citet{dodds-eden08} and \citet{Porquet08} observed a bright flare from the GC simultaneously in NIR and X-ray on April 4, 2007. The power spectrum of the NIR light curve gives indication for the existence of a quasi-periodic signal at about $23$~min. This characteristic timescale can be interpreted as the orbital period
\begin{equation}
P = \frac{2\pi \, \rg}{c}\left[\left(\frac{r}{\rg}\right)^{3/2}+ \; a\right]
\label{eq:period}\end{equation}
of matter on a circular orbit of radius $r$ around a rotating black hole (Kerr black hole with gravitational radius $\rg:=\Rs/2$) with dimensionless spin-parameter $a$ ($a~\in~[-1;1]$) \citep{Bardeen}. It is a natural timescale, if the flares involve material on  these orbits close to Sgr~A*. However, this orbit is only stable for radii larger than the innermost stable circular orbit (ISCO), which itself depends on $a$ \citep[see][]{Bardeen}. Thus, there exists a lower boundary on the orbital timescale for a given spin-parameter. In the case of Sgr~A*, this timescale lies somewhere between $27.3$~min (for $a=0$) and $3.7$~min (for $a=1$, counter-rotating orbits with $a<0$ are not considered here). For instance, \citet{genzel03nat} find a quasi-periodic signal at roughly $17$~min. Assuming this variation corresponds to the orbital period on the co-rotating ISCO around Sgr~A*, they give a lower limit for its spin-parameter ($a\ga0.52$). \citet{trippe07} argue that flares happen at various radii larger than the ISCO. From the shortest period ever observed ($13\pm2$~min), they argue that $a\ga0.70$.

In addition to this, \citet{eisenhauer05sinfoni} were able for the first time to derive the slope of the spectral energy distribution (SED) of GC-flares in the K-band.  They found a red spectrum: $\nu S_\nu\propto\nu^{-3}$. \citet{gillessen06} showed in addition that the spectral index varies significantly during a flare and is correlated with the concurrent brightness. While dim states show a red spectrum, the flare appears increasingly blue as its brightness rises. In contrast, \citet{hornstein07} find a constant SED-slope for the flaring states of Sgr~A* with a spectral index being close to zero. In any case, these observations indicate the flares being due to synchrotron emission.

Recent observations of the GC in the NIR also included polarimetric measurements of Sgr~A* \citep{eckart06,trippe07}. \citet{trippe07} find significant linear polarization in the flares with polarization degrees of $15\%$--$40\%$. While they see a correlation between the variability of polarized flux and total flux, the degree of polarization seems to remain constant on the orbital timescale. Furthermore, an increase of the polarization degree and a swing of the polarization angle of roughly $70\degr$ within $15$~min is reported in one case. It happens in the declining phase of the flare. This could be interpreted in a scenario where the emitting material leaves the plane in which is was orbiting, perhaps through a jet. These observations strongly favor the nature of the emitted light as synchrotron radiation and that the accretion disk is seen nearly edge-on \citep[compare][]{bl06b}.

In the radio domain, Sgr~A* appears as a permanent, resolved source. Scattering effects in the interstellar dust make it appear larger, with an apparent size proportional  to  $\lambda^2$ \citep{davies76}. \citet{bower04} and \citet{shen05} were able to measure the intrinsic size of the emission region at 3.5 and 7~mm, finding $0.126\pm0.017\mathrm{~mas}\simeq15\Rs$ and $0.268\pm0.025\mathrm{~mas}\simeq30\Rs$, respectively (major axis). The latest measurements were carried out by \citet{Doeleman08} at 1.3~mm that set a size of about 37~\muas\ on the intrinsic diameter of Sgr~A*. Hour-scale variability has been observed at 2 and 3~mm \citep{miyazaki04,mauerhan05}. \citet{Yusef06} have detected flaring activity at 7~mm and 13~mm, with a 20--40~min delay between the peaks of the two lightcurves. At centimeter wavelengths, periods of 57 and 106 days have been claimed \citep{townes99,zhao01}, but were later refuted \citep{macquart06}. Polarimetric measurements yielded polarization fractions around $10\%$ and highly variable polarization angles on timescales of a few hours \citep{bower05a,marrone06}.

At the currently possible imaging resolution ($\simeq1\arcsec$ for Chandra, $60$~mas for the VLT in the K-band), the flares occur at the same location as the unresolved X-ray quiescent emission. This position is consistent (within the error bars of a few mas or a few hundred \Rs) with the focus of the orbits of nearby stars orbiting the black hole \citep{schoedel02,schoedel03,ghez03,eisenhauer05sinfoni}. It is also coincident with the radio source, which appears motionless \citep{reid07}. However, on shorter timescales (between $50$ and $200$~min) a centroid wander of about $100~\muas$ has been observed by \citet{reid08} with the Very Long Baseline Array (VLBA). From their observations they can rule out hot spots with orbital radii larger than $7.5\Rs$.

The purpose of this paper is to demonstrate the feasibility of measuring the position and proper motion of NIR flares with an accuracy of $10$~\muas. In order to resolve this proper motion, one must not only reach the necessary astrometric accuracy, but also resolve flares temporally. Because the sampling must exceed the characteristic periodicity seen in the NIR light curves, this represents a very challenging experiment.

In the typical observational setup mentioned  above, this accuracy requires a 30~min  integration time in order for atmospheric residuals to be averaged out. Given Eq.~\ref{eq:accuracy  scaling}, the required  integration  time  for a fixed accuracy depends on the separation $\Theta$. In the GC a suitable $m_K=10$ reference  star can be found as close to Sgr~A* as $1.21\arcsec$ \citep[GCIRS~16NW,][]{paumard06disks}. With  such a close reference object and a projected   baseline of 100~m, the required integration time for $10$~\muas~astrometry reduces to roughly $1$~min. Therefore, the VLTI infrastructure is able to provide time-resolved astrometry  of GC-flares at a sufficient accuracy. However, this is a very demanding experiment and  requires a specifically optimized instrument. The GRAVITY experiment has been proposed to fulfill all the requirements for the GC science case \citep{eisenhauer05gravity}.

Similar attempts for the astrometric observation of Sgr~A* are being planned for the radio domain \citep{bower05b}. The VLBA is expected to improve its imaging resolution down to 20~\muas\ within the next decade \citep{Doeleman04}. However, for technical reasons the integration times in radio and submillimeter wavelengths are significantly longer than in the NIR. Thus, no dynamic picture will be accessible easily, rather a time-averaged image can be generated. This has already been studied by means of simulations by \citet{bl06b}. Generally, enhanced opacity at longer wavelengths smoothes out the strong features in the light curves and centroid tracks expected in the NIR.

\subsection{Flare-models}

Several hypotheses have been proposed to explain the origin of the flares from the GC. The most common ones can be summarized as follows:
\begin{itemize}
\item[A] Star-disk interactions rather far away from the black hole \citep[$10^2$--$10^4\Rs$,][]{nayakshin04};
\item[B] Synchrotron emission  from electrons  accelerated in  the  central part ($\lesssim10\:\Rs$), through  processes  such  as  turbulent acceleration, reconnections, and weak shocks \citep[e.g.][]{yuan04};
\item[C] Sudden  heating of hot electrons  (e.g.\  by magnetic  reconnection) in a permanent jet \citep{markoff01};
\item[D] Increase in the accretion rate due to the infall of a clump of material, being rapidly driven inwards due to a Rossby wave instability (RWI) \citep{tagger06,Falanga07}.
\end{itemize}

The star-disk interaction model appears unlikely. It is unable to explain either the observed synchrotron emission or the quasi-periodic signal on a timescale of minutes. Moreover, actual constraints on Sgr~A*'s disk size in the Radio \citep[$r_{\mathrm{disk}}\lesssim1~AU$,][]{shen08} are in conflict with the assumption of stars remaining stable on such close orbits.

No matter what the emission process producing the flares is, the matter responsible for it must move with at least the order of the Keplerian circular velocity
\begin{equation}
v\simeq\sqrt{\frac{\mathcal GM_{0}}{r}}=\frac{c}{\sqrt{2}}\sqrt{\frac{\Rs }{r}}\simeq\left(\frac r\Rs\right)^{-1/2}\times10\:\mu\mathrm{as}\:\mathrm{min}^{-1}\:
\end{equation}
This is true irrespective of whether this matter is orbiting the black hole, falling into it in free fall, or escaping in a jet or wind. Given that $r\lesssim1\:\mathrm{mas}\simeq100\:\Rs$, we have $v\gtrsim1\:\mu\mathrm{as}\:\mathrm{min}^{-1}$. Therefore, this matter moves by several times the proposed accuracy ($10$~\muas) during one flare. This typical velocity is large compared to the apparent velocity of the star proposed as interferometric phase reference.

If the flares are due to magnetic reconnection in a jet (scenario C), sufficiently high above the accretion disk, the electrons responsible for the emission would have relativistic velocities, but the emission region may not move at the same speed. This region is tied to the node in the magnetic field where the reconnection happens. If this node drifts on the hour timescale, some long term proper motion could be observed. In this scenario, the short-term variability in the lightcurves could come from  individual blobs of denser material being ejected at  a pseudo-frequency, typical for the dynamical  timescale of the inner disk. When passing through the node at a significant fraction of $c\simeq15\:\mu$as~min$^{-1}$, a short term proper motion could be expected from these individual blobs.

However, the observational data at hand \citep[see][for a summary]{trippe07} strongly favor another interpretation, namely the emission of synchrotron radiation from matter orbiting the black hole near its ISCO. This matter would be confined to a small region in the disk \citep[$r\la0.3\Rs$,][]{gillessen06} containing highly accelerated electrons in a broken power-law distribution \citep[$n_{\mathrm{e}}\approx10^{5}~\mathrm{cm}^{-3}\,\gamma^{-3}$ in the NIR,][]{yuan04}. Such a radiating blob could be produced by scenario B. Despite the very high velocity of the material (about half the speed of light), no long-term proper motion would be detected at the scale of more than one orbital period. Nevertheless, this scenario opens up extremely interesting  scientific perspectives. Time-resolved astrometry of the flares would allow the shape of the ISCO to be traced, lensed by the strong gravitational field of the black hole. This would give access to  the orientation of the orbit as well as other physical parameters of the model, and more fundamentally provide a probe for the strong gravitational field of a black hole.

This scenario is akin to the RWI scenario (scenario D). In both cases, the emission originates very close to the black hole, within a few Schwarzschild radii \citep{tagger06}. However, there is a fundamental difference: instead of a blob, scenario D produces a somewhat symmetrical spiral structure \citep{Falanga07}. It is created by a RWI in the accreting plasma producing quasi-periodic oscillations in the accretion rate which are close to the orbital period. Though, these oscillations are not of Keplerian origin and therefore do not allow the black hole spin to be constrained from the periodic timescale. They are caused by a pattern rotation in the disk, a wave phenomenon. Further studies are needed to assess whether this process could mimic the observed light curves and whether it could lead to an observable astrometric signature.

\section{SIMULATIONS}

We will hereafter always assume that flares from Sgr~A* are due to matter on Keplerian orbits around the black hole and that the periods observed in the light curves are a signature of these orbits. After a discussion of our basic methods we will first investigate the consequences of this scenario on the basis of a simplified model. Therefore, we study the appearance of emission from a small sphere on the co-rotating ISCO about a Kerr black hole and generate photometric, astrometric as well as polarimetric data. We will then address an advanced model with an extended emission region as a phenomenological model for a variable source.

The lensing effects in the strong  gravitational field of a black hole and the optical  appearance of accretion disks and close-by objects have  been computed a number of  times \citep[e.g.][]{cunningham73,luminet79,bao94,kling00}. Derivations of analytical lens-equations, giving both the position of multiple images of a source and the time of arrival of each of these images have also been attempted.  These purely analytical studies are however limited in the computation of higher order images (no more than second order) and do not account for extended sources with variable intrinsic brightness.

The apparent path on sky of a particle revolving around a black hole in a circular orbit of radius $r$ ($r=3\Rs$ for the ISCO of a Schwarzschild black hole) does not have the shape of an ellipse due to strong deflection of light rays by the gravitational field. Observed in high inclination, the orbit's near side appears as half-ellipse (the usual appearance of an inclined circle), whereas the far side is de-projected and closely resembles a semi-circle.

In addition, light rays emitted by a nearby source are bent so strongly that this source appears at several different locations on sky. In principle, infinitely many such images emerge. However, higher order images are exponentially suppressed in angular diameter and have increasingly longer light travel times compared to the primary image (see Fig.~\ref{mult_images}). Moreover, absorption increasingly diminishes the brightness of images with longer light paths.

\subsection{Phenomenological assumptions}

Unlike the Newtonian case, in both Schwarzschild and Kerr geometry bound orbits are not closed in general. Eccentric orbits in the Schwarzschild metric or in the equatorial plane of the Kerr metric are planar, but not closed \citep[see][]{bao94}.  Due  to  Lense-Thirring  precession,  circular orbits outside this equatorial  plane are not planar and can be described as helices \citep{wilkins72}. For this reason, a cloud falling onto the  black hole on a trajectory inclined with respect to the equatorial plane would soon form a geometrically thick disk, symmetrical with respect to the equatorial plane, and settle on this plane to be circularized. In the following we will therefore concentrate on circular orbits in the equatorial plane.

The power law SED observed in the GC-flares with a red spectral index in the NIR \citep{eisenhauer05sinfoni,gillessen06,Krabbe06} supports synchrotron emission. This indicates the phenomenon to be dictated by a strong magnetic field. Possible configurations of such magnetic fields are discussed by \citet{uzdensky05}. The field lines should intersect the accretion disk orthogonally at its surface. We will therefore consider the field to be poloidal near the disk. On the other hand, the magnetic field inside the accretion disk is likely frozen. The field lines must warp and align with the velocity field, leading to a toroidal geometry inside. Shear tends to make the magnetic field lines loop around the black hole, leading to magnetic reconnection events. The stronger shear near the inner edge of the disk causes more reconnection events towards this edge. We will consider a  bright sphere close to the ISCO, which should be a good approximation for the inner edge of the quasi-circular accretion disk. The sphere is just a first approximation of the emitting region geometry. Its real shape presumably follows the field lines and would therefore extend as an arc along the orbit.

\subsection{Relativistic ray tracing}

\begin{figure}
 \resizebox{\hsize}{!}{\includegraphics{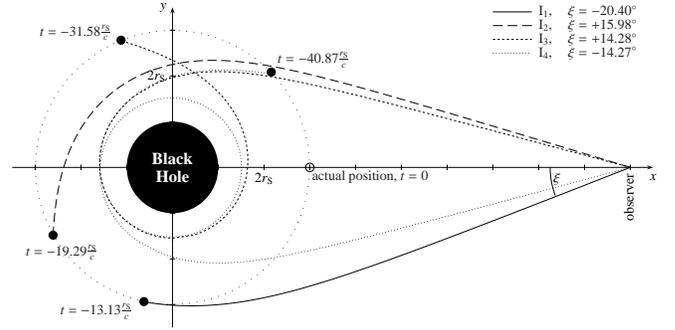}}
 \caption{Travelled light paths of the primary (solid), secondary (long-dashed), tertiary (short-dashed) and quaternary (dotted) image of the source at a given arrival time ($t=0$) at the observer's location. The higher the order $n$ of an image $\mathrm{I}_n$, the earlier its emission time $t$ and the smaller the absolute value of its deflection angle $\xi$. \label{mult_images}}
\end{figure}

We choose to use relativistic ray tracing to model the light bending effects \citep{weiskopf00,weiskopf01,mueller06}. Light rays are represented by null geodesics which are determined by integrating the equation of motion for photons in Kerr spacetime. Initial values for this ordinary differential equation are determined by the position, orientation, and field of view  of the observer's camera and by  the coordinates of the corresponding pixel in the image plane.  Starting  from the observer, the geodesics have  to be integrated back in  time until they fulfill one of the breaking conditions:  they either reach the surface of the source (as in Fig.~\ref{mult_images}), they cross the horizon of the black hole, or they leave a certain bounding box around the region of consideration. A further limit is given by a maximum length for any computed geodesic.

The  main advantage  of this  method  is that it does not yield any analytical approximation, it is exact up to the numerical precision. The fact that different null geodesics do not affect each other allows the code to be parallelized, i.e.\ the field of view can be divided into smaller sections to be computed on several CPUs and reassembled later on. We set up a very simplified configuration of a nonrotating sphere of radius $0.25\Rs$ on a circular orbit in the equatorial plane of a Kerr black hole. The sphere's surface is homogeneous and emits light isotropically. From the ray tracing code we compute the lensed images of the source (we typically resolve images up to the fourth order) with the corresponding light travel times $t$ and frequency shifts $z := \nu_\mathrm{obs}/\nu_\mathrm{em}$ computed for each pixel (`obs' for observed and `em' for emitted). In addition, the code provides us with the local tetrad $\mathbf{e}_{\alpha}$ of the observer, parallel transported to the source in each pixel. It is necessary for the polarimetric analysis of the flare model.

The main restriction to the ray tracing method is a tradeoff between resolution and computing time. In order to keep the computing time manageable, we had to consider frames with no more than $1000\times 1000$ pixels. In addition to this, again for the sake of reasonable computing time, it has been necessary to use a finite distance between the observer and  the black hole ($60\Rs$). However, this yields a good enough approximation, since we are able to resolve all multiple images up to the third order.

For our simulations we set up a grid of four inclinations ($i=20\degr$, $50\degr$, $70\degr$, $90\degr$), four orbital radii ($r/r_{\mathrm{ISCO}}=1.0$, $1.2$, $1.5$, $2.0$) and four black hole spin-parameters ($a=0$, $0.52$, $0.7$, $0.998$). All possible combinations yield a grid of $64$ different setups, allowing to analyze the impact of these parameters. For each configuration we ray traced one complete orbit, consisting of roughly $100$--$500$ frames depending on the orbital period of the source (for the $a=0.998$ case only two simulations with $r=2.0r_{\mathrm{ISCO}}$ were carried out).

\subsection{Light curves and centroid tracks}

Assuming the source to emit a power law SED in its rest frame in the NIR,
\begin{equation}
\nu_{\mathrm{em}}S_\nu^{\mathrm{em}}\propto\nu_{\mathrm{em}}^{\,\alpha}
\end{equation}
with spectral index $\alpha$, we can compute the apparent surface brightness $S_{\nu}^{\mathrm{obs}}$ (as seen by a distant observer) from the frequency maps provided by the ray tracing code. We simply have to apply Liouville's theorem \citep{Misner}, stating:
\begin{equation}
S_\nu^{\mathrm{em}}/\nu_{\mathrm{em}}^{\,3} =
S_\nu^{\mathrm{obs}}/\nu_{\mathrm{obs}}^{\,3} = \mathrm{const.}
\end{equation}
For the apparent surface brightness this yields
\begin{equation}
\nu_{\mathrm{obs}}S_{\nu}^{\mathrm{obs}}\propto z^{4-\alpha}\nu_{\mathrm{obs}}^{\,\alpha}
\end{equation}
depending on the generalized redshift factor $z$, which is responsible for relativistic beaming- and Doppler effects, as well as gravitational redshift. The result is that the source appears brighter and blueshifted on the approaching side, respectively fainter and redshifted on the receding side. We are not interested in the absolute value of $S_\nu^{\mathrm{obs}}$, since we do not model the emission process in a self-consistent manner. We only study the dependencies of the simulated light curves on the various parameters of the model.

Light curves are created by integrating the apparent surface brightness over a certain frequency range and over the field~of~view:
\begin{equation}
L:=\int\!\!\!\int
S_{\nu}^{\mathrm{obs}}\,\mathrm{d}\nu_{\mathrm{obs}}\,\mathrm{d}A_{\mathrm{fov}}
\end{equation}
Thus, every simulated frame produces one data point of the light curve. We normalize each light curve by its minimum flux and plot its magnitude ($m=-2.512\,\mathrm{log}_{10}L$) against the orbital phase~$\phi$ in radians ($\phi=0$ designates the point where the hot spot appears closest to the observer).

Similarly, the centroid of emission is computed as
\begin{equation}
\mathbf{C} := \frac{1}{L}\int\!\!\!\int
\mathbf{r} \,S_{\nu}^{\mathrm{obs}}\,\mathrm{d}\nu_{\mathrm{obs}}\,
\mathrm{d} A_ { \mathrm { fov } }
\end{equation}
with $\mathbf r$ being the vector from the origin (center of field of view) to the pixel position. The centroid track is generated by plotting $\mathbf C$ for every frame. We always use the mass and distance of Sgr~A* for the scaling of its apparent size ($1\Rs$ corresponds to about $9.3~\mu$as).

\section{RESULTS}

\subsection{Small, constant source \label{constant}}

First we study the observable properties of a constantly radiating sphere of radius $0.25\Rs$ and spectral index $\alpha=0$ on a co-rotating circular orbit around a generally rotating black hole. The sense of rotation of the black hole (and thus the direction of rotation of the source) is arbitrarily chosen to be counterclockwise. This approach is suited to investigate the main features of the hot spot model.

\subsubsection{Photometry and astrometry}

\begin{figure*}
 \resizebox{\hsize}{!}{\includegraphics{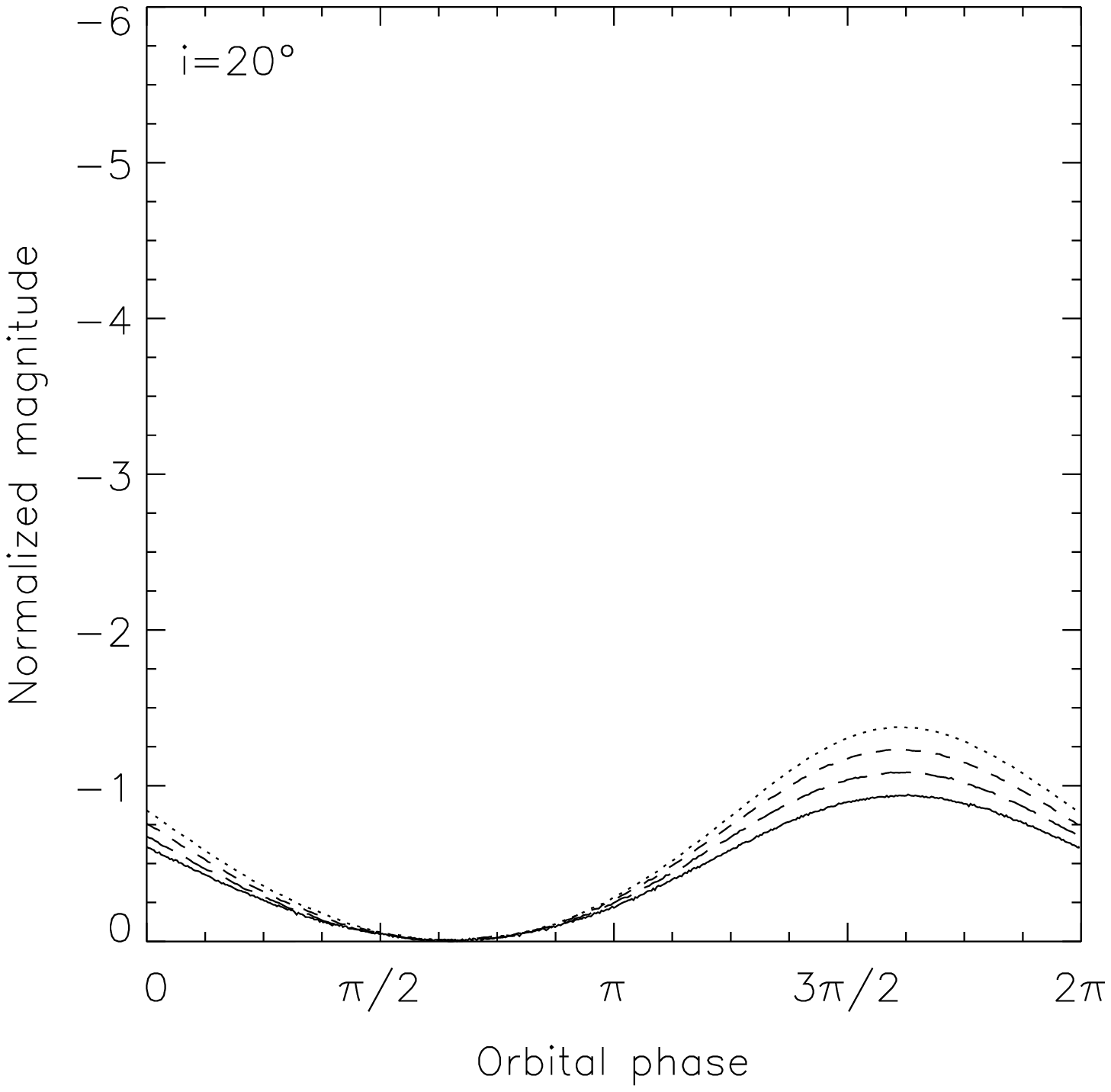}\includegraphics{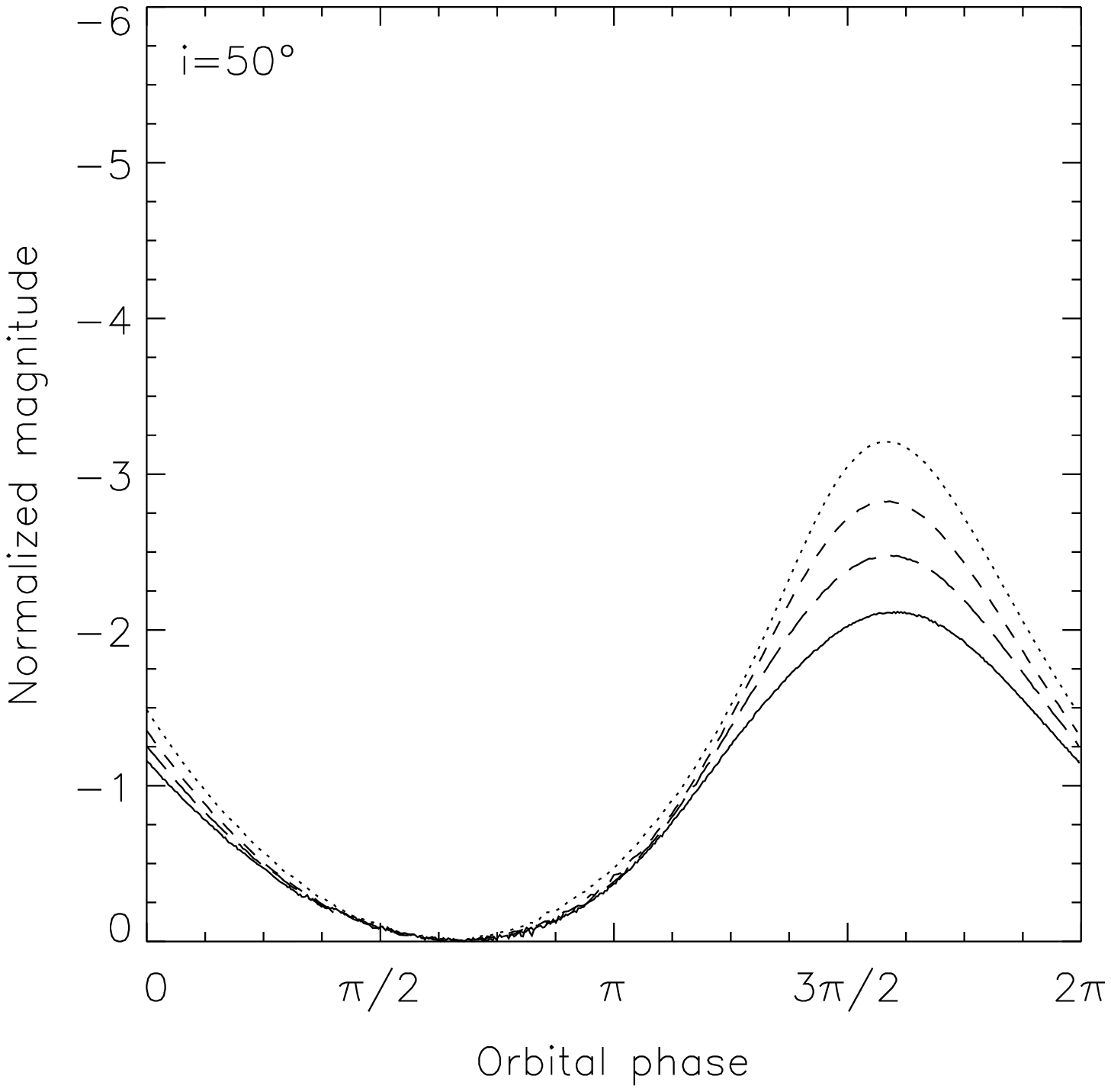}\includegraphics{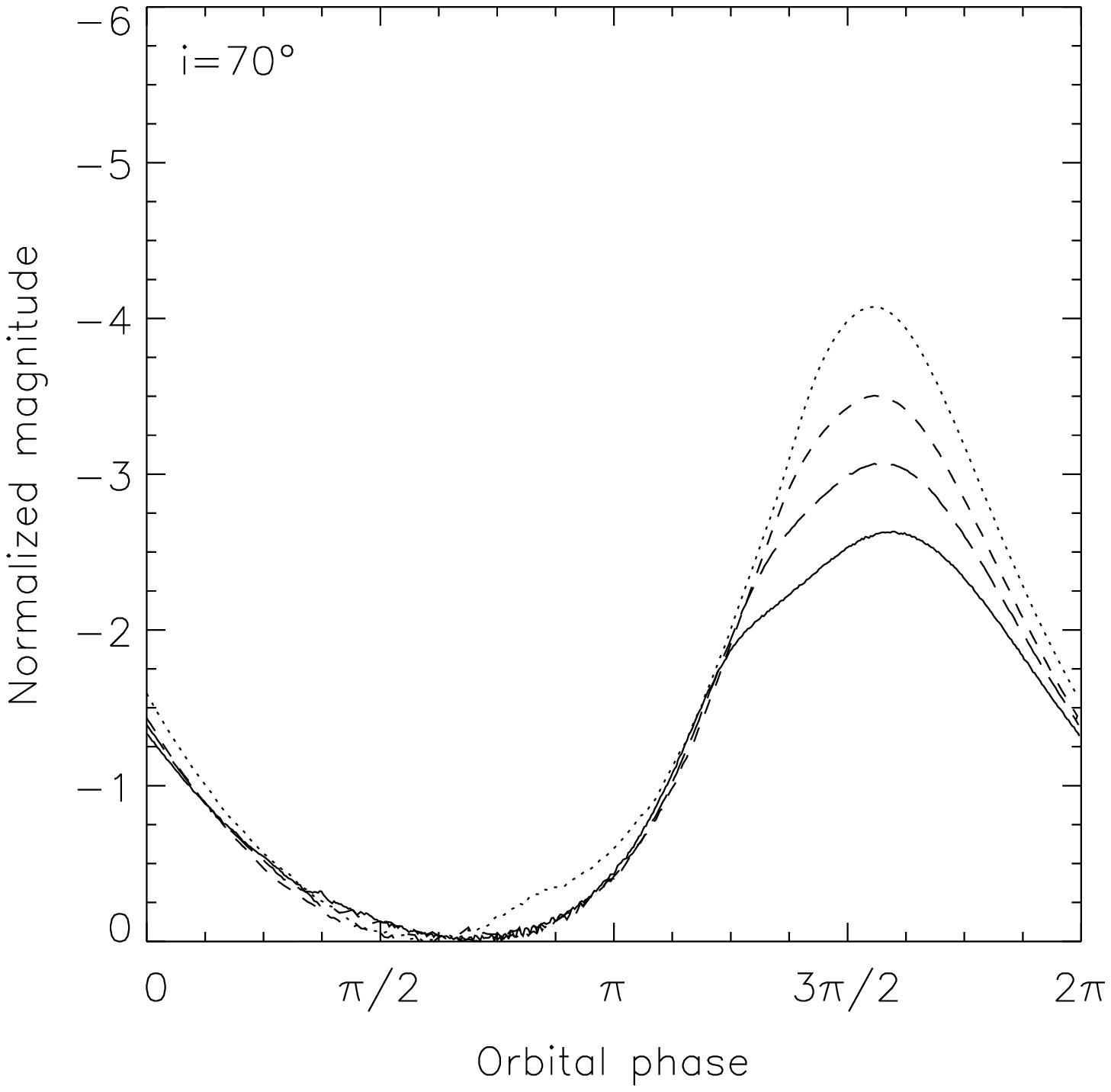}\includegraphics{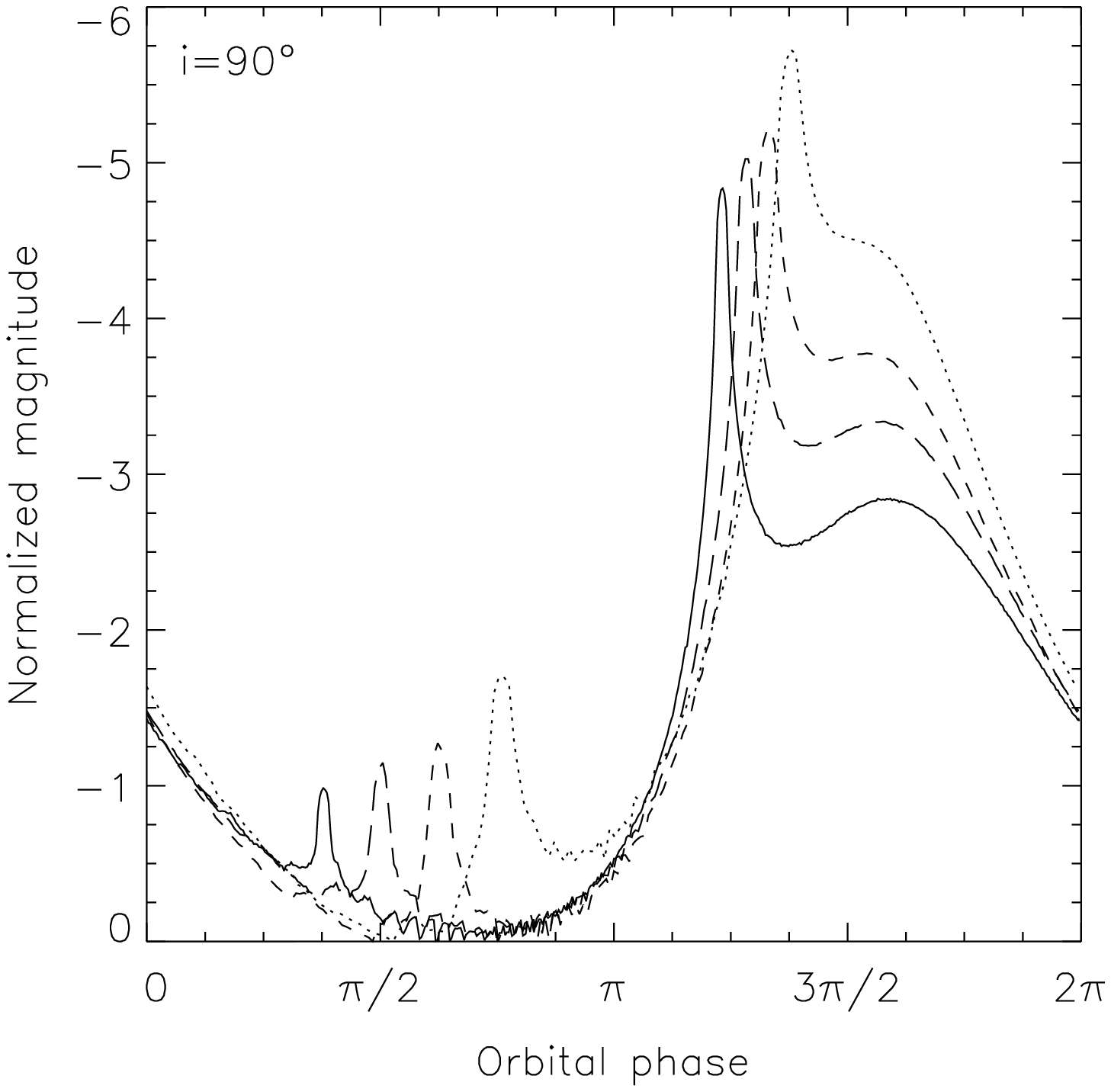}}
 \resizebox{\hsize}{!}{\includegraphics{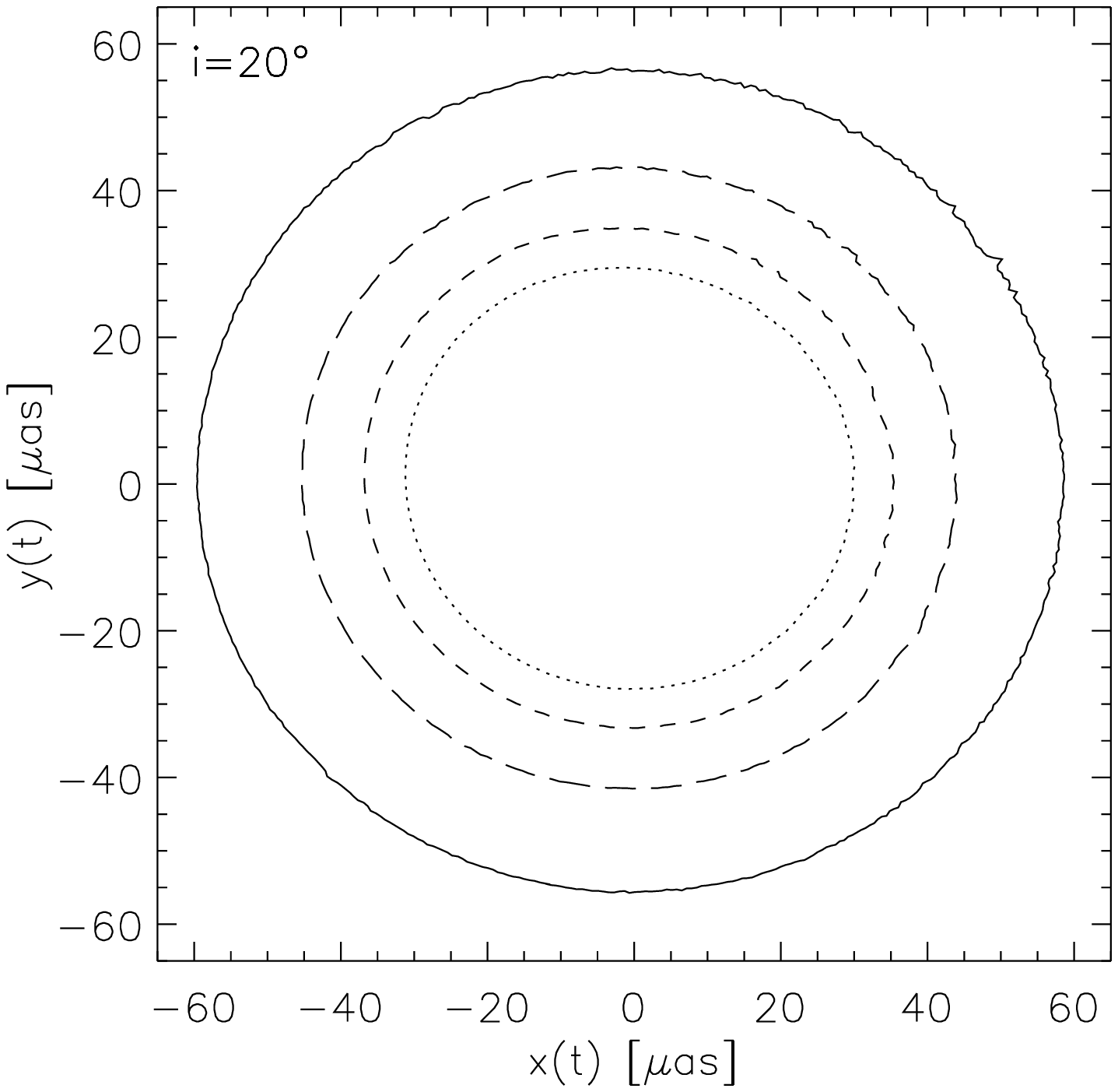}\includegraphics{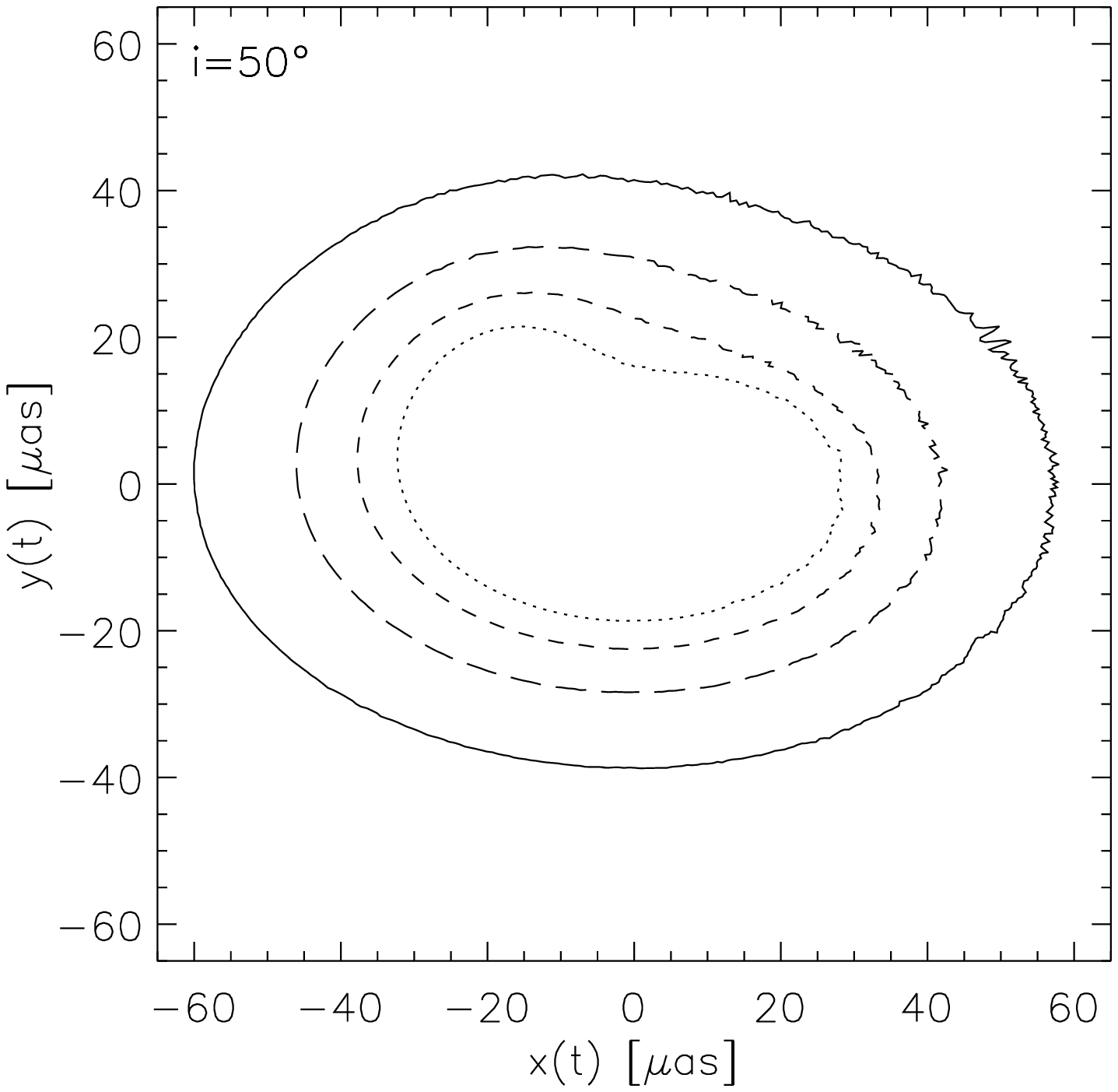}\includegraphics{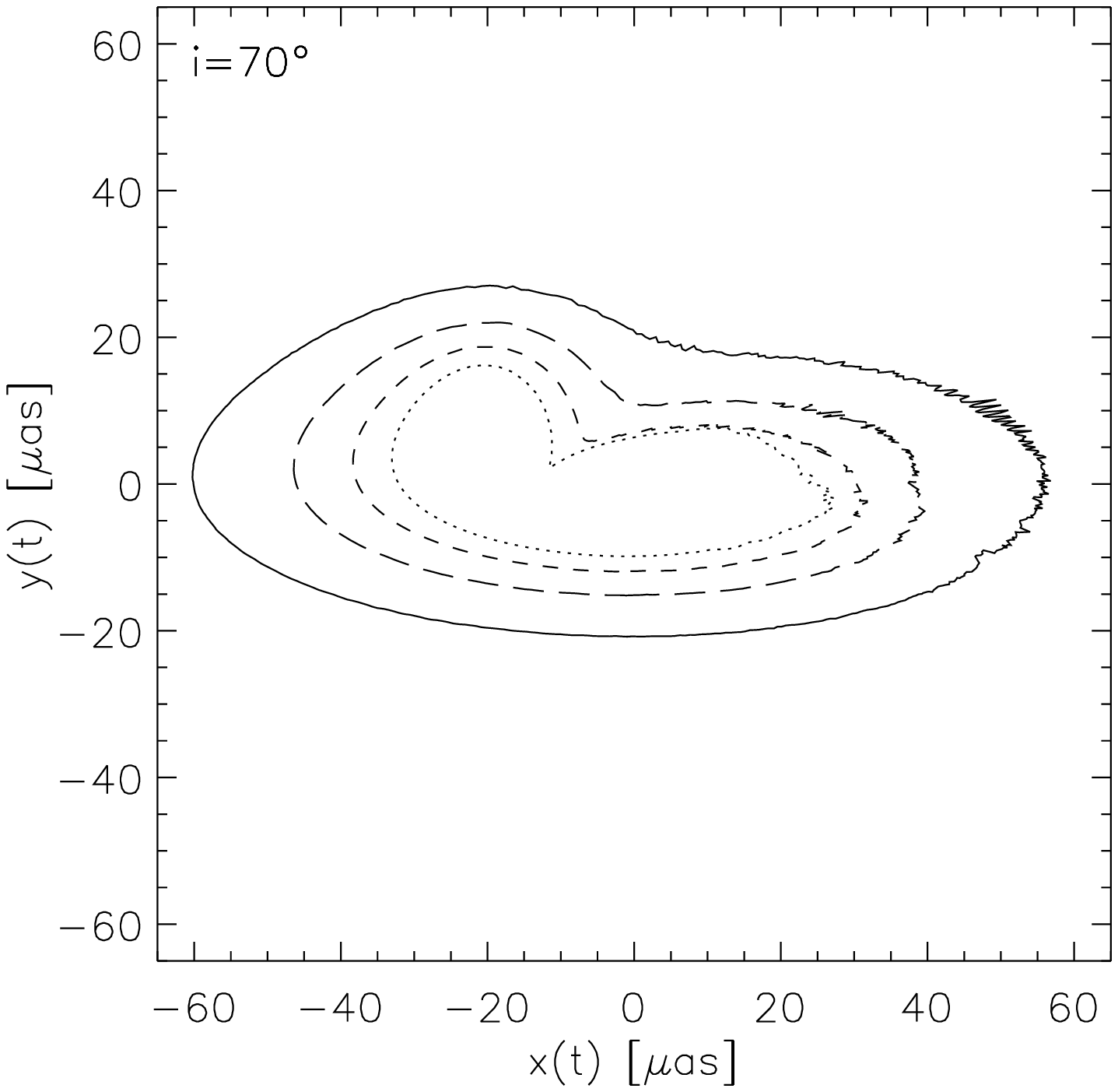}\includegraphics{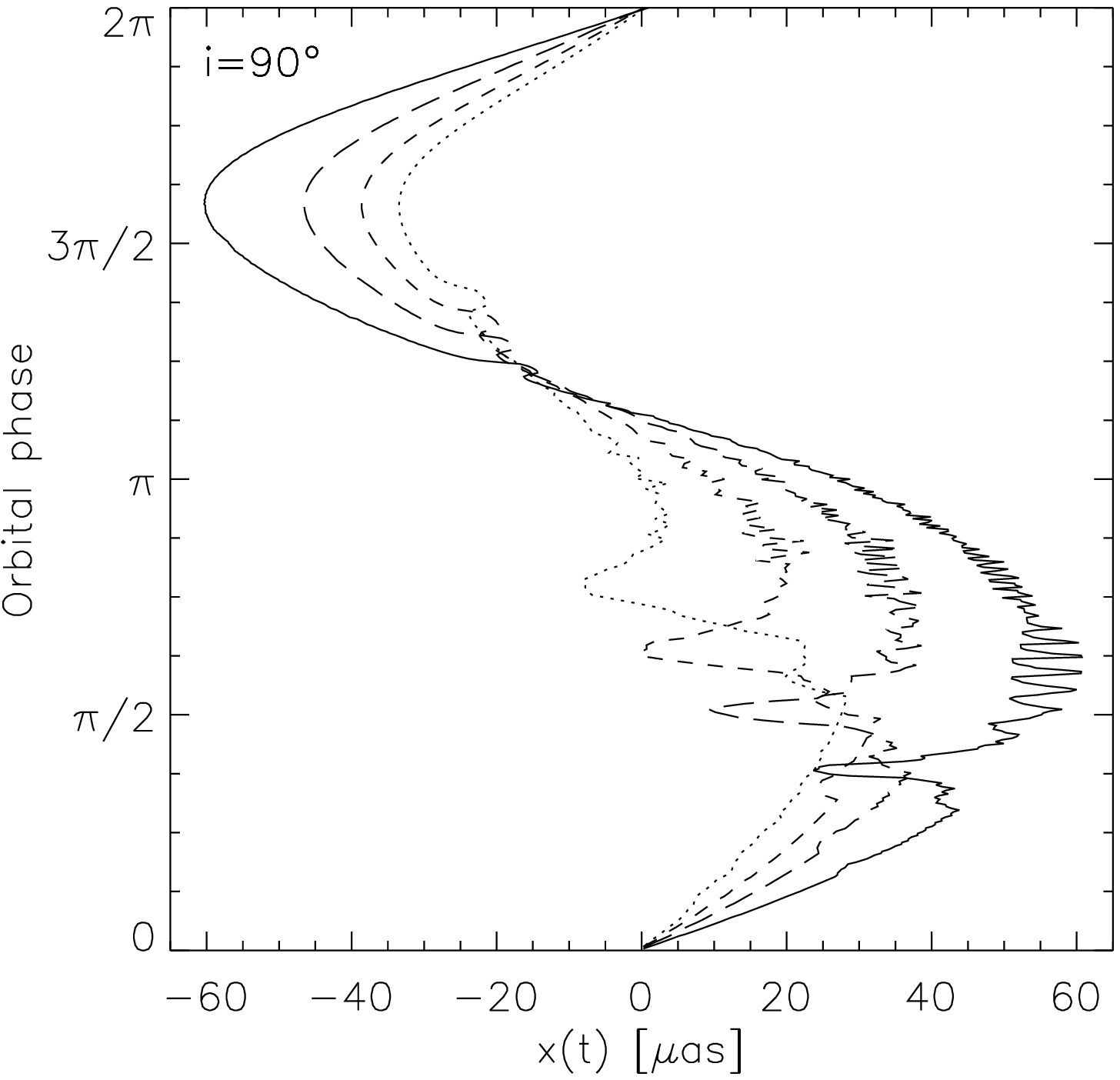}}
 \caption{Light curves (top) and centroid tracks (bottom) of a compact hot spot orbiting a Schwarzschild black hole in four different orbital radii and inclinations ($r/r_{\mathrm{ISCO}}=2.0$ (solid), $1.5$ (long-dashed), $1.2$ (short-dashed), $1.0$ (dotted), inclinations are given in the upper left corner of each plot). Since the ordinary centroid track for the $90\degr$-case just yields a horizontal line, we instead plot the orbital phase $\phi(t)$ against $x(t)$. \label{radii_incl_s0}}
\end{figure*}

\paragraph{Orbital radius and inclination}

Fig.~\ref{radii_incl_s0} displays light curves and centroid tracks of this model for different orbital radii and inclinations of the hot spot. Decreasing the orbital radius generally leads to narrower, higher peaks in the light curves. This is due to the increasing spot-velocity leading to stronger relativistic Doppler- and beaming effects. The same changes can be produced by increasing the inclination of the orbit, since this increases the line of sight velocity of the hot spot. In our chosen parameter range, changes in the inclination exhibit these relativistic effects more strongly than changes in the orbital radius.

The light curves in $90\degr$ inclination reveal an additional feature: before each of them reaches its usual peak-maximum (approaching node), a spiky second peak appears. Its peak-flux exceeds the usual maximum by far. It originates from the development of an \emph{Einstein ring}, an apparent ring-shaped image of the hot spot with the black hole in its center. It occurs when the light source, the black hole and the observer are aligned (and the black hole is in between). Due to the finite extension of the source, conditions for an Einstein ring are fulfilled in a finite (although small) range of inclinations. Thus, it is no mathematical artifact of the exact  $90\degr$-case. Higher order Einstein rings appear, when light rays revolve the black hole. For instance, a secondary Einstein ring appears when the hot spot passes in front of the black hole ($\phi=0$). We then receive photons from the far side of the spot, which have orbited the black hole once. In fact it occurs a bit later, since one has to consider the finite speed of the photons when orbiting the black hole. The corresponding peaks can also be seen in the $90\degr$-plot (at $\phi\approx\pi/2$), although they are a lot fainter than the ones caused by the primary Einstein ring.

At lower inclinations we no longer expect Einstein rings, but still multiple images of the hot spot. Obviously, their effects on the light curves are marginal. Only the light curves with $70\degr$ inclination show a slight distortion of the main peak, suggesting the existence of a secondary image. It is not only its small apparent size that prevents the secondary image from being observable in the light curves, but also the fact that its brightness is not as variable as that of the Einstein ring.

Turning to astrometry, decreasing the orbital radius obviously leads to smaller centroid tracks. Then, as with the light curves, the relativistic effects increase, i.e.\ the deviation of the centroid tracks from the Newtonian case (projected circles) becomes more significant. On the one hand, this deviation is caused by multiple images of the hot spot. When the secondary image gets brightest, the centroid of the received light lies somewhere in between the primary and the secondary image. But since the latter only brightens for a short time (it moves faster on the approaching side), this causes a cusp in the centroid track. On the other hand, deflections of the primary rays result in a further distortion of the centroid track. The rear part of the orbit appears to be folded upwards at higher inclinations.

Again, raising the inclination leads to stronger relativistic effects. The cusp in the centroid track, for instance, becomes more and more prominent. As shown in the lower right plot of Fig.~\ref{radii_incl_s0}, the classical sine-pattern of the $90\degr$-case is deformed. It is stretched when the source is receding and compressed when approaching (due to the finite propagation speed of the emitted light rays). In addition, distinct spikes appear in the pattern. These can again be associated with the Einstein~rings. Comparing the occurrence of these spikes with the occurrence of the sub-peaks in the light curves, one can ascertain that they appear at the same orbital phase.

\paragraph{Spectral index}

Changing the SED of the hot spot emission has strong implications for the light curves as well. Since we are observing in a confined range of frequencies (observing band), Doppler shifting the emitted radiation makes us observe different parts of the SED every instant of time. If, for instance, the hot spot is approaching us, the observed radiation is blue-shifted. Depending upon whether the hot spot emission follows a blue, white or red SED, the observed flux is lower, equal or higher, respectively. This means that a red SED ($\alpha<0$) amplifies the beaming effect by brightening the hot spot on the approaching side and dimming it on the receding side, while a source with a blue SED  ($\alpha>0$) behaves contrariwise. For a white SED ($\alpha=0$) the Doppler shift does not change the observed flux at all.

\begin{figure}
 \resizebox{\hsize}{!}{\includegraphics{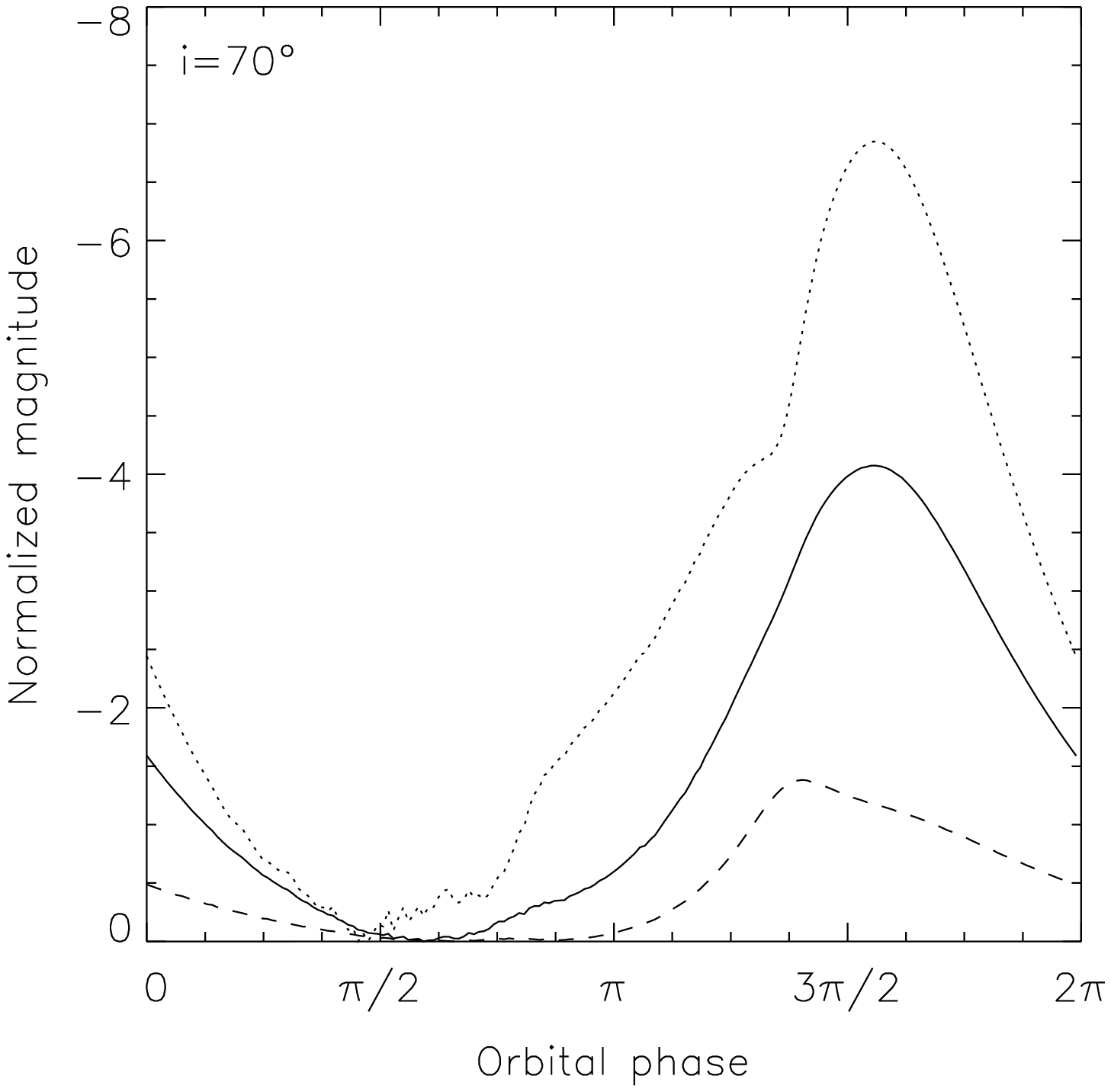}\includegraphics{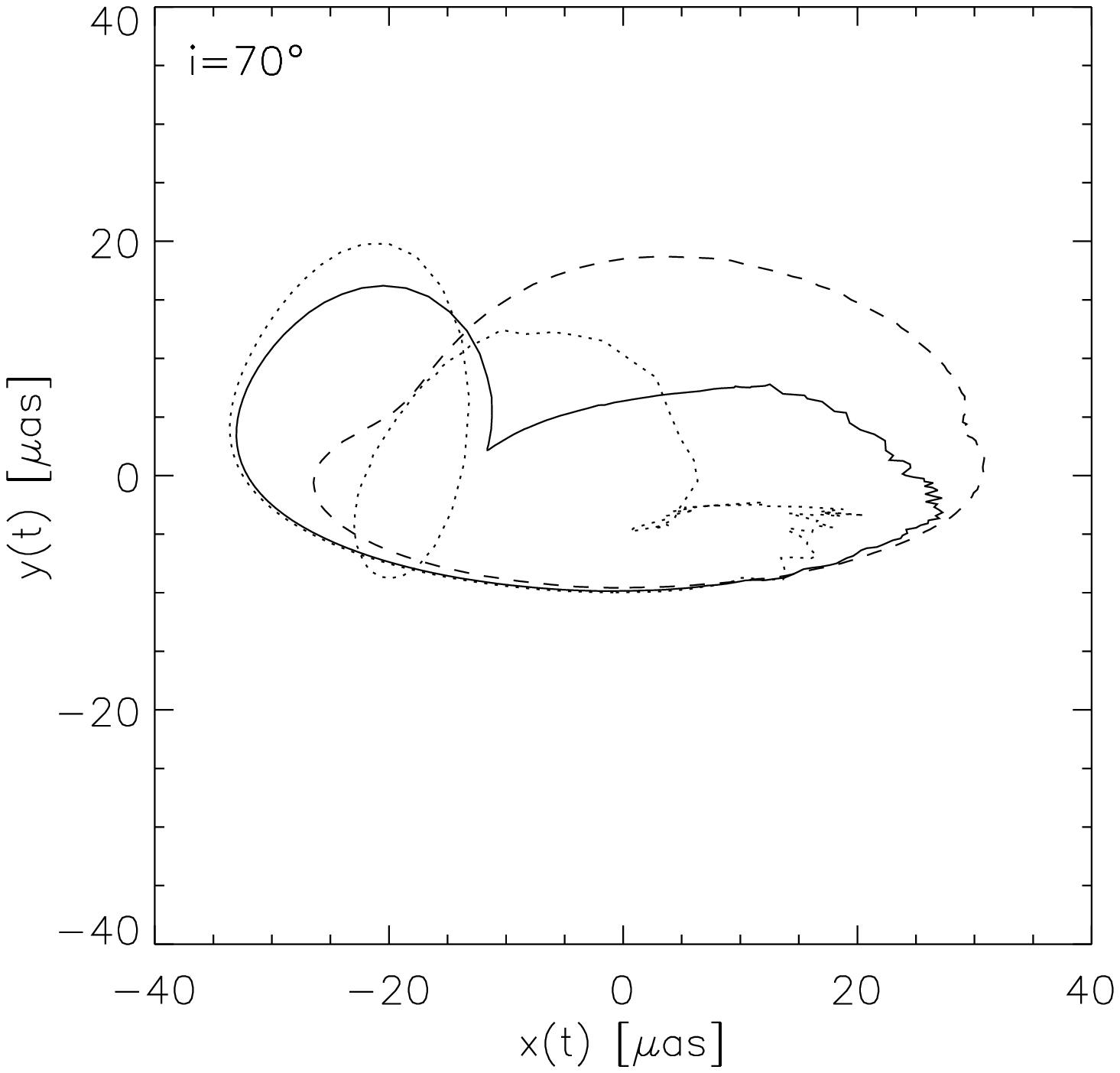}}
 \caption{Light curves (left) and centroid tracks (right) of a compact hot spot with different spectral indices ($\alpha=-3$ (dotted), $\alpha=0$ (solid), $\alpha=3$ (dashed)) orbiting a Schwarzschild black hole in $70\degr$ inclination on the ISCO. \label{SED}}
\end{figure}

Fig.~\ref{SED} displays light curves and centroid tracks for all three cases. Clearly, the light curve peaks rise with decreasing spectral index. In the case of a blue SED the observed flux of the hot spot at the approaching side is diminished to such an extent that the secondary image of the blob is revealed. Its brightness exceeds the total flux of the primary image for a short fraction of the orbital period. In the case of a red SED the opposite happens: the primary peak gets so narrow and bright that the secondary peak appears slightly before the hot spot reaches the approaching node.

For the centroid tracks, it appears that those with a red spectrum are dragged towards the approaching side, whereas those with a blue spectrum are more centered towards the receding side. The multiple images have a greater influence in the case of a red SED, since the primary image of the hot spot is dimmed more strongly on the receding side of the orbit, whereas the secondary image is amplified more strongly on the approaching side. This even makes the cusp from the secondary image turn into a loop; the tertiary image may create another cusp. The centroid tracks of hot spots with a blue SED are generally more symmetric and show less relativistic deflection.

\paragraph{Black hole spin}

\begin{figure}
 \resizebox{\hsize}{!}{\includegraphics{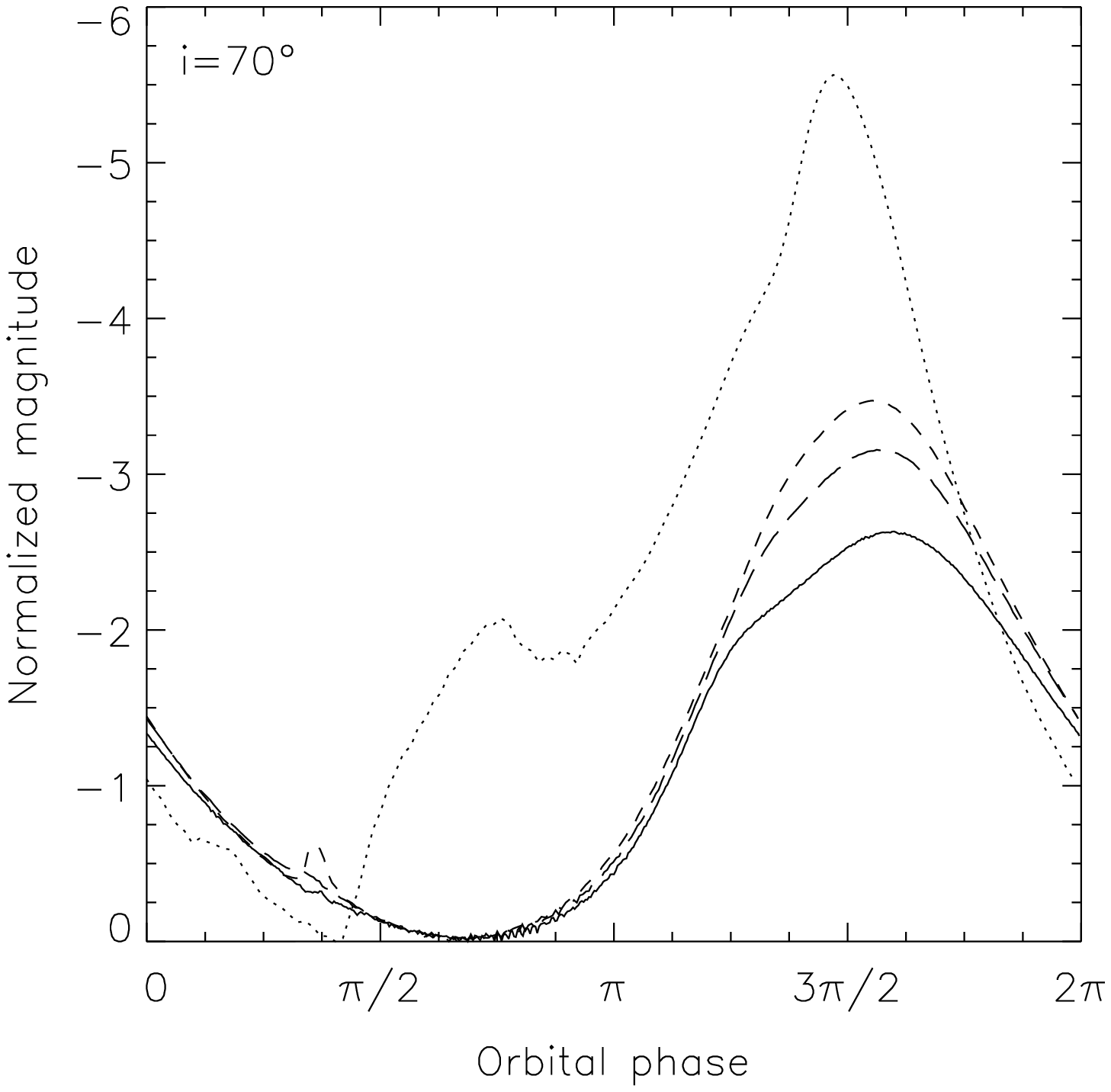}\includegraphics{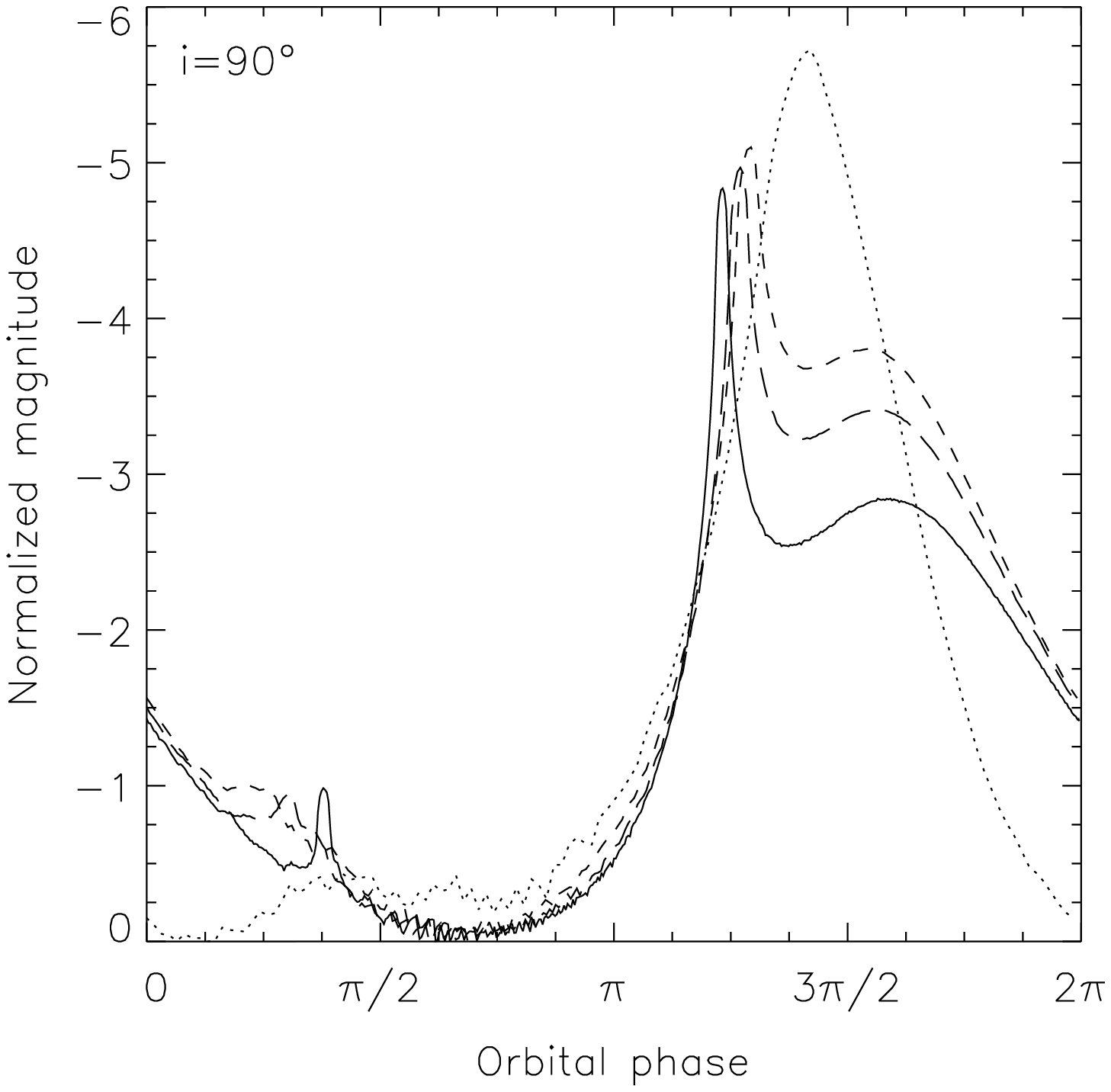}}
 \resizebox{\hsize}{!}{\includegraphics{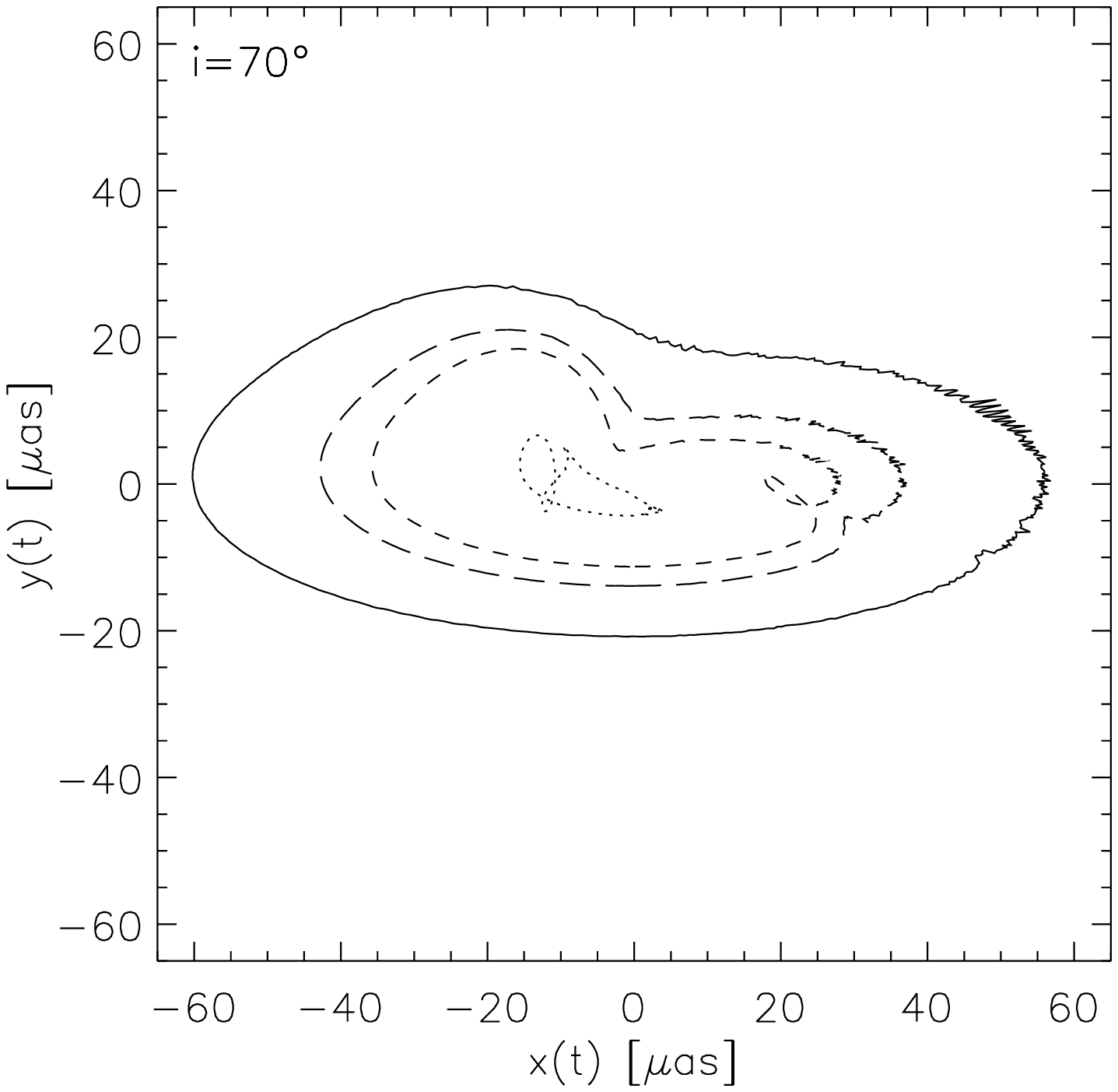}\includegraphics{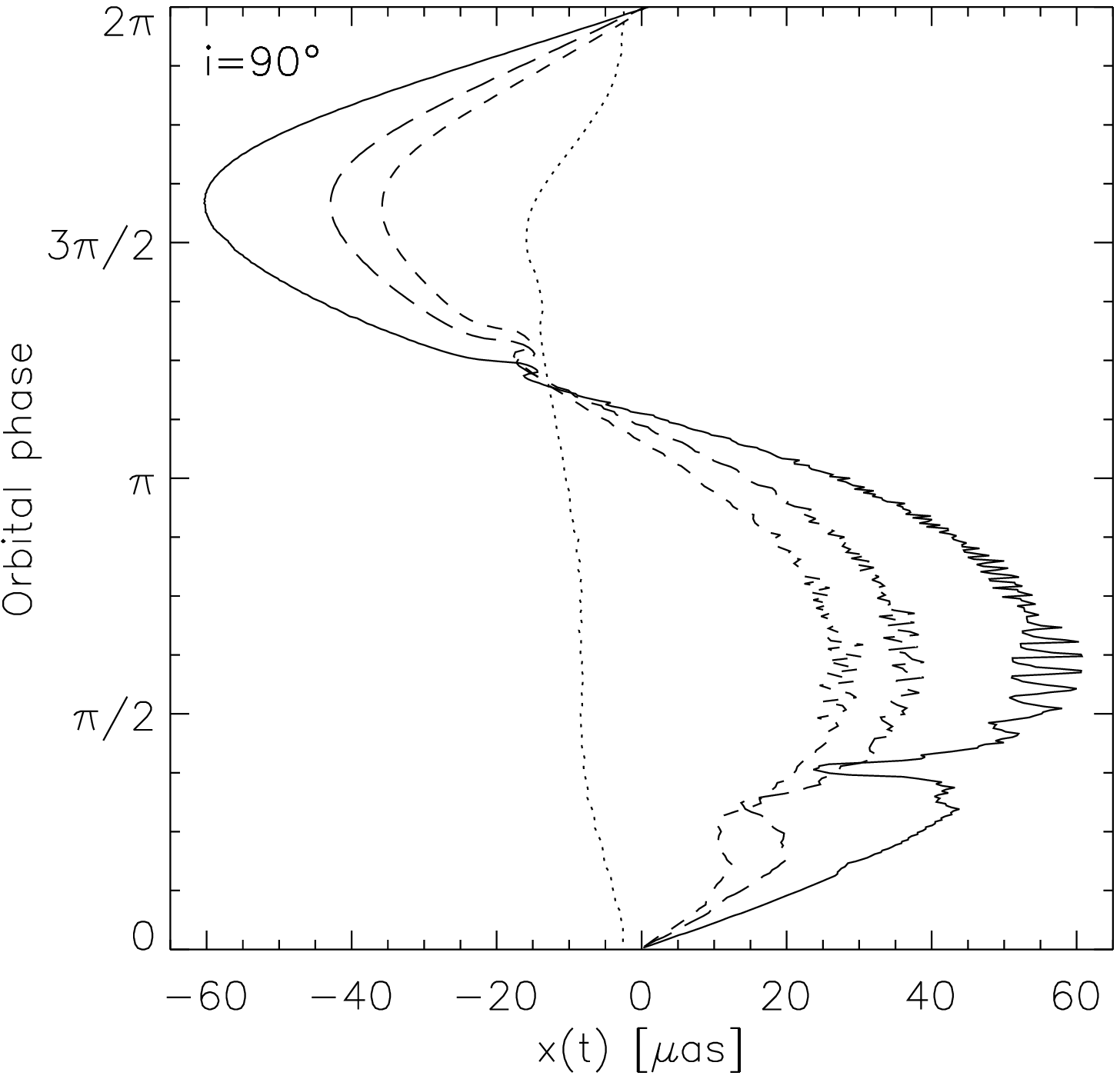}}
 \caption{Light curves (top) and centroid tracks (bottom) of a compact hot spot orbiting a Kerr black hole with four different spin-parameters ($a=0$ (solid), $0.52$ (long-dashed), $0.7$ (short-dashed), $0.998$ (dotted), inclinations are given in the upper left corner of each plot, the orbital radius is $r=2.0r_{\mathrm{ISCO}}$). \label{spin}}
\end{figure}

The impact of different black hole spin-parameters on light curves and centroid tracks is shown in Fig.~\ref{spin}. Increasing the spin-parameter essentially produces the same changes as decreasing the orbital radius of the hot spot, both in the light curves and the centroid tracks. This is due to the fact that we specify the orbital radius in units of the ISCO, which shrinks with higher black hole spin ($r_{\mathrm{ISCO}}\approx 3\Rs$, $2.1\Rs$, $1.7\Rs$, $0.6\Rs$ for $a=0$, $0.52$, $0.7$, $0.998$, respectively). By looking more closely at the Einstein ring peaks in the $90\degr$-light curves, one can find a slight broadening with higher spin. This is because the source occupies a bigger fraction of the orbit at smaller radii. The result is that the required conditions for an Einstein ring are met for a larger orbital phase (this can already be seen in Fig.~\ref{radii_incl_s0}). In the high spin case ($a=0.998$) the shape of the light curves is dominated by multiple images.

With higher spin-parameters the centroid tracks show even greater deformation, i.e.\ the cusp is enhanced and the rear part of the track is dragged up further. In addition, a second cusp emerges on the receding side, due to the tertiary image of the spot. The cusps may again turn into loops provided the multiple images are bright enough. The lower right plot in Fig.~\ref{spin} shows the $90\degr$-case. Just like the sub-peaks in the light curves, the spikes in the deformed sine-pattern are broadened with higher spin-parameters of the black hole. In the high spin case, Einstein rings dominate throughout the entire orbital motion. Thereby the motion of the hot spot is almost completely shrouded.

\begin{figure}
 \resizebox{\hsize}{!}{\includegraphics{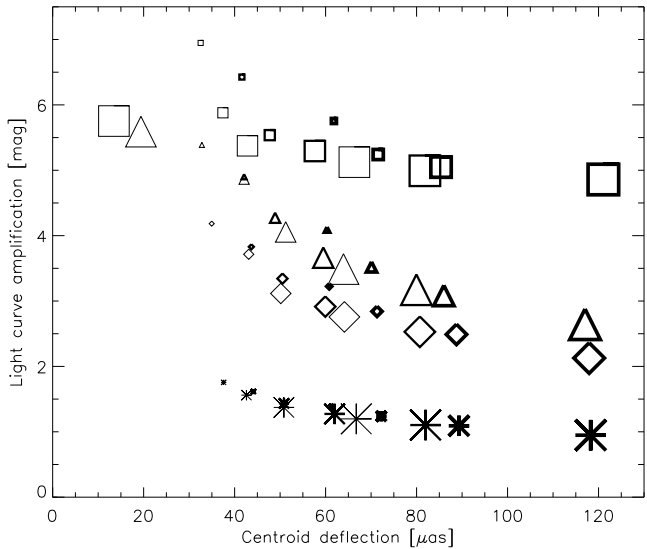}}
 \caption{Amplification of the light curve plotted against the horizontal diameter of the centroid track for each simulated grid-configuration of the compact hot spot. The symbol-shapes indicate different inclinations ($i=20\degr$ (star), $50\degr$ (diamond), $70\degr$ (triangle), $90\degr$ (square)). The symbol-size indicates the orbital radius ($r/r_{\mathrm{ISCO}}=2.0$ (large), $1.5$ (medium), $1.2$ (small), $1.0$ (tiny)). The symbol-thickness indicates the spin-parameter ($a=0$ (bold), $0.52$ (thick), $0.7$ (medium), $0.998$ (thin)). \label{gravity}}
\end{figure}

Aside from its influence on the radius of the ISCO, the black hole spin has only marginal impact on either the shape of light curves and centroid tracks \citep[see also][]{bl05,bl06a}. For this reason it is presumably impossible to deduce the spin-parameter from the light curves alone. However, including polarimetric properties of this model allows making stronger statements about the spin parameter. On the basis of observational data presented by \citet{eckart06}, \citet{Meyer06a, Meyer06b} were able to constrain the spin parameter of Sgr~A* to the range $0.4\le a \le 1$ on a $3\sigma$ level.

Another possible approach to measure the spin of Sgr~A* would be the combination of photometric and astrometric observations. This is illustrated in Fig.~\ref{gravity}, where we compare the maximum-to-minimum flux-amplification in the light curves with the maximum horizontal deflection of the centroid tracks for each grid-configuration. Obviously, in such a representation it is possible to distinguish every single configuration by the set of parameters $\left\{r,i,a,P\right\}$. This is because the inclination mainly influences the light curve peaks and does not change the horizontal centroid deflection that much, whereas the orbital radius has a stronger influence on the centroid deflection than on the light curves. Changing the spin-parameter of the black hole has no big effect on the location of a configuration with given orbital radius in this plot. Configurations with identical locations (i.e.\ same orbital radius and inclination) can be distinguished by their orbital period $P$, which is known from the light curves. With this the spin-parameter is determined via Eq.~(\ref{eq:period}).

\subsubsection{Polarimetry}

Recent data show that the NIR flare emission is polarized \citep{eckart06,Meyer06a,trippe07}. It is not yet certain whether the magnetic field in the accretion disk is mostly poloidal or toroidal, but there is evidence both from simulations by \citet{devilliers03} and observations by \citet{trippe07} in support of a toroidal configuration. Assuming the magnetic field within the hot spot (compact sphere) to follow this geometry one can investigate the polarimetric properties (temporal evolution of polarization fraction and angle) of the rotating hot spot scenario.

A detailed treatment was carried out by \citet{bl06a}. They find that the polarized flux follows the evolution of the total flux with a short delay. Additionally, small dips in the polarized flux light curve are produced by the multiple images of the source. The variation of the polarization angle is mainly influenced by the inclination of the hot spot orbit. Within one orbital period it rotates continuously, but is punctuated by rapid rotations (around $90\degr$) at times where the dips in the polarized light curve appear. In case of an edge-on view, the angle remains mostly constant on account of axial symmetry, but the polarization fraction drops significantly during Einstein ring events.

\citet{Meyer07} used polarimetric data together with hot spot simulations to estimate the orientation of the Sgr~A* system. They investigated two different magnetic field geometries (sunspot-like and toroidal) and derive the position angle of the Sgr~A* equatorial plane normal to be between $60\degr$ and $108\degr$ (east of north) on a $3\sigma$ level. Besides this, a high inclination angle of the hot spot orbit is favoured.

In the context of the VLTI we also want to study the effect of polarization on the centroid motion of the flares. Astrometric measurements using polarizing filters may help filter out physically different components of the emission region (e.g.\ sphere and arc, see section \ref{arc}) and thus increase the apparent proper motion of the single components. Furthermore, the signatures of multiple images may be easier to detect, since they generally possess different polarization angles with respect to the primary image. To this end we have simulated centroid tracks produced by a small, constant source, observed through two perpendicular polarizing filters in the NIR.

\begin{figure*}[t]
 \resizebox{\hsize}{!}{\includegraphics{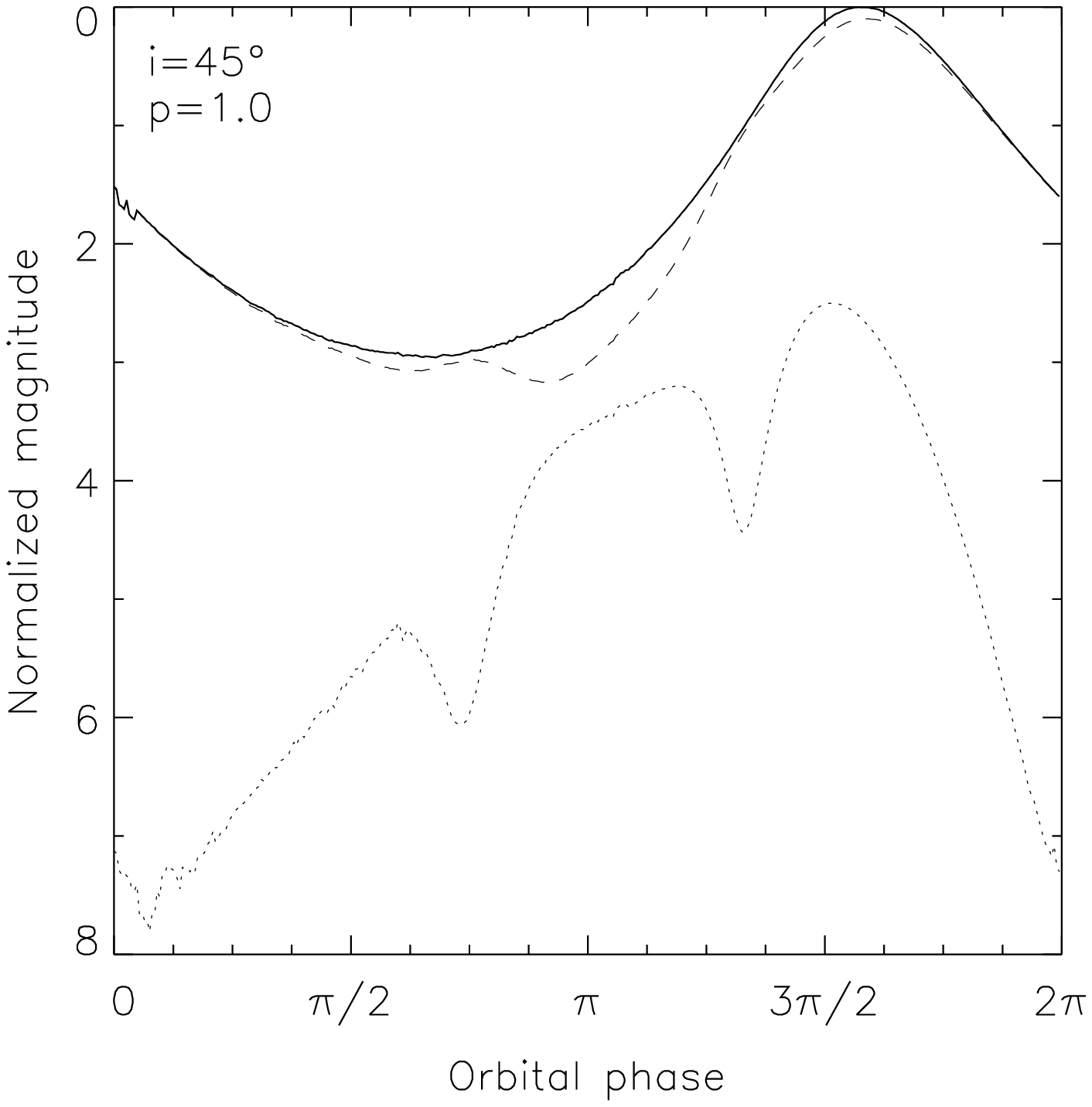}\includegraphics{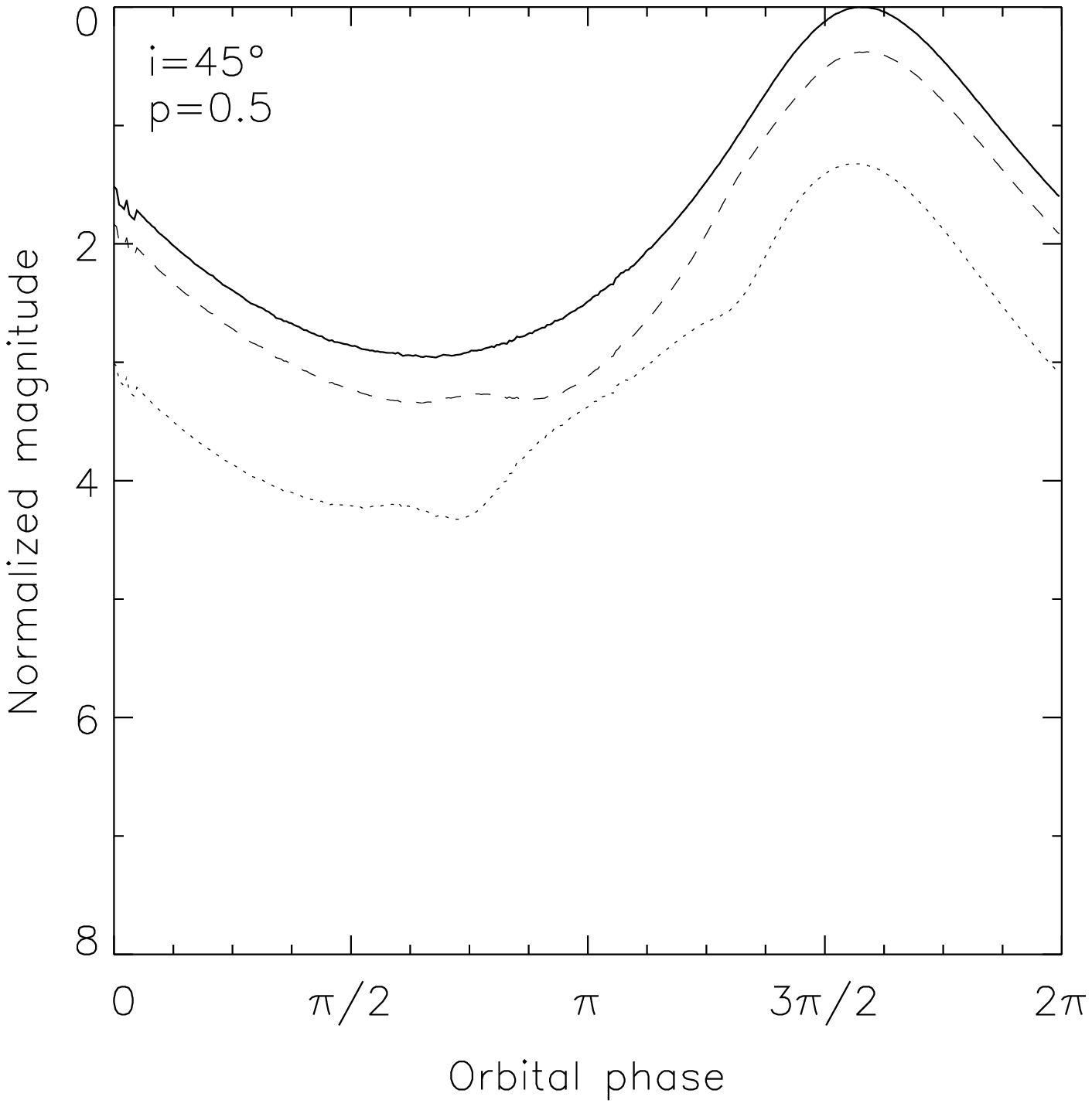}\includegraphics{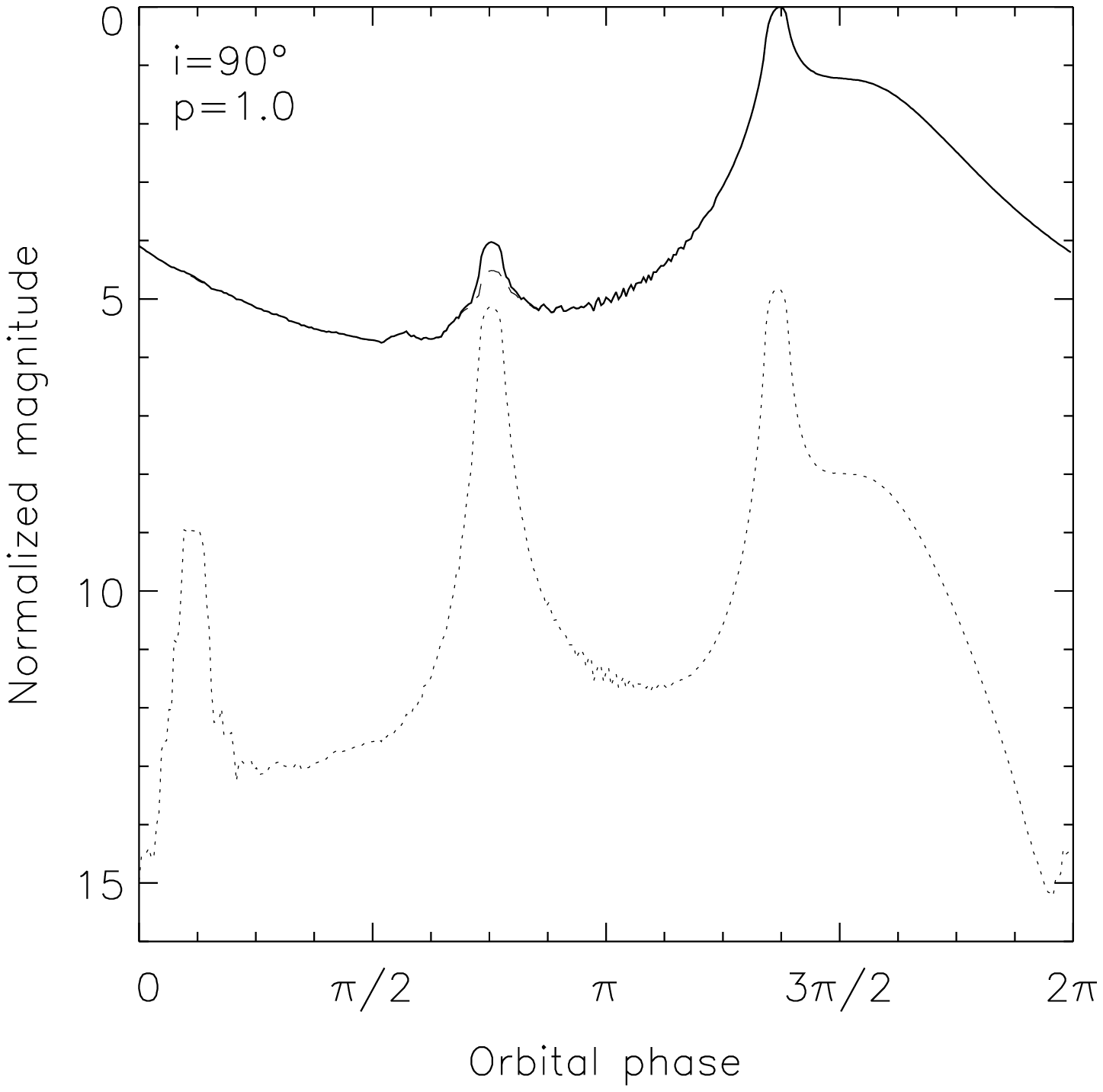}\includegraphics{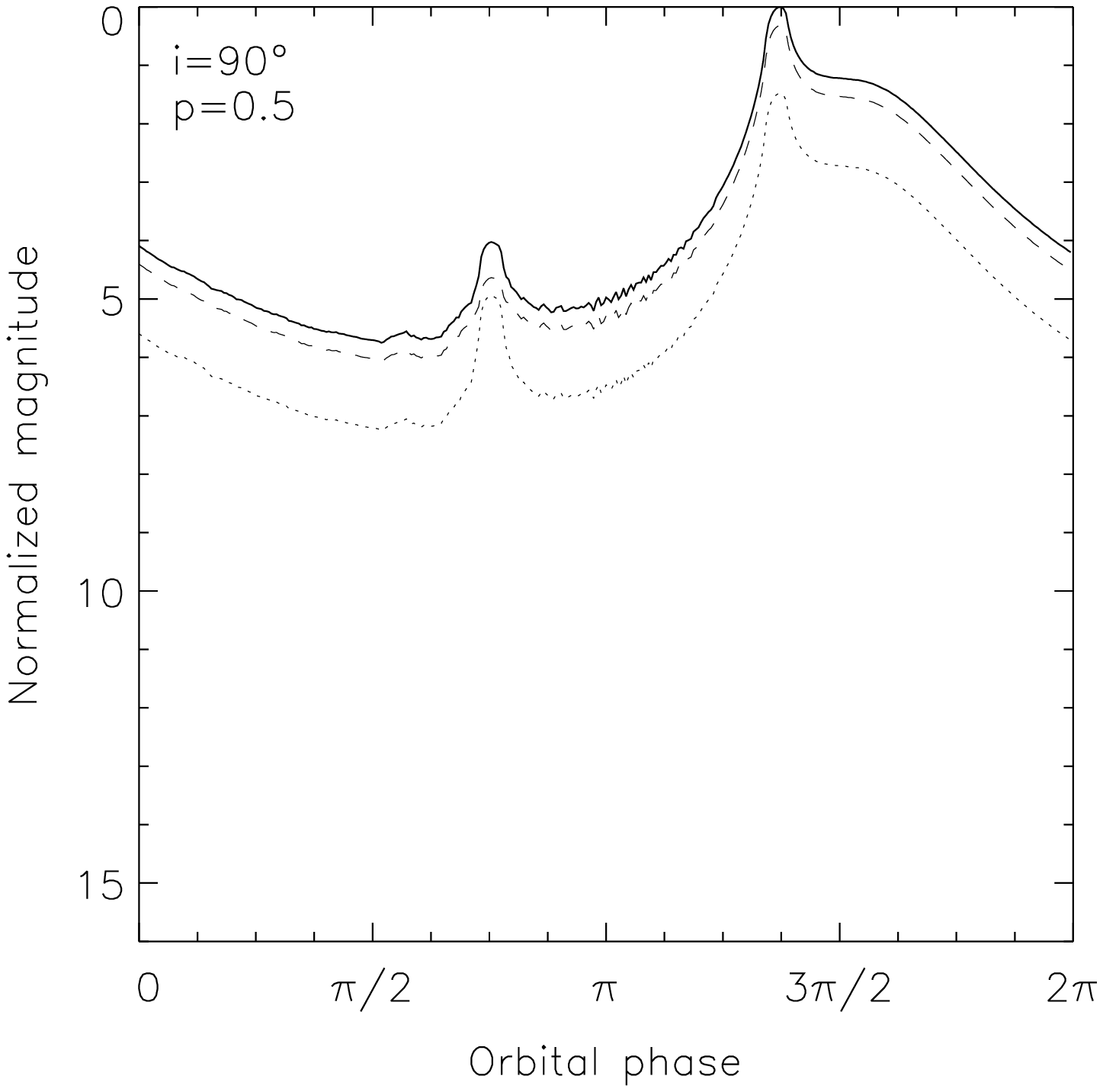}}
 \resizebox{\hsize}{!}{\includegraphics{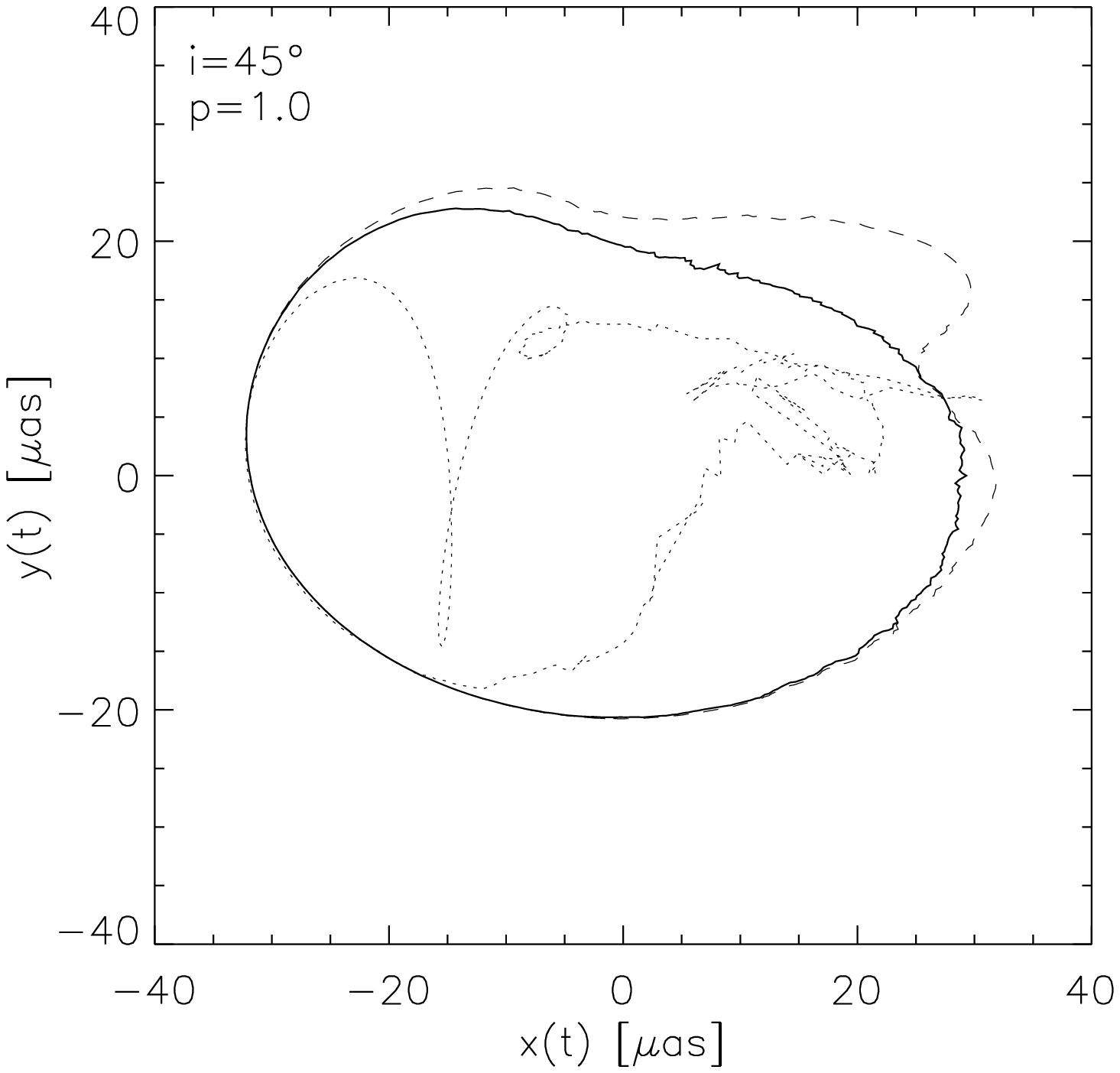}\includegraphics{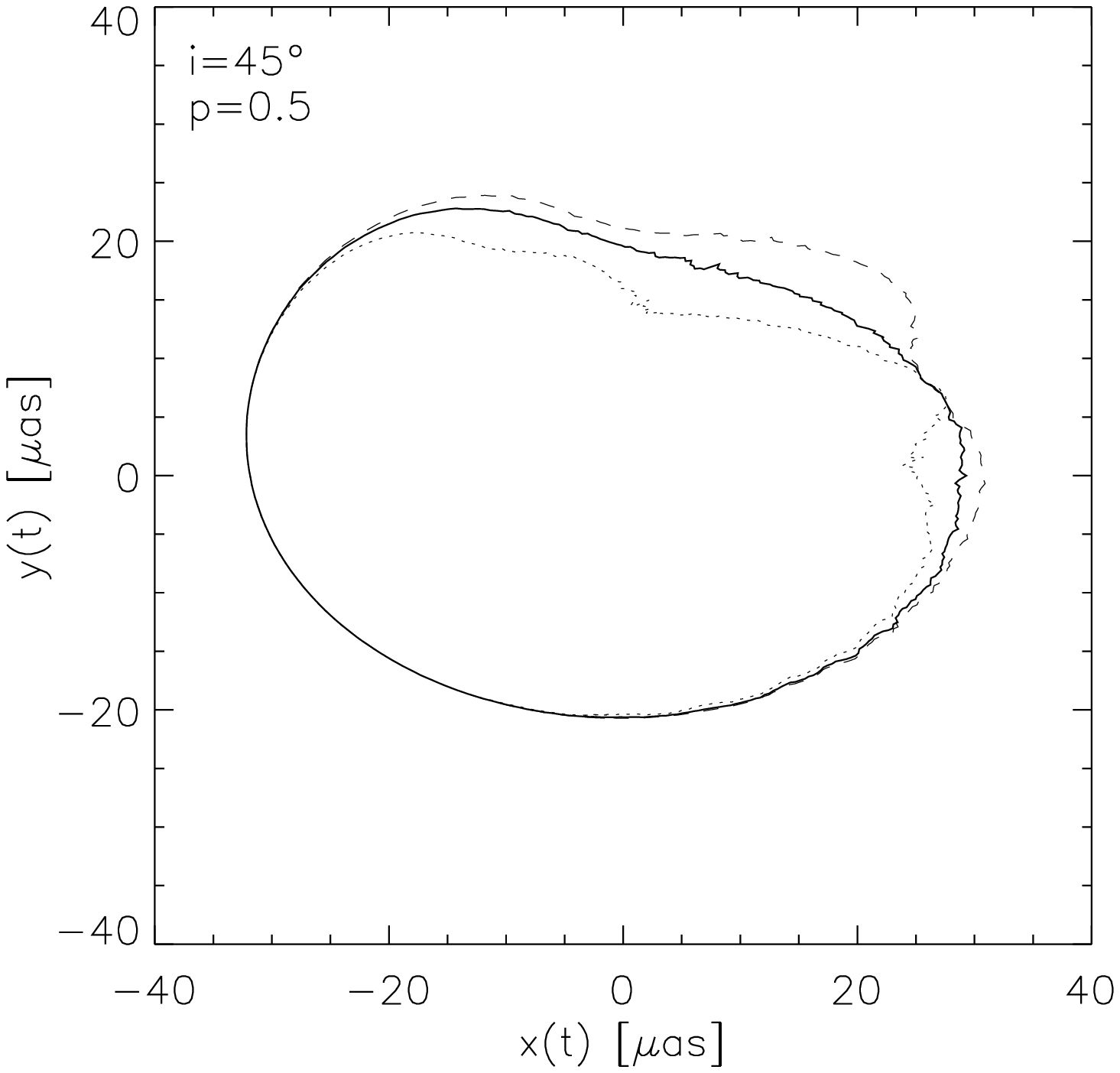}\includegraphics{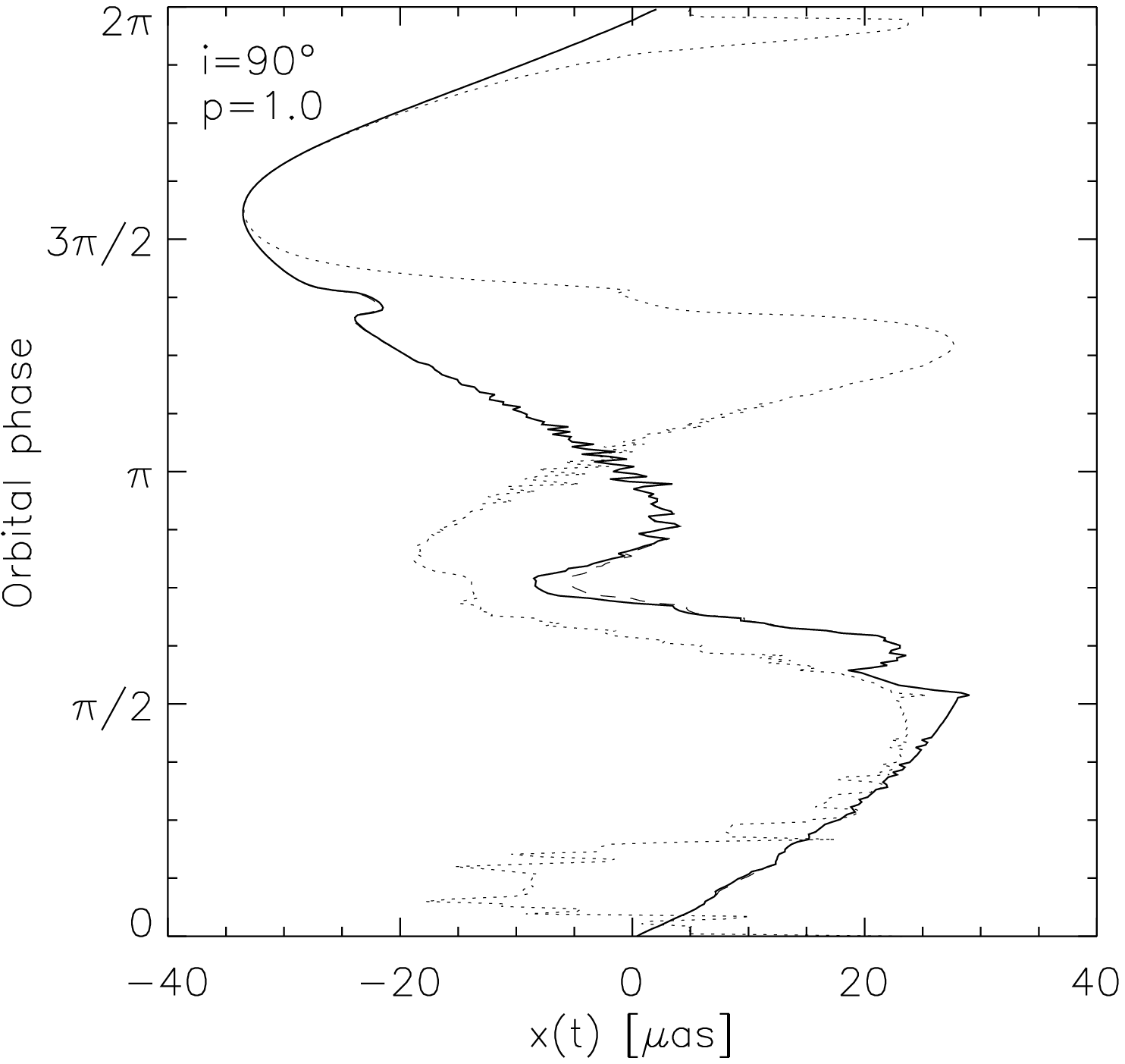}\includegraphics{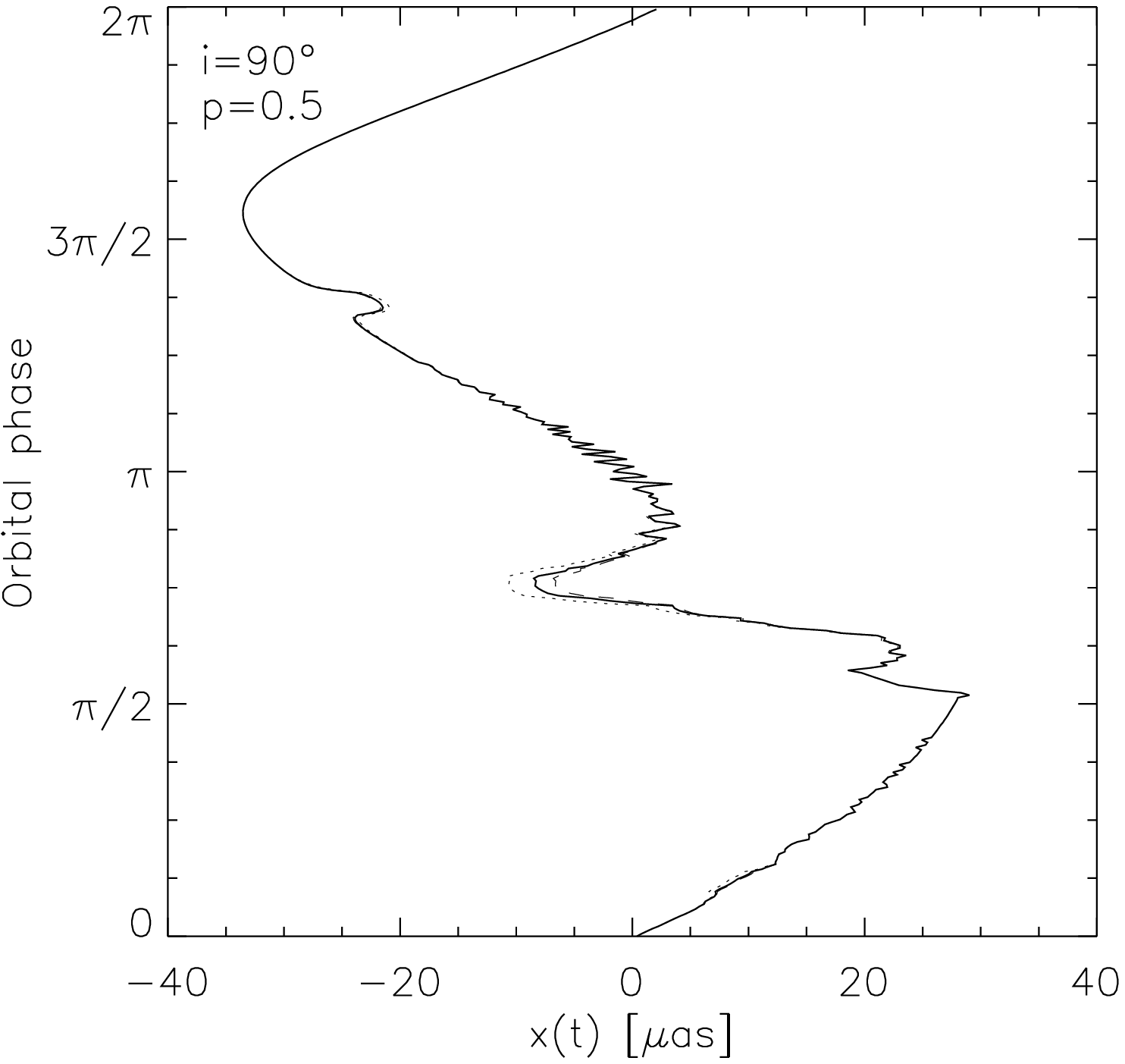}}
 \caption{Light curves (top) and corresponding centroid tracks (bottom) in polarized light, produced by a small sphere on the ISCO around a Schwarzschild black hole seen in $45\degr$ and $90\degr$ inclination. For the source we assume a spectral index of $\alpha=0$, a toroidal magnetic field, and a polarization fraction of $100\%$, respectively $50\%$ (polarization fraction and inclination are given in the upper left corner of each plot). The polarizing filter axis is oriented along the orbital line of nodes (dotted lines) or along the orbital spin axis (dashed lines). The solid curves indicate the total flux. \label{polarimetry}}
\end{figure*}

The results are shown in Fig.~\ref{polarimetry}. The effect on the astrometry can be very important when the polarization fraction~$p$ of the source is high. The centroid tracks show pronounced features caused by the multiple images of the hot spot, especially when observed in the polarization component along the orbital line of nodes. However, most of the flux is blocked by the filter in this observing mode, as can be seen in the light curves. It only rises up shortly when the multiple images (respectively Einstein rings) are amplified most strongly, causing the swing in the polarizing angle. Thus, an auspicious observation of this kind might be hard to achieve. Moreover, observational data support a rather low polarization fraction (\citet{trippe07} mostly see $10\%$--$15\%$). As apparent from Fig.~\ref{polarimetry}, a polarization fraction of $50\%$ already reduces the polarimetric deviations from the common centroid tracks to less than the projected accuracy of order $10$~\muas.

\subsection{Extended, variable source \label{arc}}

The conception of a single compact object revolving on a close orbit about a supermassive black hole, maintaining a constant shape and brightness is somewhat artificial and has presumably little physical relevance. In order to improve this model we implement, in addition to the emitting sphere, an elongated arc that spreads from this sphere along the orbit. A similar model was already adopted by \citet{Meyer06a, Meyer06b}, it is mainly motivated by the broad shape of the observed light curve peaks from Sgr~A*. By fitting their model to observed flare data they find a constant elongation of the arc troughout the flare. This can also be inferred from the fact that the width of the major peaks in the flare light curves does not change significantly during one flare. A steady broadening of the peaks, for instance, would indicate gravitational shear of the emitting region. However, this does not seem to happen given the data at hand.

The crucial assumption for considering gravitational shear is to neglect any bonding forces within the plasma around the black hole. However, a strong magnetic field is likely to be present and may prevent hot spots from being sheared on the Keplerian timescale. The characteristic shear-time is then increasing with higher magnetic field strengths, due to a \emph{magneto-rotational instability} (MRI) \citep{Balbus}. It acts against the tidal forces caused by the gravitational field and creates an attractive force among adjacent elements of the accretion disk, trying to enforce a rigid rotation. Thus, in the case of a very strong magnetic field, the hot spot can be regarded as a rigid source, retaining its shape during the whole duration of a flare.

We vary the intrinsic brightness of the two components (sphere and arc) with time to emulate heating and cooling sequences. We assume that the flares are due to a heating process (e.g.\ magnetic reconnection) taking place in a small region of the inner disk. This heating process has its own lifecycle that can be described with a rise time and a decay time of each component.

The timescales for these processes are determined by the synchrotron cooling time $\tau_{\mathrm{syn}}$, which depends on the magnetic field strength $B$ and the wavelength $\lambda$ of the observed synchrotron radiation. For the case of Sgr~A* it is estimated as \citep{gillessen06}:
\begin{equation}\tau_{\mathrm{syn}}\approx 8\left(B/30\mathrm{G}\right)^{-3/2}\left(\lambda/2\mu\mathrm{m}\right)^{1/2}\mathrm{min}\end{equation}
The magnetic field strengths predicted by radiatively inefficient accretion flow (RIAF) models for Sgr~A* lie around $20\mathrm{G}$, yielding cooling times that are less than the duration of a flare \citep{Quataert03,yuan04}. Computational models in general relativistic magnetohydrodynamics (GRMHD) suggest that in the close vicinity of a Kerr black hole the magnetic field strength within the accretion disk decreases with radial distance. In contrast, higher spin-parameters lead to stronger magnetic fields \citep{devilliers03}. These dependencies are fairly complicated and presumably cannot be derived analytically. Since a full GRMHD simulation exceeds the scope of this work, we empirically determine the heating and cooling timescales from the observed light curves of Sgr~A*.

It must also be mentioned that up to now, self-consistent GRMHD simulations can only produce hot spots with very short life times, amounting to fractions of the orbital period \citep[$\tau\approx0.3P$,\ ][]{Schnittman06}. Hence, there have been attempts to explain the observed light curves with multi-spot models \citep{eckart08a}. Several hot spots distributed radially around the black hole can produce similar quasi-periodic signals in the light curves as a single source. However, with the number of spots the flux amplification in the light curves and the deflection of the emission centroid is increasingly reduced. Unknown mechanisms that may stabilize hot spots in the accretion flow may cause longer life times, though.

\begin{figure}[t]
 \resizebox{\hsize}{!}{\includegraphics{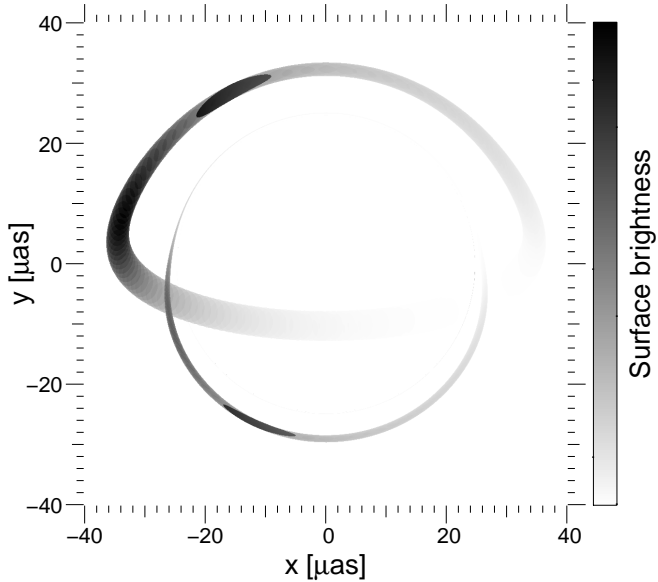}}
 \caption{A composition of ray traced images of a sphere on the ISCO about a Schwarzschild black hole in $70\degr$ inclination. The resulting hot spot is composed of a bright compact sphere in its center and an extended arc that follows a Gaussian brightness profile. Due to relativistic beaming, this profile appears modified.\label{blob_arc_image}}
\end{figure}

We implement the arc by adding up all the sphere's ray traced images of one complete orbit. This results in a lensed ring around the black hole, including its multiple images. Assigning a spatial and temporal brightness distribution to this ring allows us to simulate any desired hot spot. We use simple Gaussian functions to describe the azimuthal brightness distribution of the arc and the temporal evolution of the entire hot spot. We leave the radial brightness distribution of the hot spot unaltered, i.e.\ it remains confined to a radial region of $0.5\Rs$ diameter (see Fig. \ref{blob_arc_image}).

The spectral indices of sphere and arc remain to be specified. Observations of the GC-flares suggest a connection between brightness and color \citep{gillessen06}. When the flare is dim it shows a red spectrum, whereas in bright states it is rather white. Since in our model the sphere is brighter than the arc, motivated by the observations, we use $\alpha^{\mathrm{sphere}}=0$ and $\alpha^{\mathrm{arc}}=-3$, thus covering the range of observed values \citep{eisenhauer05sinfoni,gillessen06,Krabbe06,hornstein07}.

\subsection{Fitting observational data}

In this section we fit the parameters of our hot spot model to observational data from Sgr~A*. To this end a $\chi^2$-minimization is applied to fit the simulated light curves to the observed ones. However, for the fit of a flare light curve with a given periodicity we have to restrict ourselves to those model configurations with nearly the same orbital timescale. Thus, only grid-configurations with fixed $r$ and $a$ can be considered and leave us with four different inclinations per fit (our grid does not contain configurations with different values for $r$ and $a$ for a given orbital period). However, it is not the objective of this paper to estimate the spin-parameter of Sgr~A* (this has already been done by \citet{Meyer06a, Meyer06b}). We rather want to focus on the astrometric study of the hot spot model. Hence, for the parameters describing the hot spot itself, we can calculate optimized values.

We used photometric measurements of two bright GC-flares, namely the flare from April 4 and from July 22, 2007. A detailed discussion of the fitting procedure is given in the following subsections.

\subsubsection{L-band flare from April 4, 2007}

\paragraph{Photometry}

\begin{figure*}
 \resizebox{\hsize}{!}{\includegraphics{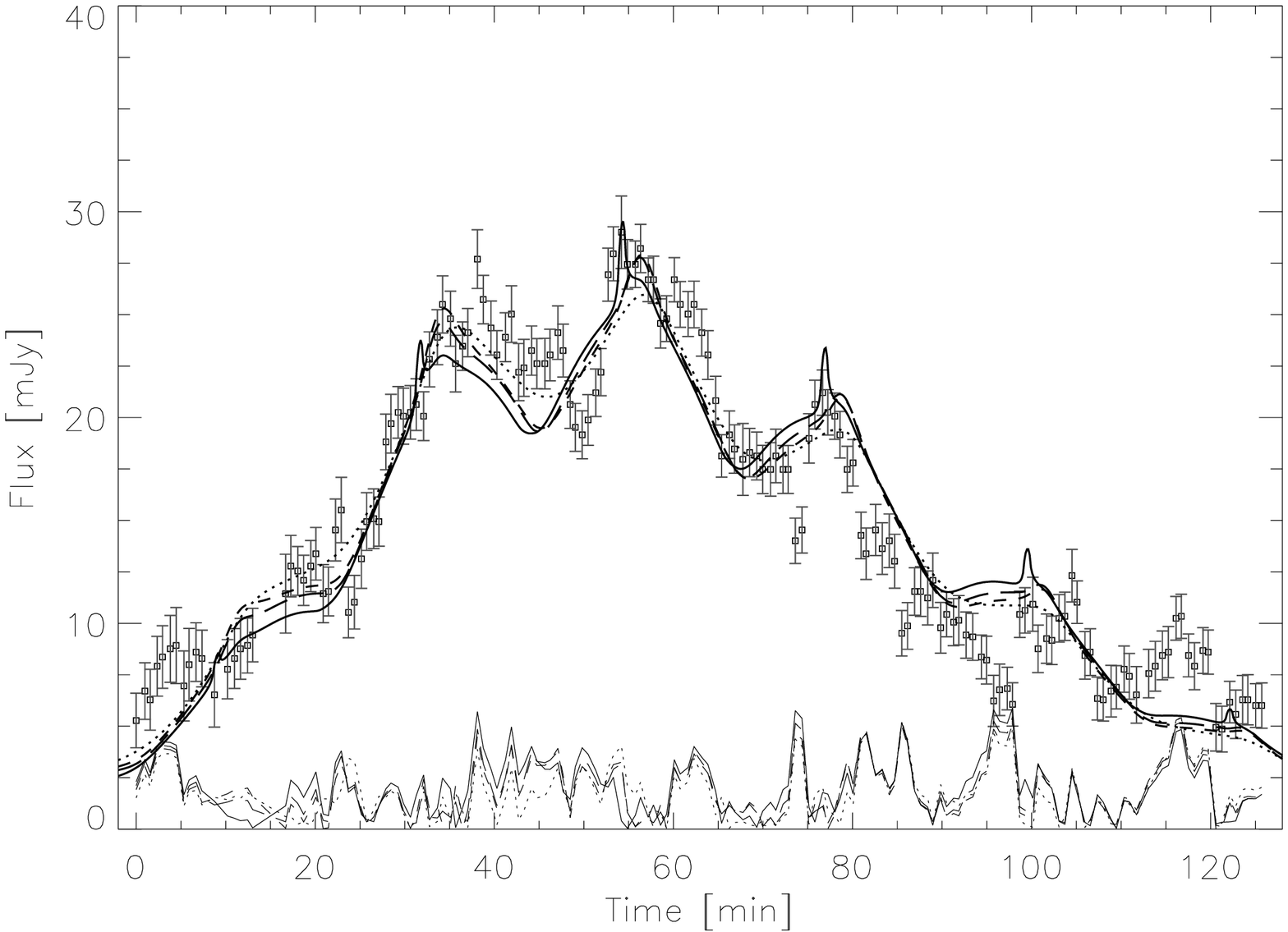}\includegraphics{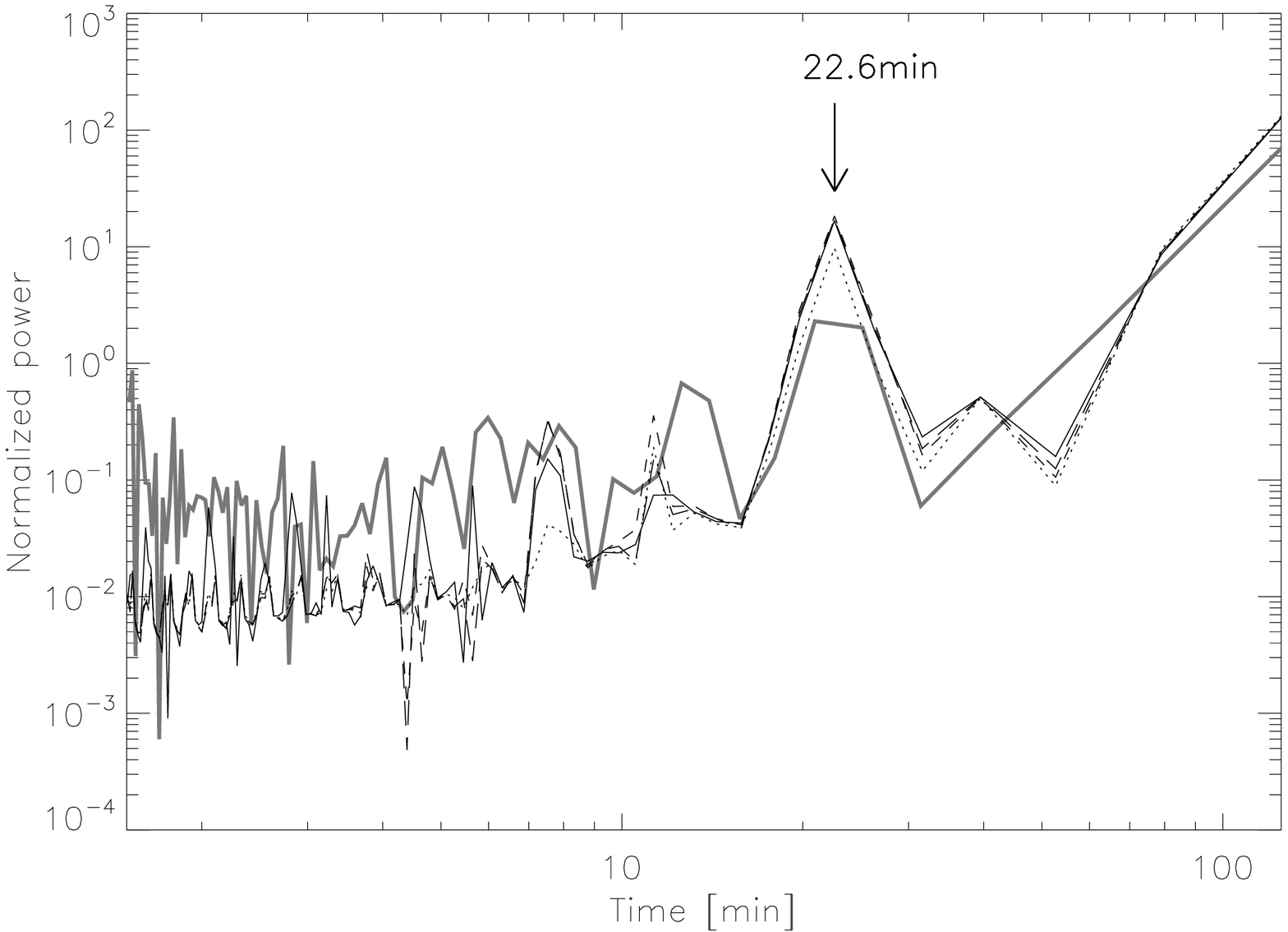}}
 \caption{Left: Best fit light curves for the L-band flare from April 4, 2007 for four different inclinations ($i=20\degr$ (dotted), $50\degr$ (short-dashed), $70\degr$ (long-dashed), $90\degr$ (solid)). The gray error bars represent the measurement, at the bottom the residuals of the fits are drawn. Right: Power spectrum of these light curves (measurement in gray). \label{fit_0404}}
\end{figure*}

The light curve of this flare has three major, fairly broad peaks with quite a substructure (see Fig. \ref{fit_0404}). Additionally, four smaller peaks are found on the wings of the light curve. The power spectrum indicates a periodic timescale of about $23$~min. Any details about the observation of this flare and the used instrumentation can be found in \citet{dodds-eden08}. For the fit we choose grid-configurations with $r=1.5r_{\mathrm{ISCO}}$ and $a=0.7$, since they yield the closest orbital timescale of $P=22.6$~min.

In Fig.~\ref{fit_0404} the best fits for four different inclinations are shown. Each light curve constitutes seven orbital motions (according to the number of peaks) of a compact sphere ($\alpha^{\mathrm{sphere}}=0$) with an extended arc ($\alpha^{\mathrm{arc}}=-3$). For the parameters of the hot spot we find the values given in Table~\ref{parameter_0404}. The decay time of the hot spot is slightly larger than its rise time, because the light curve is more flattened on the decaying flank. The arc provides quite a strong contribution to the total flux: The surface brightness at its center is only four times less than the surface brightness of the sphere. It is spread along the orbit, possessing a full width at half maximum ($\mathrm{FWHM}\approx2.355\sigma_{\mathrm{arc}}$) of more than one complete orbit. This value is required by the broad peaks in this flare light curve.

\begin{table}[h]
\newcommand\T{\rule{0pt}{2.6ex}}
\begin{center}
\begin{tabular}{lccc}
\tableline\tableline
Parameters: \T & Sphere & & Arc  \\
\tableline
Rise time \T $\sigma_{\mathrm{rise}}$ & $1.1P$ & & $1.1P$ \\
Decay time \T $\sigma_{\mathrm{decay}}$ & $1.8P$ & & $1.8P$  \\
Brightness ratio \T & 4 & / & 1 \\
Extension (FWHM) \T & - & & $1.41C$ \\
\tableline
\end{tabular}
\end{center}
\caption{Best fit parameters for the L-band flare from April 4, 2007. $P$~is the orbital period and $C=2\pi r$ the circumference of the hot spot orbit. \label{parameter_0404}}
\end{table}

The reduced $\chi^2$-values favor lower inclinations. However, one has to be careful with drawing conclusions from this for the following reason: The arc-width $\sigma_{\mathrm{arc}}$ and the inclination~$i$ are, to a certain extent, degenerate parameters that both determine the width and the height of the light curve peaks. Nevertheless, the implementation of an extended arc is necessary in order to fit the broad light curve peaks in the flare. The inclination only moderatly influences the flare shape, by far weaker than the arc-width does.

The fit worsens towards the wings of the flare. By checking the spacing between the major peaks one finds that the outer peaks do not match the $23$~min timescale as well as the inner ones. One could argue for either an acceleration or a deceleration of the orbital period, corresponding to shrinking or growing orbits, respectively. But, a close look at the spacing between the outer peaks does not support this reasoning, since their occurrence deviates from the orbital periodicity in both ways.

\begin{figure*}
 \resizebox{\hsize}{!}{\includegraphics{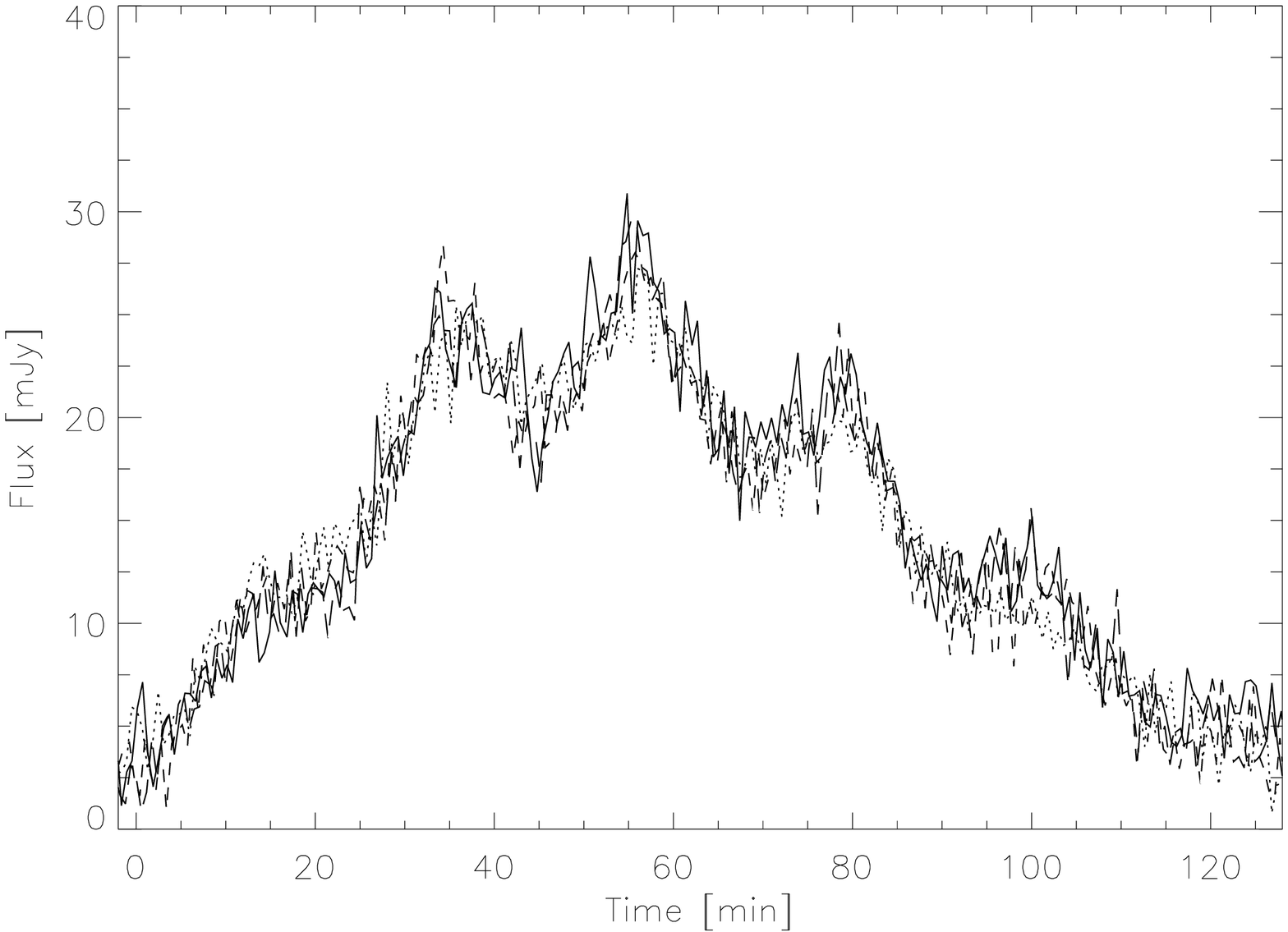}\includegraphics{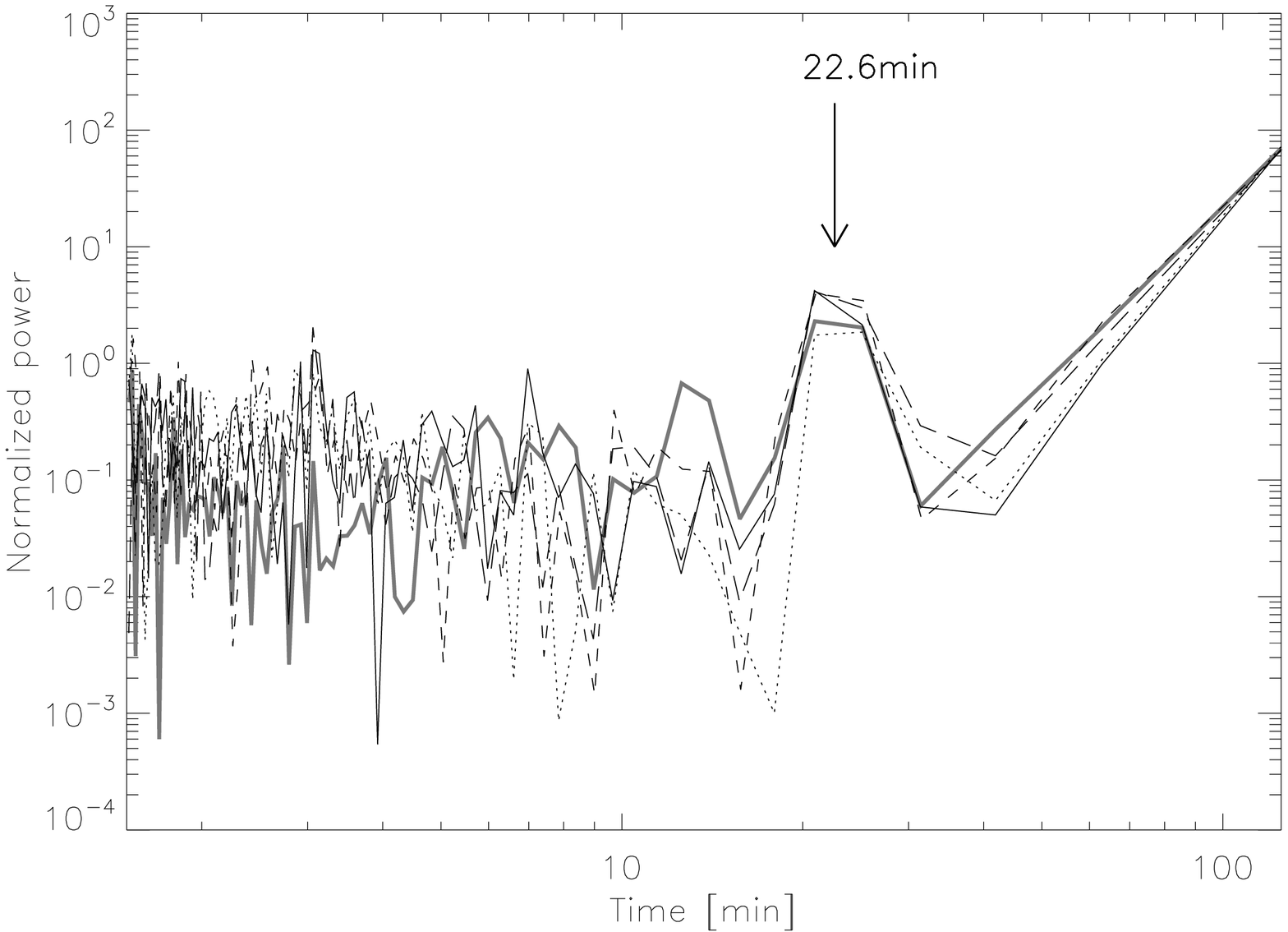}}
 \caption{Left: Best fit light curves from Fig.~\ref{fit_0404} with Gaussian noise added. Right: Power spectrum of these light curves (measurement in gray). \label{fit_0404_noise}}
\end{figure*}

Higher inclinations produce sharper sub-peaks on top of the major peaks. These are generated by the compact sphere and reproduce some substructure in the observed flare quite well. The Einstein ring peaks in the $90\degr$-case are still compatible with the data, however, the sampling of the observed light curve is too sparse to constrain them. An unambiguous detection of such a peak could directly suggest the occurrence of an Einstein ring, a high-order general relativistic effect.

Having found a reasonable fit to the observed flare light curve it is of interest to extract its periodic properties by computing a power spectrum and comparing it to the one obtained from the experimental data. To this end we use a \emph{Lomb normalized periodogram} \citep{lomb76,scargle82} which is a standard tool for analyzing unevenly sampled time series. It is normalized by the variance of the overall flare light curve flux.

The power spectra for the four best fit light curves from above are shown in the right inset of Fig.~\ref{fit_0404}. Obviously, the $23$~min feature is nicely reproduced in the simulated data. Higher inclinations show more power in the periodogram due to the fact that the peaks in the corresponding light curves are more pronounced. The rise towards longer timescales is caused by the overall structure of the flare light curves, yielding a periodic timescale of the whole flare duration. Higher frequencies, though, are suppressed in comparison to the observed data. This is due to the fact that the source as well as the observing instrument generate intrinsic noise. While the power spectral density (PSD) of the instrumental noise is usually Gaussian (i.e.\ constant), the one generated by the source may have a more complicated profile (like a broken power law, for instance). In this case the overall power spectrum of the observed flare shows a slight rise to longer timescales.

In order to include the observational process into our simulations we artificially add random noise to our data. We can also mimic the flux-integration at the detector of the telescope by averaging over several successive data points. Fig.~\ref{fit_0404_noise} shows the light curves from above, treated with the said procedure. First, the light curves were integrated with an integration time of 45 seconds, which is the average value from the observation. Then, a normally (Gaussian) distributed noise with a mean of zero and a standard deviation of 5 percent of the flare-peak-flux was superimposed. This value can be obtained by comparison of the mean error in the flare-data to its maximum flux-value.

\begin{figure*}
 \resizebox{\hsize}{!}{\includegraphics{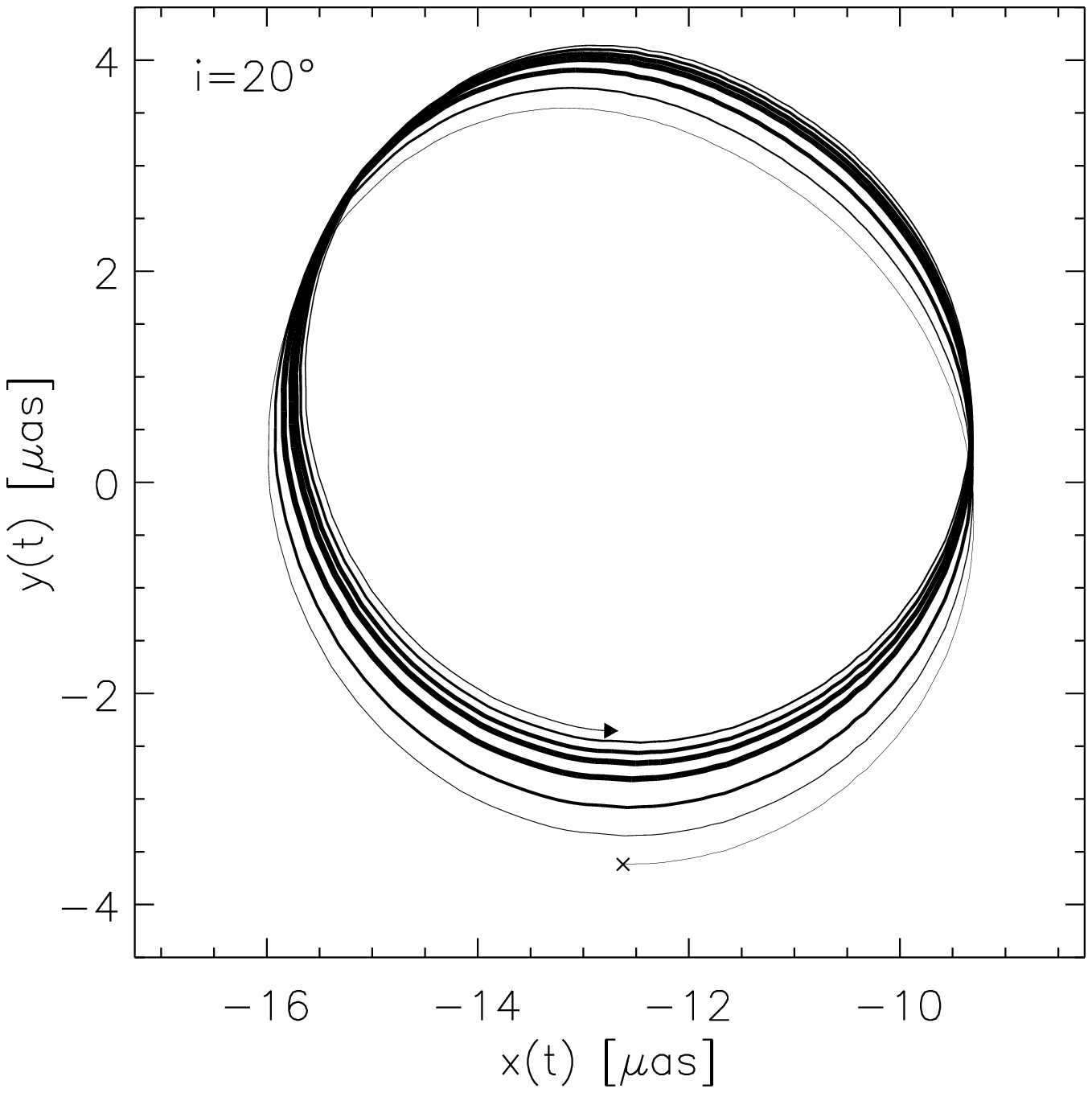}\includegraphics{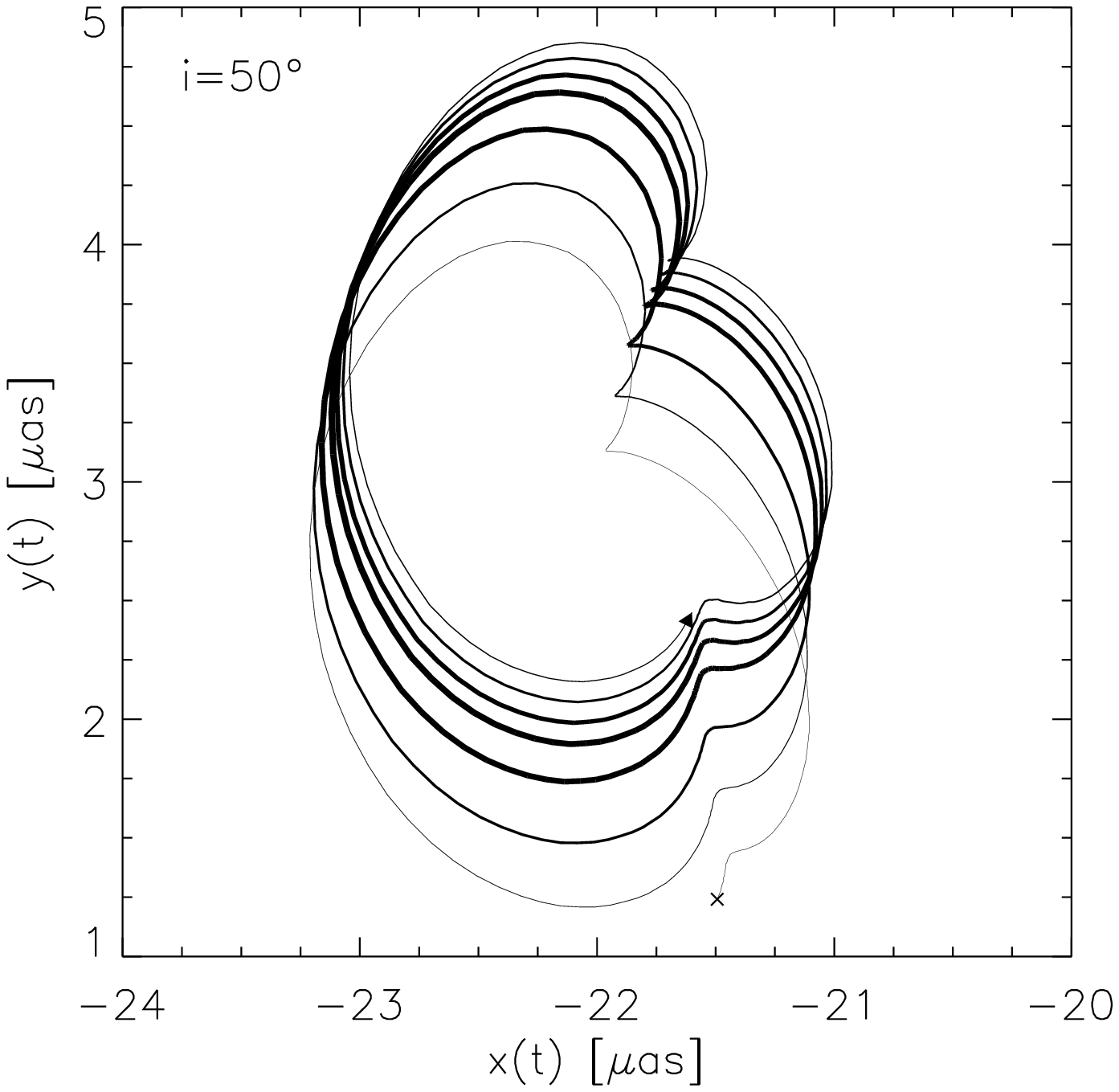}\includegraphics{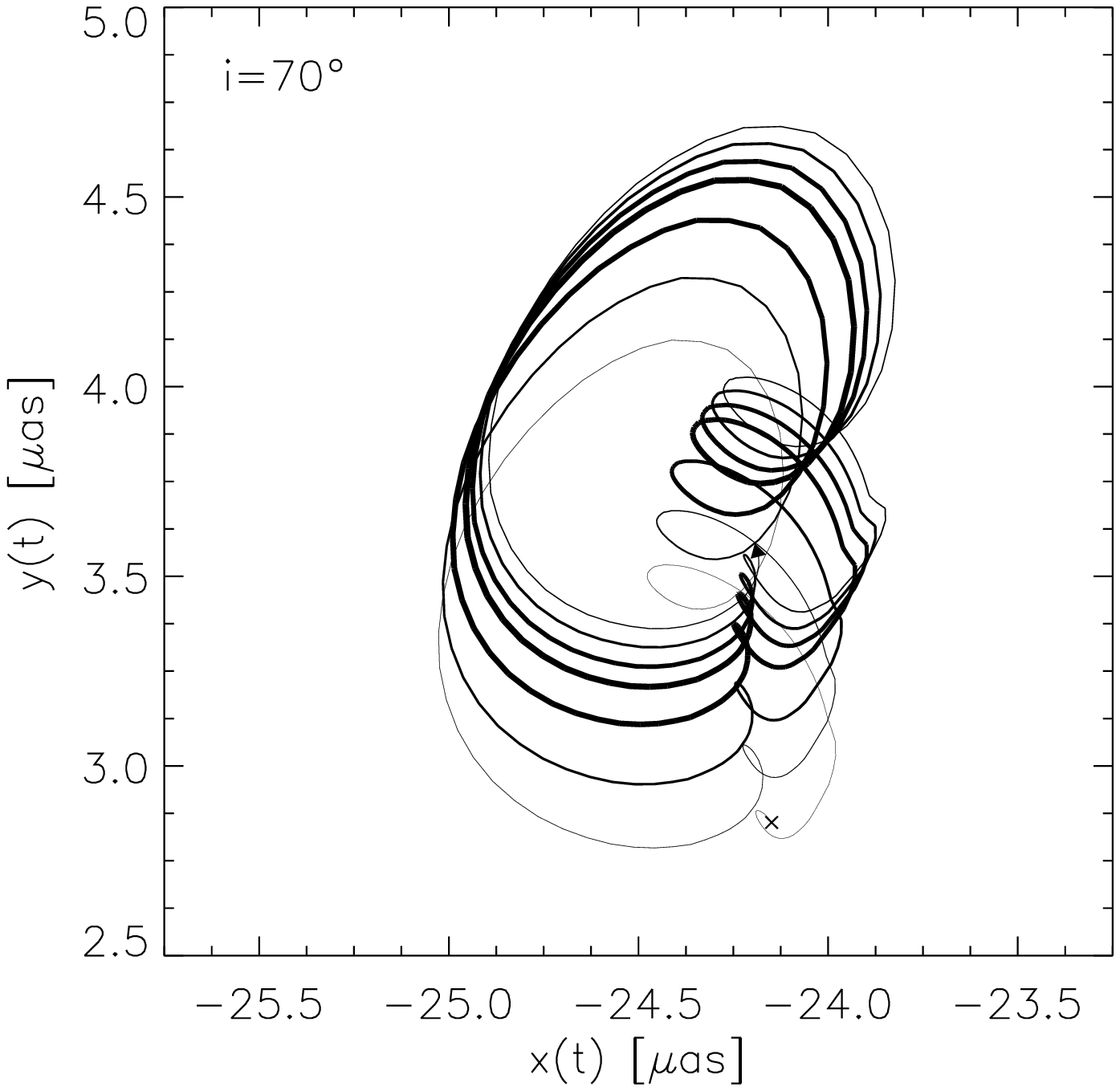}\includegraphics{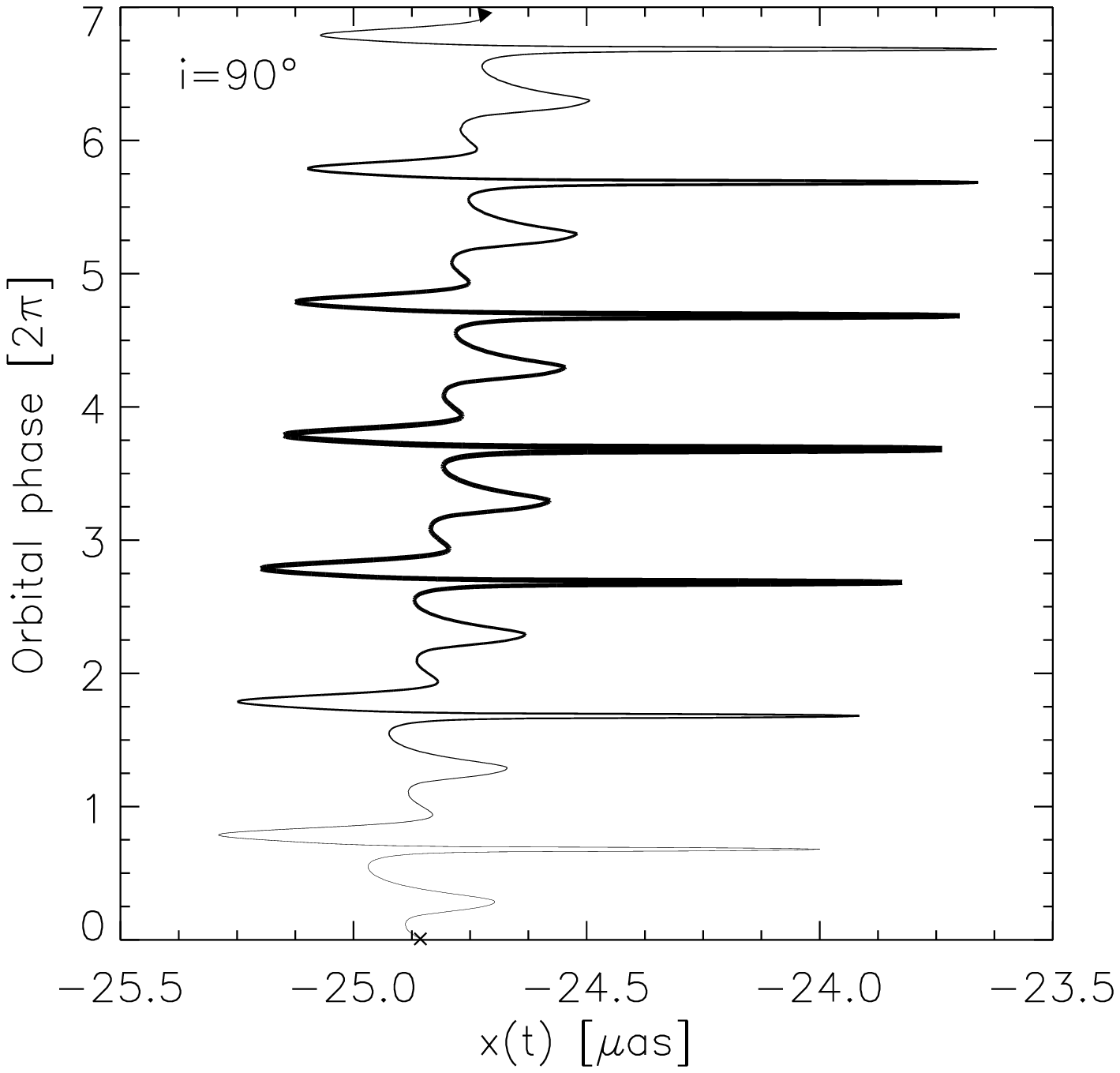}}
 \caption{Centroid tracks of the best fit model for the L-band flare from April 4, 2007 for four different inclinations. The brightness evolution of the flare is indicated by the line thickness (the thicker the line the brighter the flare). A cross marks the beginning, an arrow the end of the flare. \label{fit_centroid_0404}}
\end{figure*}

The corresponding power spectra are presented in the right panel of Fig.~\ref{fit_0404_noise}. The power in the high frequencies is raised and has a flat shape, pretty similar to the observed data (the fact that there is more power in the simulated periodograms at short timescales may be due to an overestimation of the observational errors). The peaks associated with the $23$~min timescale now match up very well. The fact that we add Gaussian noise to the simulated data makes the power in this characteristic frequency decline. Thus, the observed light curves are consistent with the assumption of a clearly deterministic physical process (orbital motion) plus a superposed instrumental Gaussian noise.

\paragraph{Astrometry}

Our code additionally provides simulations for the centroid motion of the flare. This is a direct prediction of the hot spot model and may be checked by future observations. In Fig.~\ref{fit_centroid_0404} the centroid tracks of the four best fit models from above are displayed. They seem to show strong relativistic effects, however the axis labelling reveals the scaling. Due to the rather strong contribution of the extended arc, the centroid of emission does not move far, yielding deflections of about $1.5$ to $6~\mu$as during one orbit.

The centroid track is dragged to the approaching side because of relativistic beaming; lensing effects shift it upwards. The higher the inclination the stronger are these effects, except for the $90\degr$-case where symmetry avoids a y-direction shift. Due to the delay between primary and secondary image, the centroid track slowly moves upwards, following the emission centroid of the primary image. A longer flare duration would cause the centroid track to move down again, one can see that at the end of the simulation the motion is just at its turning point.

Although it may be impossible to astrometrically resolve a flare like this in the near future, it is noteworthy that the smearing of the light source relatively strengthens the relativistic features in the centroid tracks. The motion of the overall centroid track is an indication of the secondary image, however, the shift is marginal.

\subsubsection{L-band flare from July 22, 2007}

\paragraph{Photometry}

Another fairly bright flare from Sgr~A* was caught on July 22, 2007 at the VLT under good observing conditions. There were a few short interruptions because of sky observations, causing some small gaps in the light curve data. The details about this observation can be found in \citet{Haubois}. Qualitatively, this light curve appears quite different from that of April 4. The major peaks are more prominent and the overall rise and decay of the whole flare light curve is suppressed. In particular, a wider separation of the peaks is noticeable, amounting to roughly $45$ minutes. In order to fit this flare one needs to assume a larger orbital radius of the orbiting material and a higher contribution of a compact emission region. To this end we use a similar hot spot model as before and find the best fit parameters, reported in Table~\ref{parameter_2207}. We used configurations with $r=2.0r_{\mathrm{ISCO}}$ and $a=0.52$, yielding an orbital period of $45.4$~min and simulated $5$ orbits.

\begin{table}[ht]
\newcommand\T{\rule{0pt}{2.6ex}}
\begin{center}
\begin{tabular}{lccc}
\tableline\tableline
Parameters: \T & Sphere & & Arc  \\
\tableline
Rise time \T $\sigma_{\mathrm{rise}}$ & $0.5P$ & & $4.0P$ \\
Decay time \T $\sigma_{\mathrm{decay}}$ & $1.0P$ & & $7.0P$  \\
Brightness ratio \T & 86 & / & 1 \\
Extension (FWHM) \T & - & & $1.24C$ \\
\tableline
\end{tabular}
\end{center}
\caption{Best fit parameters for the L-band flare from July 22, 2007. $P$~is the orbital period and $C=2\pi r$ the circumference of the hot spot orbit. \label{parameter_2207}}
\end{table}

\begin{figure}
 \resizebox{\hsize}{!}{\includegraphics[trim=15 0 0 0,clip]{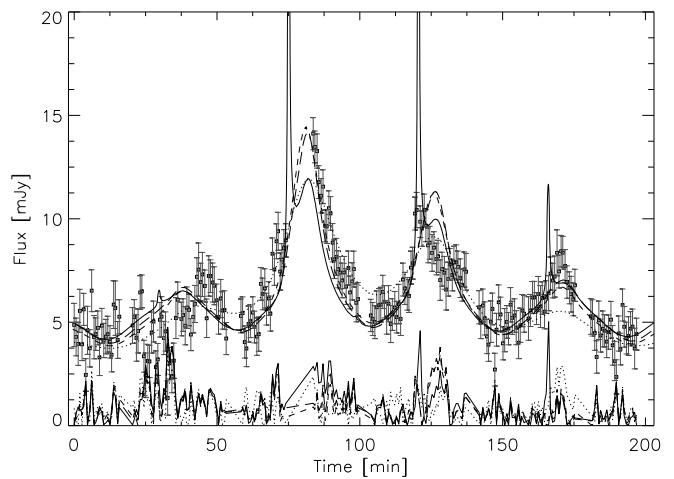}}
 \caption{Best fit light curves for the L-band flare from July 22, 2007 for four different inclinations ($i=20\degr$ (dotted), $50\degr$ (short-dashed), $70\degr$ (long-dashed), $90\degr$ (solid)). The gray error bars represent the measurement, at the bottom the residuals of the fits are drawn. \label{fit_2207}}
\end{figure}

The resulting best fit light curves are superimposed on the flare-plot in Fig.~\ref{fit_2207}. Because the sphere is dominant in this fit, the $90\degr$-case shows very pronounced Einstein ring peaks. Unfortunately, the strongest peak is not covered with data. The arc only makes a small contribution to the light curve flux and has the effect of broadening the wings of the major peaks. The arc's rise and decay time is so long that it barely varies in brightness. In contrast, the sphere must show stronger variations in order to reproduce the observed peaks. Although its rise and decay times seem to be very short in terms of the orbital period, they are comparable to the corresponding timescales of the April flare. This concordance indicates a fundamental heating- and cooling-process of the flares which does not change from one event to the other (such as a magnetic reconnection).

\begin{figure*}
 \resizebox{\hsize}{!}{\includegraphics{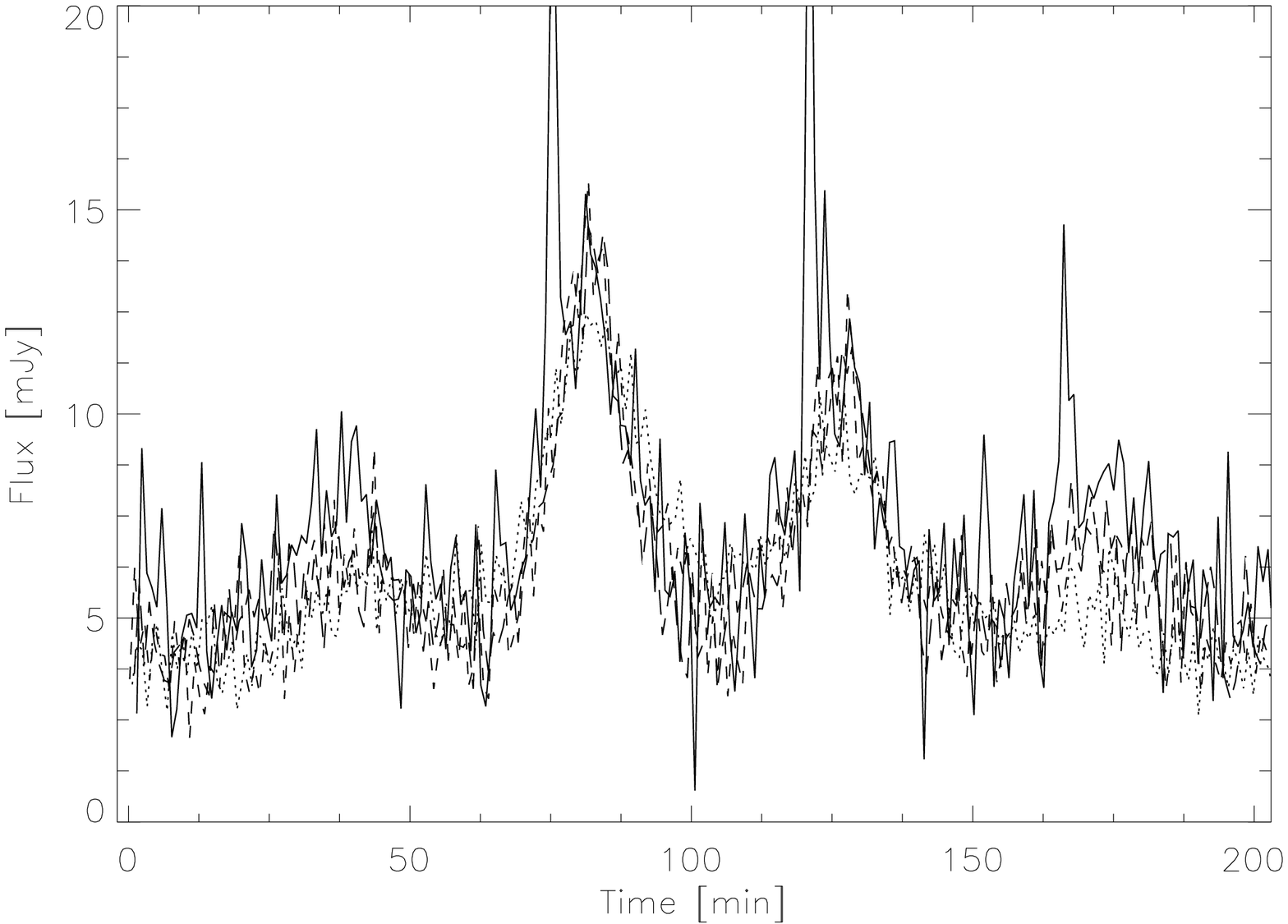}\includegraphics{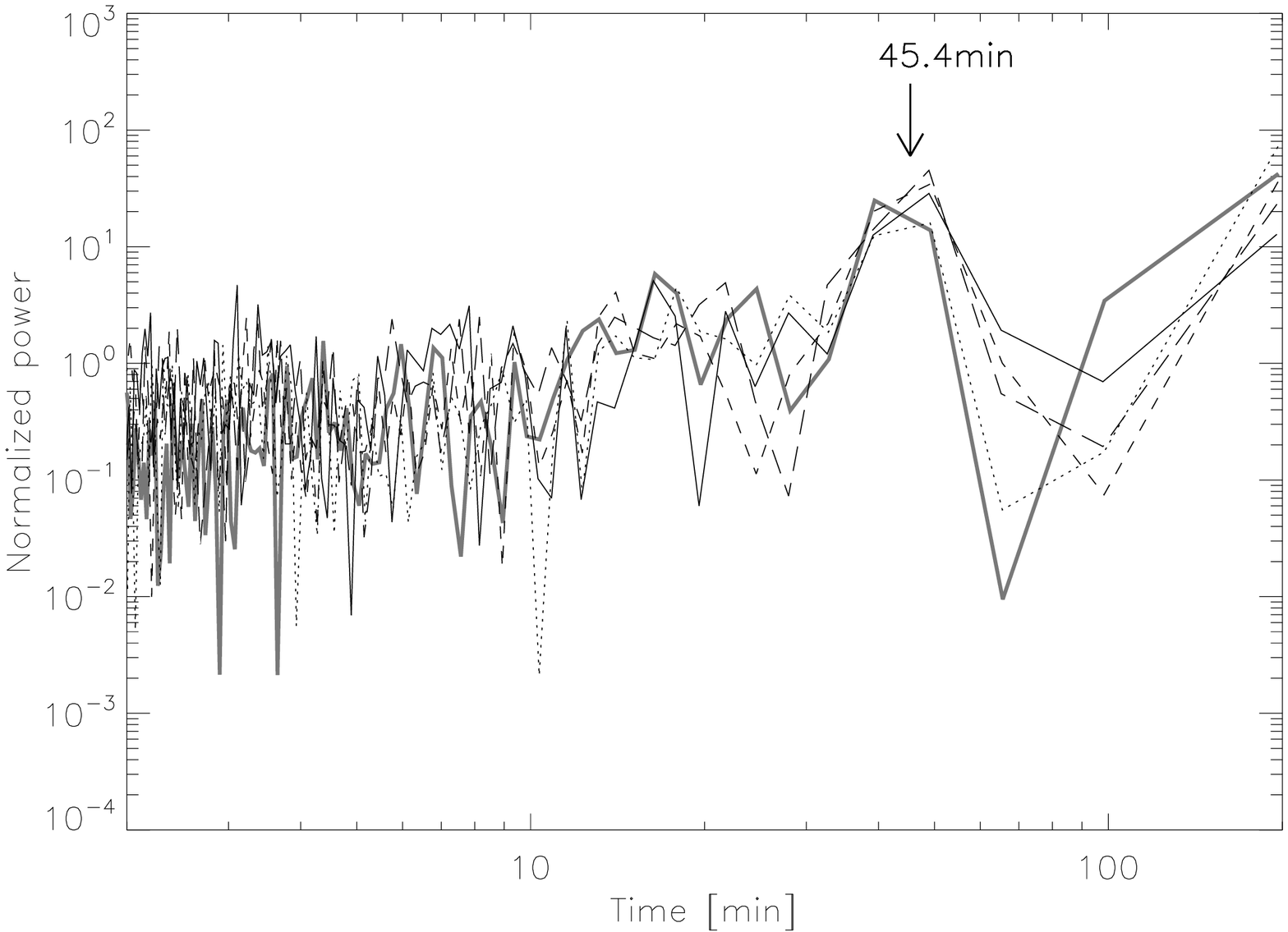}}
 \caption{Left: Best fit light curves from Fig.~\ref{fit_2207} with Gaussian noise added. Right: Power spectrum of these light curves (measurement in gray). \label{fit_2207_noise}}
\end{figure*}

A realistic simulated observation is shown in Fig.~\ref{fit_2207_noise} using an integration time of 50 seconds and a Gaussian noise with a standard deviation of 6 percent of the flare-peak-flux (average values from the data). The corresponding power spectrum is also given in that figure. There is quite a good agreement between the simulations and the observation. The power in the characteristic timescale (orbital period) exceeds the one in the April flare. This is due to the concentrated emission of the compact sphere in the center of the emission region, which produces more accentuated peaks in the light curves. Here, also the intermediate timescales yield a good agreement. The slightly too high power in the upper frequencies again suggests an overestimation of the errors in the measurement.

\paragraph{Astrometry}

\begin{figure*}
 \resizebox{\hsize}{!}{\includegraphics{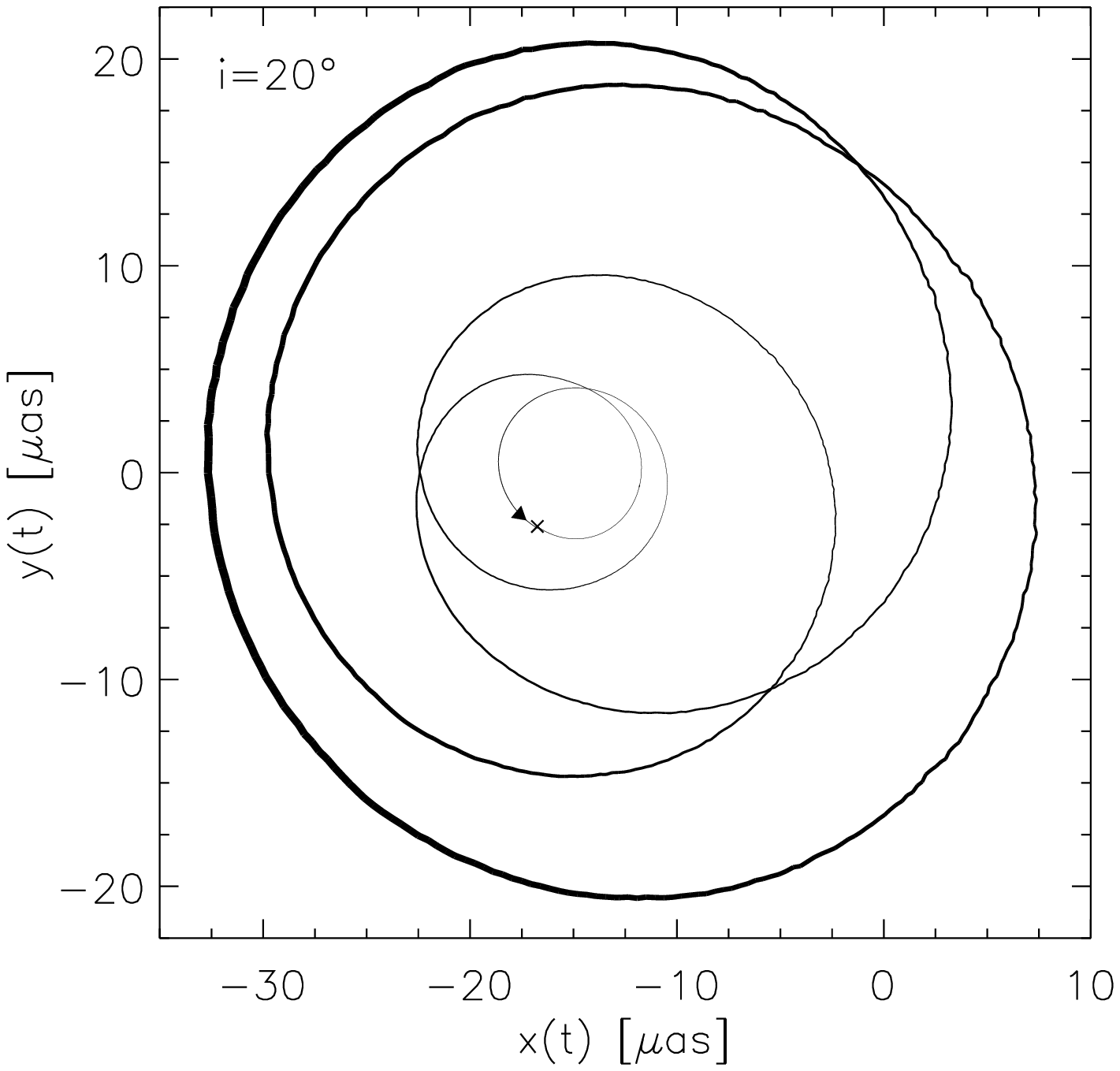}\includegraphics{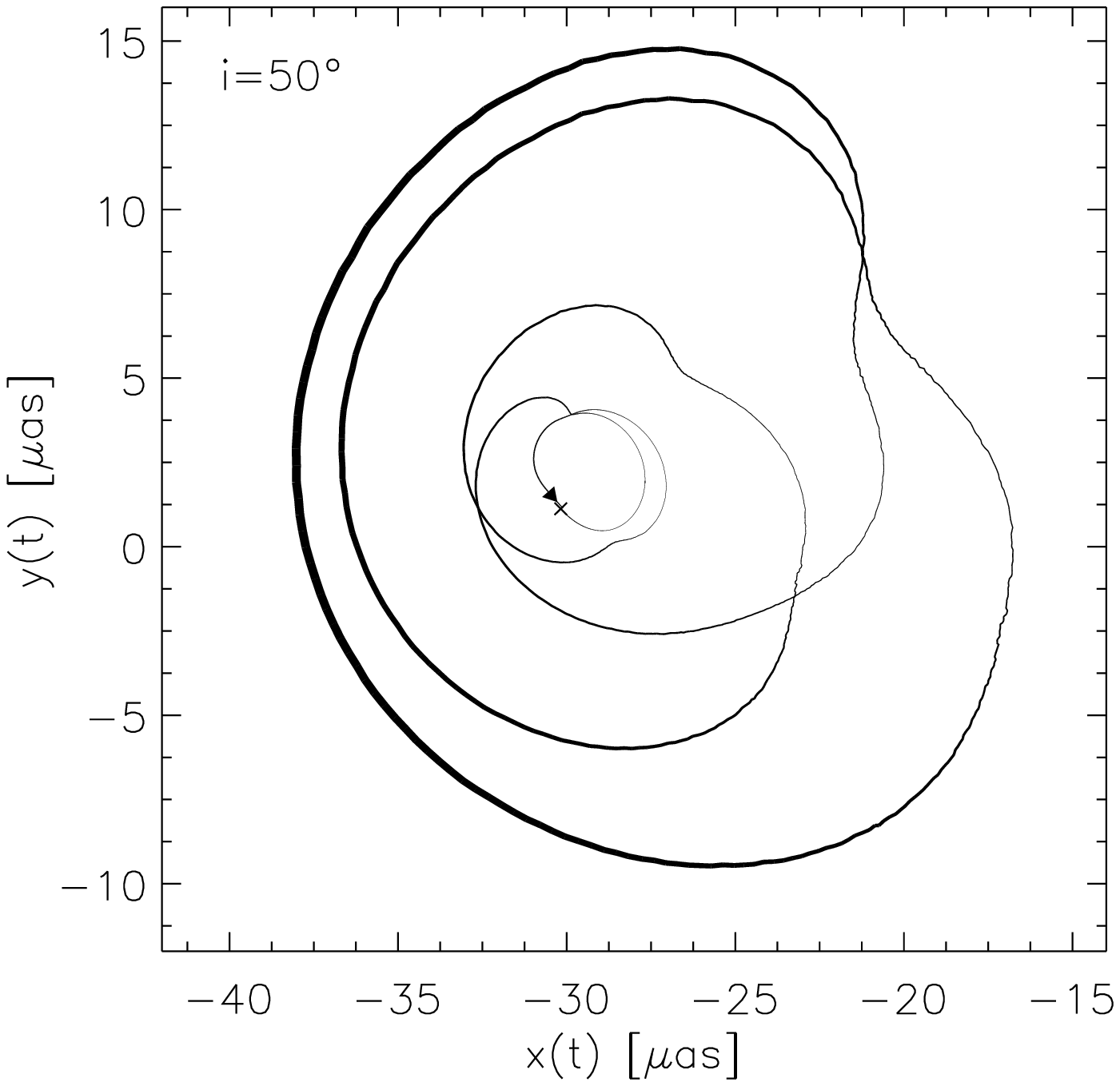}\includegraphics{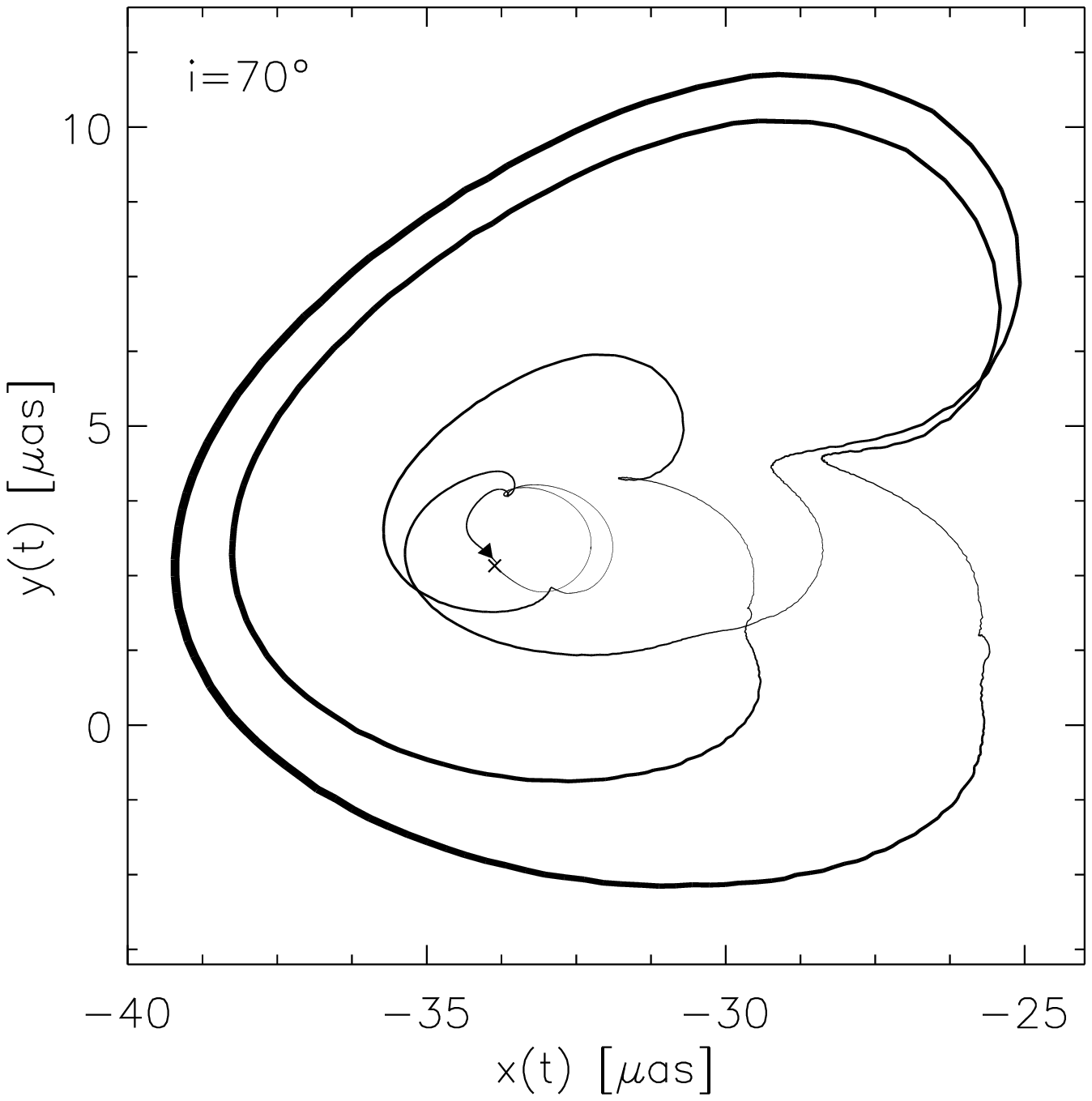}\includegraphics{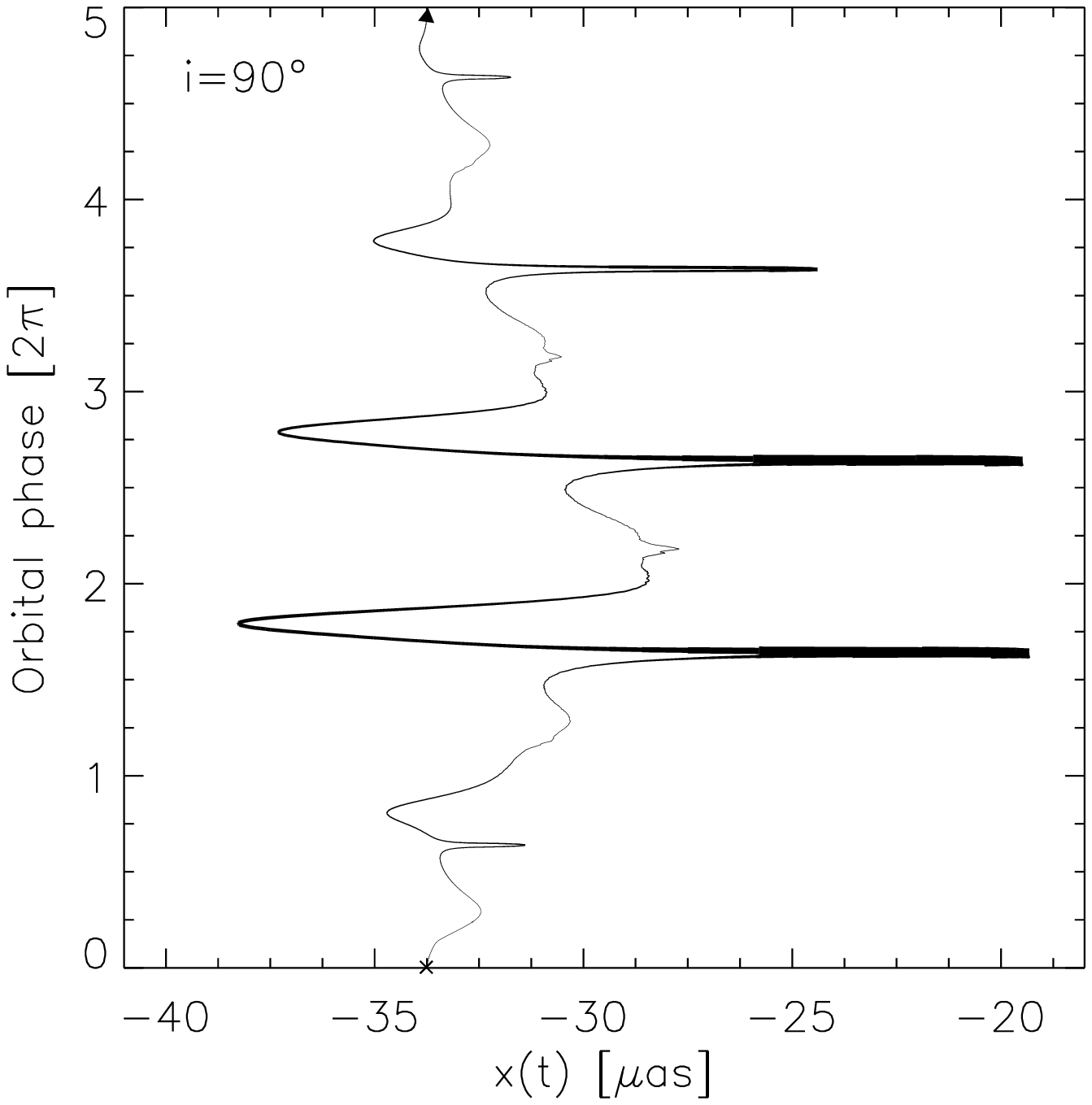}}
 \caption{Centroid tracks of the best fit model for the L-band flare from July 22, 2007 for four different inclinations. The brightness evolution of the flare is indicated by the line thickness (the thicker the line the brighter the flare). A cross marks the beginning, an arrow the end of the flare. \label{fit_centroid_2207}}
\end{figure*}

The astrometric motion of the centroid, as it results from this fit, yields better prospects regarding future measurements than the April flare. As can be seen in Fig.~\ref{fit_centroid_2207}, diameters of up to $40~\mu$as are reached in the centroid tracks. This is due to the bright compact component in this model and the larger orbital radius of the hot spot. Because the sphere's brightness varies strongly, the centroid deflection grows first and then shrinks again, until it finally reaches its starting point.

As for the light curves, observations of the centroid motion as they might be possible in future experiments can be simulated. For these we again integrate every 50 seconds and add Gaussian noise to both the x- and y-component of the centroid. Its standard deviation is chosen to be $10~\mu$as, the desired resolution of ground-based interferometric instruments like GRAVITY.

\begin{figure*}
 \resizebox{\hsize}{!}{\includegraphics{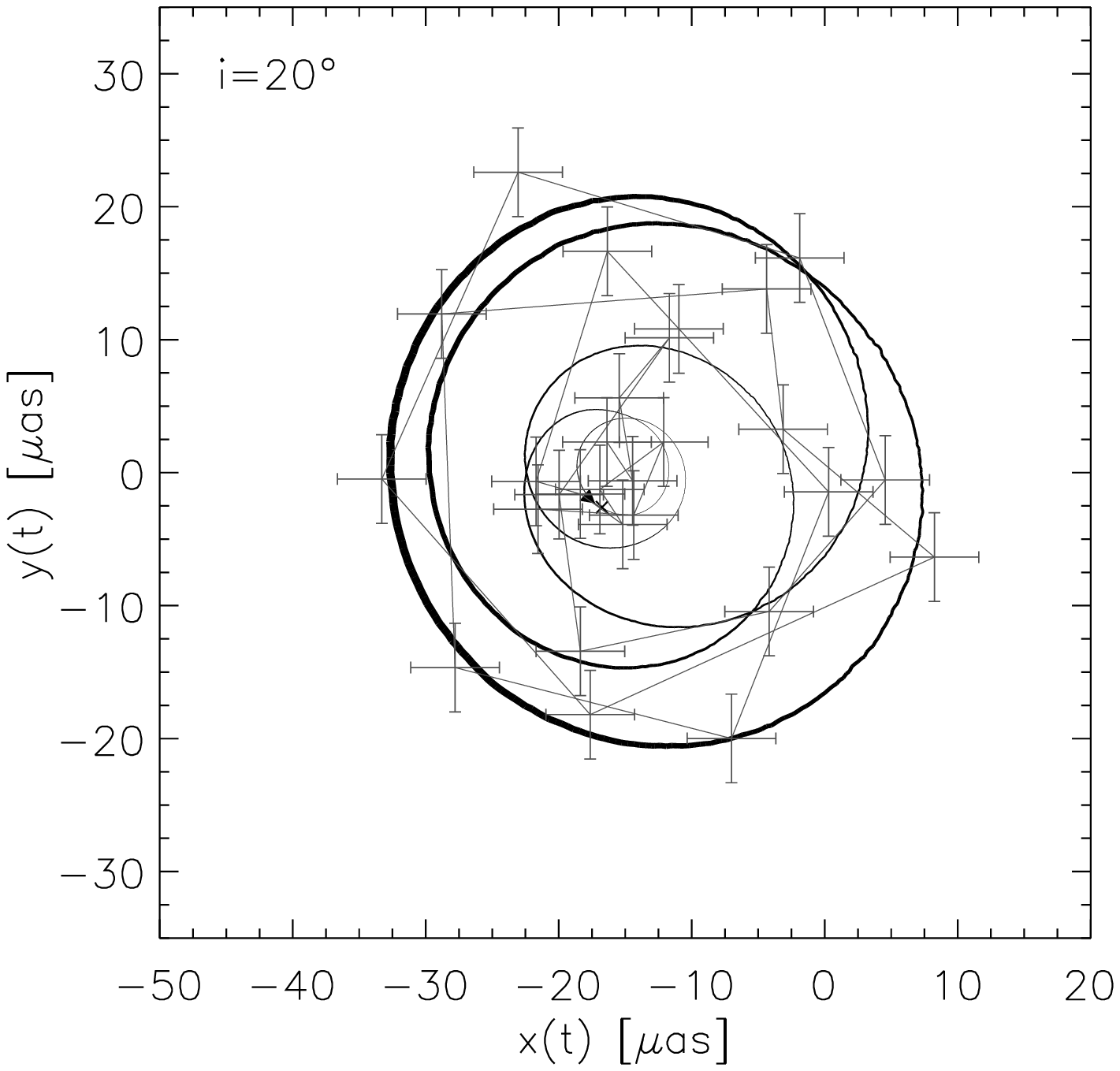}\includegraphics{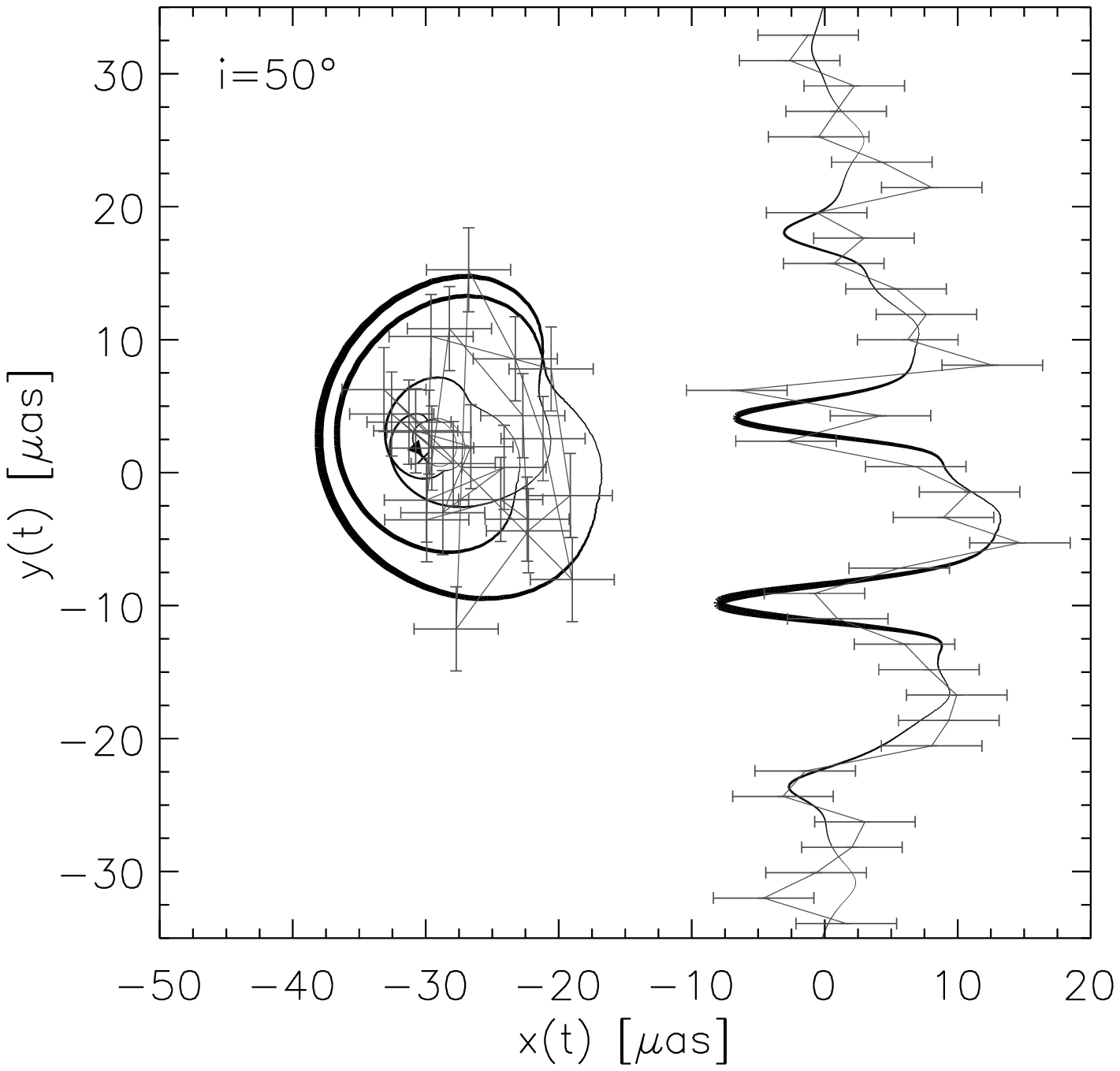}}
 \resizebox{\hsize}{!}{\includegraphics{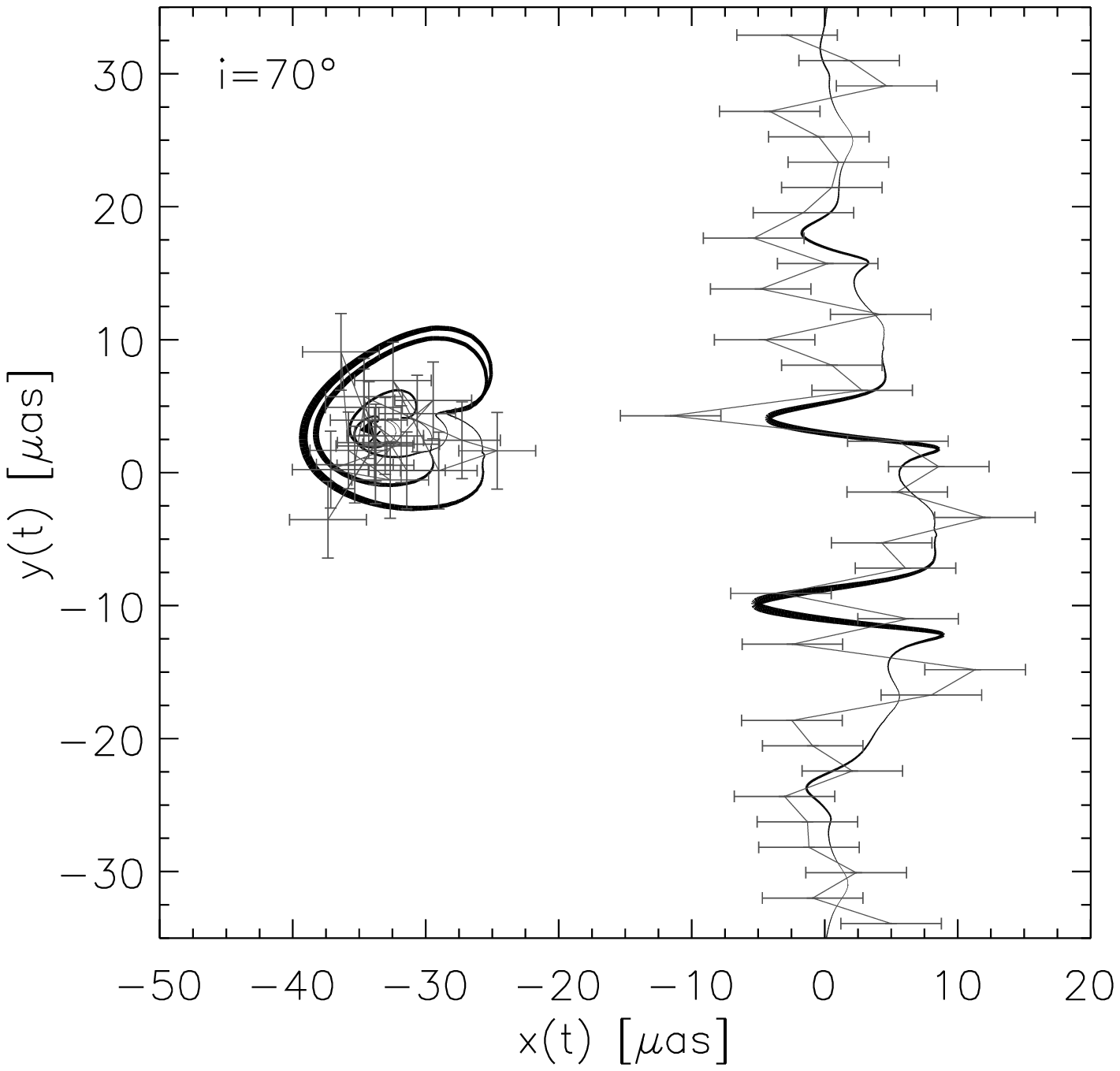}\includegraphics{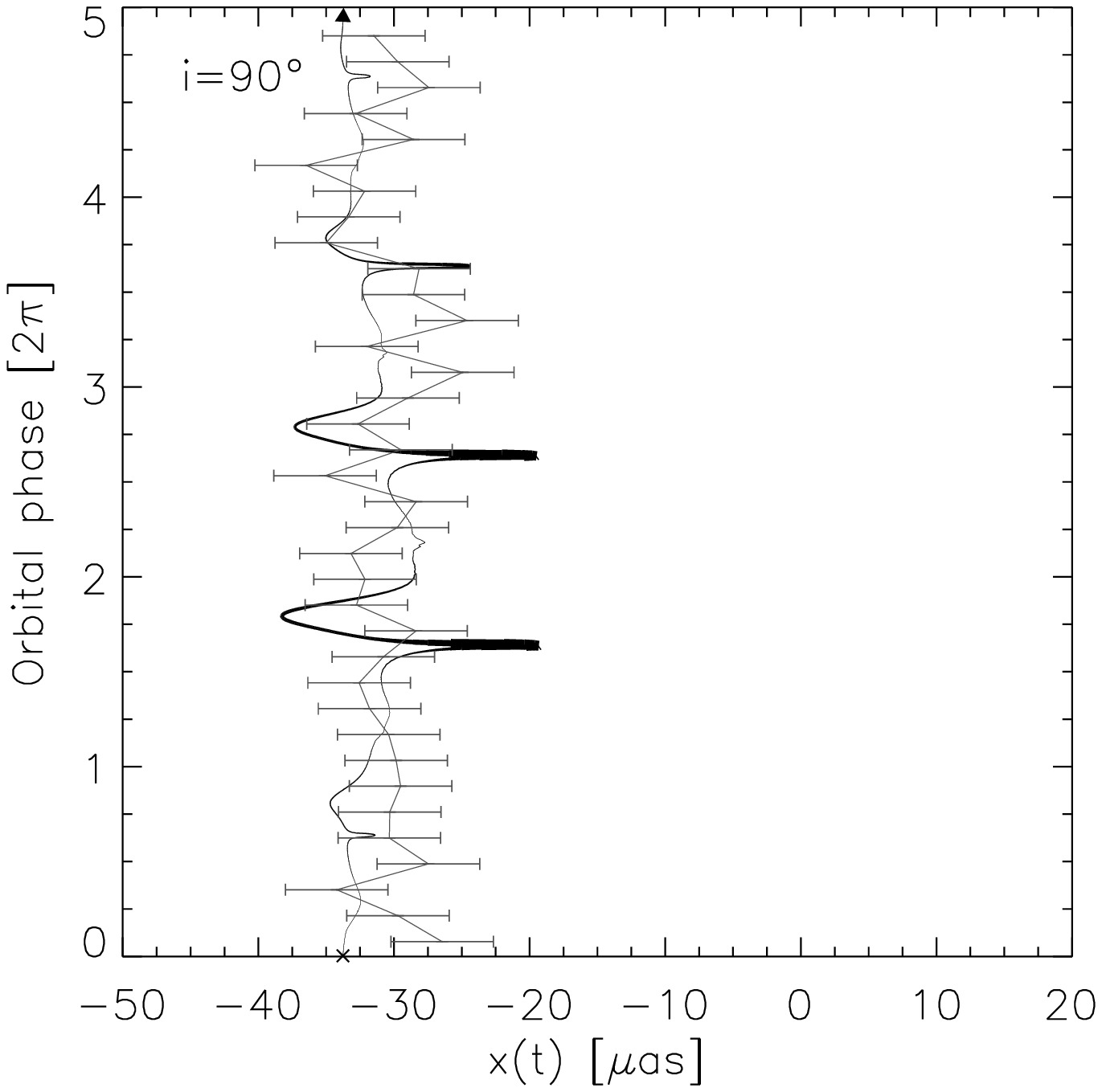}}
 \caption{Centroid tracks from Fig.~\ref{fit_centroid_2207}, superimposed by simulated data points as expected to be obtained with GRAVITY. The cases of $50\degr$ and $70\degr$ inclination additionally contain the $x(t)$ versus $\phi(t)$ motion of the centroid shifted to $x(0)=0$. \label{fit_centroid_2207_noise}}
\end{figure*}

Fig.~\ref{fit_centroid_2207_noise} displays the data points obtained from this procedure. In order to shrink the error bars, it is useful to average over several successive data points (here about 10). In the cases of $50\degr$ and $70\degr$ inclination the plots additionally contain the $x(t)$ versus $\phi(t)$ motion of the centroid (shifted to $x(0)=0$) to visualize its deflection more clearly.

It appears that the centroid tracks at higher inclinations are harder to catch. This is due to the increasing beaming effects, dragging the centroid towards the approaching node. Nevertheless, significant deflections of the centroid can be measured in most cases. A comparison of many different flare events could further constrain the configuration of hot spot and black hole.

\section{DISCUSSION}

The hot spot model we applied in this study is in good agreement with the observed light curves and periodograms from Sgr~A* and bears good prospects for the successful detection of a centroid motion with the GRAVITY-experiment. However, some assumptions of this model still need to be justified.

Compared to other authors \citep{bl06a,Meyer06b} we used a rather small and distinct hot spot component for the central emission region (following the estimate of $r<0.3\Rs$ by \citet{gillessen06}). The resulting light curves and centroid tracks with only this component compare well to the results obtained by \citet{bl06a,bl06b}, except for the fact that the features are more distinct due to the smaller spot size.

The implementation of an arc is motivated by the broad sub-flare structures, which cannot be fitted well with a single spot, but an extended emission region. Thus, only the combination of these two components allows a reasonable fit to the observed data. We did not implement any disk component, since it does not contribute any dynamical information to the simulated data and can be subtracted in the optically thin regime \citep[see e.g.][]{bl06b}.

Although relativistic hot spot models and positional displacements in the context of Sgr~A* have been studied deeply in the recent past, observational data has never been applied to constrain the astrometry. In this respect, this work is unique.

Furthermore it needs to be mentioned that the VLTI measurements will presumably be carried out in K-band. Since this study only covers L-band observations it might lead to a different picture. However, observations by \citet{hornstein07} have shown that light curves from Sgr~A* are well correlated at both wavelengths. Also theoretical SED-models of the flares suggest a strong correlation between K- and L-band flux due to the rather broad synchrotron bump created by a population of heated electrons in the magnetic reconnection \citep{yuan03}. Therefore, the predictions for the astrometric deflections from the light curves are independent of the wavelength band (K or L).

It is being argued in recent publications which kind of physical processes could generate the observed profiles in the PSD and how to distinguish between a pure red noise process and a distinct Keplerian signal. \citet{Do08} investigate the NIR flare light curves observed with the Keck telescope using Monte Carlo simulations. They claim to find no significant periodicity at any timescale probed and that periodic oscillations with amplitudes greater than 20 percent of the flare peak flux do not exist. Instead, their observations appear to be consistent with a red noise process with a power law slope between 2.0 and 3.0. Similar results are attained by \citet{Meyer08} who use data taken with both the Keck and the VLT, whereas \citet{eckart08b} interpret the variations of sub-structures in their recently observed flare data as variation of the hot spot structure due to differential rotation.

The quality of the latest flare data is still not considered good enough to unambiguously rule out a random noise process in favor of a rotating hot spot model or vice versa. Nevertheless, it is remarkable how well our simulated power spectra fit the measured ones. The basic assumption of a periodically moving object in our model does not produce a significantly higher power in the orbital timescale than the observed characteristic timescale. Also the rise towards longer timescales below the orbital period is present. The reason for this is presumably the slightly non-sinusoidal nature of the flux-variations, as apparent from Fig.~\ref{radii_incl_s0}.

Although our model is in a way idealized and more complex scenarios such as inwards spiraling spots and hydrodynamic instabilities in the accretion flow might also exist, there may still be a fair chance that single bright hot spots can dominate the emission from Sgr~A* for several hours. In this case the hot spot model could be tested with astrometric observations as planned by the GRAVITY-experiment.

\section{CONCLUSIONS}

We have used Sgr~A* flare observations in the L-band to test the hot spot model as a physical origin for the observed quasi-periodic oscillations. In order to predict astrometric properties of the flares we developed a model for the hot spot emission and applied general relativistic ray tracing simulations to visualize its appearance for a remote observer.

In the case of a confined spherical emission-region (small, constant source) various relativistic effects can be studied in detail. Both light curves and centroid tracks show strong deviations from the expectations of Newtonian physics, especially if the hot spot is close to the black hole and its orbit is highly inclined. In such configurations the occurrence of Einstein rings and multiple images can be inferred from substructure in the data. Furthermore, we presented a new approach to estimate the parameters of the hot spot model using astrometric information. However, this model is rather suited to study the fundamental properties of these parameters in a qualitative manner, not to fit the observations quantitatively.

The use of polarizing filters for the analysis of centroid tracks in principle enhances the signatures caused by multiple images. However, due to the low degree of polarization in the flares and the little resulting flux in the individual polarizing channels, a measurement of this type is probably not very promising yet.

A more advanced model for the hot spot assumes an elongation of the compact source with a certain temporal evolution (extended, variable source). The parameters of this model can be adjusted in such a manner that the observed light curves from Sgr~A* can be reproduced fairly well. An analysis of the simulated light curves in frequency space yields a good agreement with the observed data.

In view of the observational data at hand (light curves) our model can only constrain the orbital period and the heating and cooling timescales unambiguously. Other parameters of the hot spot model, such as the orbital radius, the inclination, the arc-extension, or the spin of the black hole, are degenerate. Some of these degeneracies can be broken with astrometric data, such as the degeneracy in orbital radius, inclination and black hole spin. Nevertheless, the impact of the arc-extension and the brightness-ratio between sphere and arc remains to be disentangled from the orbital parameters. Future high quality polarimetric measurements could help in this respect.

The centroid tracks from the best fit models are encouraging for future high resolution measurements. In particular the July 2007 flare announces the possibility to detect a centroid motion. The future interferometer GRAVITY at the VLTI will most likely be able to resolve proper motion of individual flares. Such a measurement would not only further constrain the hot spot model, but also probe spacetime in the strong field limit of general relativity. Thus, our simulations provide theoretical predictions from fundamental physics that can be tested in the near future.

\acknowledgements
This work received the support of PHASE, the high angular resolution partnership between ONERA, Observatoire de Paris, CNRS and University Denis Diderot Paris 7. T.P. wants to thank all his friends and colleagues, who, often unknowingly, helped him pass peribothron.


\begin{thebibliography}

\bibitem[Baganoff et al.(2001)]{baganoff01} Baganoff, F.~K., et al.\ 2001, \nat, 413, 45 

\bibitem[Baganoff et al.(2003)]{baganoff03} ---------. 2003, \apj, 591, 891

\bibitem[Balbus \& Hawley(1991)]{Balbus} Balbus, S.~A., \& Hawley, J.~F.\ 1991, \apj, 376, 214

\bibitem[Bao et al.(1994)]{bao94} Bao, G., Hadrava, P., \& Ostgaard, E.\ 1994, \apj, 425, 63

\bibitem[Bardeen et al.(1972)]{Bardeen} Bardeen, J.~M., Press, W.~H., \& Teukolsky, S.~A.\ 1972, \apj, 178, 347

\bibitem[Bower et al.(2004)]{bower04} Bower, G.~C., Falcke, H., Herrnstein, R.~M., Zhao, J.-H., Goss, W.~M., \& Backer, D.~C.\ 2004, Science, 304, 704

\bibitem[Bower et al.(2005{\natexlab{a}})]{bower05a} Bower, G.~C., Falcke, H., Wright, M.~C., \& Backer, D.~C.\ 2005{\natexlab{a}}, \apjl, 618, L29

\bibitem[Bower(2005{\natexlab{b}})]{bower05b} Bower, G.~C.\ 2005{\natexlab{b}}, Future Directions in High Resolution Astronomy, 340, 247

\bibitem[Broderick \& Loeb(2005)]{bl05} Broderick, A.~E., \& Loeb, A.\ 2005, \mnras, 363, 353

\bibitem[Broderick \& Loeb(2006{\natexlab{a}})]{bl06a} ---------. 2006{\natexlab{a}}, \apjl, 636, L109

\bibitem[Broderick \& Loeb(2006{\natexlab{b}})]{bl06b} ---------. 2006{\natexlab{b}}, \mnras, 367, 905

\bibitem[Cl{\'e}net et al.(2005)]{clenet05} Cl{\'e}net, Y., Rouan, D., Gratadour, D., Marco, O., L{\'e}na, P., Ageorges, N., \& Gendron, E.\ 2005, \aap, 439, L9

\bibitem[Cunningham \& Bardeen(1973)]{cunningham73} Cunningham, J.~M., \& Bardeen, C.~T.\ 1973, \apj, 183, 237

\bibitem[Davies et al.(1976)]{davies76} Davies, R.~D., Walsh, D., \& Booth, R.~S.\ 1976, \mnras, 177, 319

\bibitem[De Villiers et al.(2003)]{devilliers03} De Villiers, J.-P., Hawley, J.~F., \& Krolik, J.~H.\ 2003, \apj, 599, 1238

\bibitem[Delplancke et al.(2004)]{primaRM} Delplancke, F. et al.\ 2004, Reference Missions for PRIMA, ESO STC-362
    
\bibitem[Do et al.(2008)]{Do08} Do, T., Ghez, A.~M., Morris, M.~R., Yelda, S., Meyer, L., Lu, J.~R., Hornstein, S.~D., \& Matthews, K.\ 2008, arXiv:0810.0446   

\bibitem[Dodds-Eden et al.(in preparation)]{dodds-eden08}
{Dodds-Eden, K. et al.}, in preparation

\bibitem[Doeleman \& Bower(2004)]{Doeleman04} Doeleman, S., \& Bower, G.\ 2004, Galactic Center Newsletter, 18, 6

\bibitem[Doeleman et al.(2008)]{Doeleman08} Doeleman, S., et al.\ 2008, \nat, 455, 78

\bibitem[Eckart et al.(2004)]{eckart04} Eckart, A., et al.\ 2004, \aap, 427, 1

\bibitem[Eckart et al.(2006)]{eckart06} Eckart, A., Sch{\"o}del, R., Meyer, L., Trippe, S., Ott, T., \& Genzel, R.\ 2006, \aap, 455, 1 
  
\bibitem[Eckart et al.(2008{\natexlab{a}})]{eckart08a} Eckart, A., et al.\ 2008{\natexlab{a}}, \aap, 479, 625 

\bibitem[Eckart et al.(2008{\natexlab{b}})]{eckart08b} ---------. 2008{\natexlab{b}}, arXiv:0810.0168 

\bibitem[Eisenhauer et al.(2005{\natexlab{a}})]{eisenhauer05sinfoni} Eisenhauer, F., et al.\ 2005{\natexlab{a}}, \apj, 628, 246

\bibitem[Eisenhauer et al.(2005{\natexlab{b}})]{eisenhauer05gravity} Eisenhauer, F., Perrin, G., Rabien, S., Eckart, A., Lena, P., Genzel, R., Abuter, R., \& Paumard, T.\ 2005{\natexlab{b}}, Astronomische Nachrichten, 326, 561

\bibitem[Falanga et al.(2007)]{Falanga07} Falanga, M., Melia, F., Tagger, M., Goldwurm, A., \& B{\'e}langer, G.\ 2007, \apjl, 662, L15

\bibitem[Genzel et al.(2003)]{genzel03nat} Genzel, R., Sch{\"o}del, R., Ott, T., Eckart, A., Alexander, T., Lacombe, F., Rouan, D., \& Aschenbach, B.\ 2003, \nat, 425, 934

\bibitem[Ghez et al.(2003)]{ghez03} Ghez, A.~M., et al.\ 2003, \apjl, 586, L127

\bibitem[Ghez et al.(2004)]{ghez04} ---------. 2004, \apjl, 601, L159

\bibitem[Gillessen et al.(2006)]{gillessen06} Gillessen, S., et al.\ 2006, Journal of Physics Conference Series, 54, 411

\bibitem[Glindemann \& L{\'e}v{\^e}que(2000)]{glindemann00} Glindemann, A., \& L{\'e}v{\^e}que, S.\ 2000, From Extrasolar Planets to Cosmology: The VLT Opening Symposium, 468

\bibitem[Haubois et al.(in preparation)]{Haubois}
{Haubois, X. et al.}, in preparation

\bibitem[Hornstein et al.(2007)]{hornstein07} Hornstein, S.~D., Matthews, K., Ghez, A.~M., Lu, J.~R., Morris, M., Becklin, E.~E., Rafelski, M., \& Baganoff, F.~K.\ 2007, \apj, 667, 900

\bibitem[Kling et al.(2000)]{kling00} Kling, T.~P., Newman, E.~T., \& Perez, A.\ 2000, \prd, 61, 104007

\bibitem[Krabbe et al.(2006)]{Krabbe06} Krabbe, A., Iserlohe, C., Larkin, J.~E., Barczys, M., McElwain, M., Weiss, J., Wright, S.~A., \& Quirrenbach, A.\ 2006, \apjl, 642, L145

\bibitem[Lomb(1976)]{lomb76} Lomb, N.~R.\ 1976, \apss, 39, 447

\bibitem[Luminet(1979)]{luminet79} Luminet, J.-P.\ 1979, \aap, 75, 228

\bibitem[Macquart \& Bower(2006)]{macquart06} Macquart, J.-P., \& Bower, G.~C.\ 2006, \apj, 641, 302

\bibitem[Markoff et al.(2001)]{markoff01} Markoff, S., Falcke, H., Yuan, F., \& Biermann, P.~L.\ 2001, \aap, 379, L13

\bibitem[Marrone et al.(2006)]{marrone06} Marrone, D.~P., Moran, J.~M., Zhao, J.-H., \& Rao, R.\ 2006, Journal of Physics Conference Series, 54, 354

\bibitem[Mauerhan et al.(2005)]{mauerhan05} Mauerhan, J.~C., Morris, M., Walter, F., \& Baganoff, F.~K.\ 2005, \apjl, 623, L25

\bibitem[Meyer et al.(2006{\natexlab{a}})]{Meyer06a} Meyer, L., Sch{\"o}del, R., Eckart, A., Karas, V., Dov{\v c}iak, M., \& Duschl, W.~J.\ 2006{\natexlab{a}}, \aap, 458, L25 

\bibitem[Meyer et al.(2006{\natexlab{b}})]{Meyer06b} Meyer, L., Eckart, A., Sch{\"o}del, R., Duschl, W.~J., Mu{\v z}i{\'c}, K., Dov{\v c}iak, M., \& Karas, V.\ 2006{\natexlab{b}}, \aap, 460, 15

\bibitem[Meyer et al.(2007)]{Meyer07} Meyer, L., Sch{\"o}del, R., Eckart, A., Duschl, W.~J., Karas, V., \& Dov{\v c}iak, M.\ 2007, \aap, 473, 707 

\bibitem[Meyer et al.(2008)]{Meyer08} Meyer, L., Do, T., Ghez, A., Morris, M.~R., Witzel, G., Eckart, A., Belanger, G., \& Schoedel, R.\ 2008, arXiv:0809.2580 

\bibitem[Misner et al.(1973)]{Misner} Misner, C.~W., Thorne, K.~S., \& Wheeler, J.~A.\ 1973, San Francisco: W.H.~Freeman and Co., 1973

\bibitem[Miyazaki et al.(2004)]{miyazaki04} Miyazaki, A., Tsutsumi, T., \& Tsuboi, M.\ 2004, \apjl, 611, L97

\bibitem[M{\"u}ller(2006)]{mueller06} M{\"u}ller, T. 2006, PhD thesis, Eberhard-Karls-Universit\"at T\"ubingen

\bibitem[Nayakshin et al.(2004)]{nayakshin04} Nayakshin, S., Cuadra, J., \& Sunyaev, R.\ 2004, \aap, 413, 173

\bibitem[Paumard et al.(2005)]{paumard05vltiws} Paumard, T., et al.\ 2005, Astronomische Nachrichten, 326, 568

\bibitem[Paumard et al.(2006)]{paumard06disks} ---------. 2006, Journal of Physics Conference Series, 54, 199

\bibitem[Porquet et al.(2003)]{Porquet03} Porquet, D., Predehl, P., Aschenbach, B., Grosso, N., Goldwurm, A., Goldoni, P., Warwick, R.~S., \& Decourchelle, A.\ 2003, \aap, 407, L17

\bibitem[Porquet et al.(2008)]{Porquet08} Porquet, D., et al.\ 2008, \aap, 488, 549 

\bibitem[Quataert(2003)]{Quataert03} Quataert, E.\ 2003, 
Astronomische Nachrichten Supplement, 324, 435 

\bibitem[Reid et al.(2007)]{reid07} Reid, M.~J., Menten, K.~M., Trippe, S., Ott, T., \& Genzel, R.\ 2007, \apj, 659, 378

\bibitem[Reid et al.(2008)]{reid08} Reid, M.~J., Broderick, A.~E., Loeb, A., Honma, M., \& Brunthaler, A.\ 2008, \apj, 682, 1041

\bibitem[Scargle(1982)]{scargle82} Scargle, J.~D.\ 1982, \apj, 263, 835

\bibitem[Schnittman et al.(2006)]{Schnittman06} Schnittman, J.~D., Krolik, J.~H., \& Hawley, J.~F.\ 2006, \apj, 651, 1031 

\bibitem[Sch{\"o}del et al.(2002)]{schoedel02} Sch{\"o}del, R., et al.\ 2002, \nat, 419, 694

\bibitem[Sch{\"o}del et al.(2003)]{schoedel03} Sch{\"o}del, R., Ott, T., Genzel, R., Eckart, A., Mouawad, N., \& Alexander, T.\ 2003, \apj, 596, 1015

\bibitem[Shao \& Colavita(1992)]{shao92} Shao, M., \& Colavita, M.~M.\ 1992, \aap, 262, 353

\bibitem[Shen et al.(2005)]{shen05} Shen, Z.-Q., Lo, K.~Y., Liang, M.-C., Ho, P.~T.~P., \& Zhao, J.-H.\ 2005, \nat, 438, 62

\bibitem[Shen(2008)]{shen08} Shen, Z.-Q.\ 2008, Astrophysics of Compact Objects, 968, 340

\bibitem[Tagger \& Melia(2006)]{tagger06} Tagger, M., \& Melia, F.\ 2006, \apjl, 636, L33

\bibitem[Townes et al.(1999)]{townes99} Townes, C.~H., Stolovy, S.~R., \& Falcke, H.\ 1999, The Central Parsecs of the Galaxy, 186, 611

\bibitem[Trippe et al.(2007)]{trippe07} Trippe, S., Paumard, T., Ott, T., Gillessen, S., Eisenhauer, F., Martins, F., \& Genzel, R.\ 2007, \mnras, 375, 764

\bibitem[Uzdensky(2005)]{uzdensky05} Uzdensky, D.~A.\ 2005, \apj, 620, 889

\bibitem[Weiskopf(2000)]{weiskopf00} Weiskopf, D. 2000, IEEE Visualization 2000 Proceedings (ACM Press), 445--448

\bibitem[Weiskopf(2001)]{weiskopf01} Weiskopf, D. 2001, PhD thesis, Eberhard-Karls-Universit\"at T\"ubingen

\bibitem[Wilkins(1972)]{wilkins72} Wilkins, D.~C.\ 1972, \prd, 5, 814

\bibitem[Yuan et al.(2003)]{yuan03} Yuan, F., Quataert, E., 
\& Narayan, R.\ 2003, \apj, 598, 301 

\bibitem[Yuan et al.(2004)]{yuan04} Yuan, F., Quataert, E., \& Narayan, R.\ 2004, \apj, 606, 894

\bibitem[Yusef-Zadeh et al.(2006)]{Yusef06} Yusef-Zadeh, F., Roberts, D., Wardle, M., Heinke, C.~O., \& Bower, G.~C.\ 2006, \apj, 650, 189

\bibitem[Zhao et al.(2001)]{zhao01} Zhao, J.-H., Bower, G.~C., \& Goss, W.~M.\ 2001, \apjl, 547, L29

\end{thebibliography}
\end{document}